\documentclass[fleqn,usenatbib]{mnras}

\usepackage[T1]{fontenc}

\DeclareRobustCommand{\VAN}[3]{#2}
\let\VANthebibliography\thebibliography
\def\thebibliography{\DeclareRobustCommand{\VAN}[3]{##3}\VANthebibliography}

\usepackage{graphicx}	
\usepackage{amsmath}	
\usepackage{amssymb}	

\usepackage{newtxtext,newtxmath}

\newcommand{\OIII}{[\ion{O}{III}]}
\newcommand{\Ha}{H$\alpha$}
\newcommand{\Hb}{H$\beta$}
\newcommand{\Hg}{H$\gamma$}

\newcommand{\NeIII}{[\ion{Ne}{III}]}
\newcommand{\OII}{[\ion{O}{II}]}

\title[Smouldering galaxies at $5 < z < 8$]{Like a candle in the wind: The embers of once aflame, now smouldering galaxies at $5 < z < 8$}

\author[J.\@A.\@A.\@ Trussler et al.]{James A.\@ A.\@ Trussler,$^{1,2}$\thanks{E-mail: james.trussler@cfa.harvard.edu}
Christopher J. Conselice,$^{1}$
Nathan Adams,$^{1}$
Duncan Austin,$^{1}$
Joseph Caruana$^{3,4}$
\newauthor
Tom Harvey,$^{1}$
Qiong Li,$^{1}$
Christopher C.\@ Lovell,$^{5}$ 
Louise T.\@ C.\@ Seeyave,$^{6}$
Aswin P.\@ Vijayan$^{7,8}$
\newauthor 
and Stephen M.\@ Wilkins$^{6,3}$
\\
$^{1}$Jodrell Bank Centre for Astrophysics, University of Manchester, Oxford Road, Manchester M13 9PL, UK\\
$^{2}$Center for Astrophysics $|$ Harvard \& Smithsonian, 60 Garden St., Cambridge MA 02138 USA\\
$^{3}$Institute of Space Sciences and Astronomy, University of Malta, Msida MSD 2080, Malta\\
$^{4}$Department of Physics, University of Malta, Msida MSD 2080, Malta\\
$^{5}$Institute of Cosmology and Gravitation, University of Portsmouth, Burnaby Road, Portsmouth PO1 3FX, UK\\
$^{6}$Astronomy Centre, University of Sussex, Falmer, Brighton BN1 9QH, UK\\
$^{7}$Cosmic Dawn Center (DAWN)\\ 
$^{8}$DTU-Space, Technical University of Denmark, Elektrovej 327, DK-2800 Kgs. Lyngby, Denmark\\
}

\date{Accepted XXX. Received YYY; in original form ZZZ}

\pubyear{2025}

\begin{document}
\label{firstpage}
\pagerange{\pageref{firstpage}--\pageref{lastpage}}
\maketitle

\begin{abstract}
We develop a photometric search method for identifying smouldering galaxies at $5< z < 8$, which are defined to have weak emission lines and thus generally have low specific star formation rates and may even be in a state of (temporary) quiescence. The deep NIRCam imaging (${\sim}29.5$~AB~mag, 5$\sigma$) from the JADES second data release is essential for finding these systems, as they are faint, relatively quiescent dwarf galaxies ($M_* \sim 10^{8\text{--}9}~\mathrm{M}_\odot)$ in the Epoch of Reionisation (EoR). Moreover, medium band imaging is key, enabling a clear identification of the lack of emission lines in these galaxies, thus betraying their dormant flame. Owing to the young age of the Universe, combined with the likely bursty star formation in these first dwarf galaxies, conventional colour-selection methods like the UVJ diagram likely miss a large fraction of the quiescent population in the EoR. Indeed, we find that smouldering galaxies constitute a considerable fraction (0.05--0.35) of the EoR dwarf galaxy population ($M_* \sim 10^{8\text{--}9}~\mathrm{M}_\odot$). As predicted by simulations, these first dwarf galaxies are fragile, the star formation in their shallow potential wells easily snuffed out by feedback-driven winds triggered by secular or merger-driven starbursts, with the smouldering fraction increasing with decreasing stellar mass. Finally, we provide observational constraints on the smouldering galaxy comoving number density (${\sim}10^{-4}\text{--}10^{-5}$~dex$^{-1}$~Mpc$^{-3}$), which, although hampered by incompleteness, should aid in our understanding of the primordial baryon cycle, as current simulations greatly disagree on whether these systems are rare (${\sim}1\%$) or common (${\sim}50\%$) in the EoR.

\end{abstract}




\section{Introduction} \label{sec:intro}

\emph{JWST} is ushering in a new golden age of discovery, made possible by its excellent sensitivity, resolution and extensive instrumentation in the infrared. Indeed, the sensitivity of the Near Infrared Camera \citep[NIRCam,][]{Rieke2005}, building on \emph{Hubble}'s Wide Field Camera 3 (WFC3) and greatly surpassing \emph{Spitzer}'s Infrared Array Camera (IRAC), is now enabling the detection and characterisation of the rest-frame optical light in faint dwarf galaxies in the Epoch of Reionisation (EoR). It is precisely this light, complementary to the rest-frame ultraviolet light already reasonably well-traced by \emph{HST}, that encodes so much valuable additional information about the properties and star formation (or lack thereof) in high-redshift galaxies. Capturing both the nebular line emission \citep[tracing young stellar populations, e.g.\@][]{Cameron2023, Trussler2023, Curti2024, Roberts-Borsani2024}, as well as the Balmer break \citep[tracing older stellar populations, e.g.\@][]{Laporte2023, Desprez2024, Trussler2024, Vikaeus2024}, \emph{JWST} is transforming our understanding of the rise \citep[e.g.\@][]{Endsley2023, Rinaldi2023, Boyett2024, Cameron2024, Caputi2024, Simmonds2024, Cole2025} and fall \citep[e.g.\@][]{Looser2023b, Looser2024, Strait2023} of star formation in the first dwarf galaxies.

In addition to the great increase in sensitivity, NIRCam photometry can also provide additional information content compared to the \emph{Hubble Space Telescope} (\emph{HST}) + \emph{Spitzer} imaging that preceded it. This comes about because of the introduction of medium band filters in the NIR, which span a narrower wavelength range than the traditional wide band filters. These medium band filters provide two main benefits \citep[see e.g.\@][]{Davis2024}. First, they provide greater emission-line sensitivity, as the bandpass-averaged flux density in the narrower filters is more strongly affected by emission lines, i.e.\@ the average is more affected by outliers (emission lines) if there are less data points (narrower spectral range). Second, the medium band filters can provide a cleaner measurement of the continuum level as they can more effectively fit in the gaps between the prominent rest-frame optical lines (e.g.\@ \OIII\ $\lambda 5007$ and \Ha) than the corresponding wide bands, whose bandpass-averaged flux densities are generally affected by one or more prominent emission lines. Thus the inclusion of medium bands helps to break the degeneracy in the wide band photometry seen for e.g.\@ strong line-emitting galaxies and Balmer-break galaxies \citep{Laporte2023, Trussler2024}, providing a clear, unambiguous measurement of the emission line strength. In this way, extreme emission-line galaxies have been identified through the substantial differential in photometry between medium bands and their neighbouring wide band filters \citep[e.g.\@][]{Withers2023, Simmonds2024}.

On top of identifying strong line-emitting galaxies, \emph{JWST} has also been instrumental in uncovering the first quiescent galaxies in the Universe. First, the extensive infrared filter set and sensitive NIRCam and MIRI photometry greatly aid in identifying traditional massive, old quiescent galaxies, with the sensitivity of NIRSpec allowing for efficient spectroscopic confirmation of these systems \citep[e.g.\@][]{Alberts2023, Carnall2023, Carnall2023b, Valentino2023, Nanayakkara2024, Wang2024}. Second, the near-infrared sensitivity of NIRCam and NIRSpec enable the detection and characterisation of the newly-discovered population of quiescent dwarf galaxies in the EoR. This new observational era begins with the serendipitous spectroscopic discovery of a recently-quenched galaxy at $z = 7.29$ \citep[JADES-GS-z7-01-QU hereafter,][]{Looser2024} which was selected for deep (28~h) NIRSpec PRISM follow-up by the JADES GTO team based solely off its \emph{HST} + NIRCam photometry indicating that it was a robust high-redshift Lyman-break candidate. The subsequent spectroscopy revealed that it was much more than that, actually being the first observational example of a quiescent dwarf galaxy ($M_* \sim 10^{8.7}~\mathrm{M}_\odot$) in the EoR with no emission lines (the otherwise typically strong \OIII\ + \Hb\ being non-detected) and a minor Balmer break (${\sim}0.3$~mag). As pointed out by \citet{Looser2024}, this young quiescent galaxy (which they refer to as ``(mini-)quenched'', depending on if it is a temporarily/permanently quenched galaxy) would not be identified as passive using the traditional rest-frame UVJ colour diagram \citep{Williams2009} due to its weak Balmer break. Hence it is possible that there is a significant population of quiescent dwarf galaxies in the EoR \citep[see also the post-starburst, nearly quenched $z=5.2$, gravitationally-lensed $M_* = 10^{7.6}~\mathrm{M}_\odot$ dwarf galaxy,][]{Strait2023}, that traditional colour-based methodologies are unable to identify as such. 

Building on this, \citet{Looser2023b} investigate the star-formation histories of their sample of high-redshift galaxies with JADES NIRSpec PRISM continuum spectroscopy. By comparing the average star formation rates over the past 10~Myr (usually traced by Balmer line emission) to that over the past 100~Myr (usually traced by the UV continuum), they find that the star formation in high-redshift, low-mass dwarf galaxies is likely bursty, with some systems displaying elevated current star formation (likely analogous to some of the extreme emission-line galaxies), some with similar star formation over these timescales, and others displaying relatively low ongoing star formation (``lulling'' galaxies, with perhaps some being similar to JADES-GS-z7-01-QU). This is in line with the predictions from simulations \citep[SERRA and IllustrisTNG from][respectively]{Gelli2023, Dome2024}, which find that high-redshift dwarf galaxies are fragile, the star formation in their shallow potential wells being easily snuffed out by feedback-driven winds triggered by secular or merger-driven starbursts. While these simulations tend to agree that the quiescent fraction increases with decreasing stellar mass in the dwarf galaxy regime, they currently tend to greatly disagree on whether these systems are rare \citep[${\sim}1\%$,][]{Dome2024} or common \citep[${\sim}50\%$,][]{Gelli2023} in the EoR. Thus observational constraints on the abundance of this quiescent dwarf galaxy population can improve our understanding of the primordial baryon cycle, informing the next-generation of simulations of galaxy formation and evolution in the EoR.

Furthermore the deep NIRSpec continuum spectroscopy of JADES-GS-z7-01-QU also places valuable constraints on its star-formation history (SFH). This can yield further insights on star formation and feedback processes in these first dwarf galaxies, as theoretical models are required to reproduce the inferred histories within a physically motivated framework that describes the gas flows in these systems. Utilising the SED-fitting code {\tt Bagpipes} \citep{Carnall2018}, \citet{Looser2024} infer JADES-GS-z7-01-QU to have undergone a short ${\sim}$20~Myr burst of star formation, followed by abrupt quenching ${\sim}$20~Myr prior to the epoch of observation. Using the SERRA cosmological zoom-in simulations and an analytical model, respectively, \citet{Gelli2023} and \citet{Gelli2024} find that supernova feedback operates too slowly, releasing insufficient energy to reproduce the observed SFH in this ${\sim}10^{8.7}~\mathrm{M}_\odot$ galaxy. They argue that the observed rapid quenching is therefore due to either (dusty) radiation-driven winds powered by young massive stars in a high specific star formation rate (sSFR) system or active galactic nucleus (AGN) feedback. Likewise, \citet{Dome2024} find that high-redshift galaxies in the VELA and IllustrisTNG simulations are not bursty enough on small time-scales (${\sim}$40~Myr) to reproduce the photometry (and thus the SFH) of JADES-GS-z7-01-QU, perhaps requiring further calibration of sub-grid models. Though \citet{Faisst2024} find that 60--80 per cent of all galaxies in the SPHINX$^{20}$ cosmological radiation hydrodynamic simulations exhibit similar observed properties (lack of emission lines, blue UV to optical $\lambda f_\lambda$ flux ratio) to JADES-GS-z7-01-QU at some point during their lifetime until $z = 7$, due to stochastic variations in their SFH (i.e.\@ bursts), remaining in this state for less than 20~Myr. Such studies would naturally benefit from a greater sample of quiescent dwarf galaxies, which ideally can be reliably identified from photometry, allowing for targeted, optimised follow-up \emph{JWST}spectroscopy to constrain their SFHs.

Thus we develop a photometric search method for identifying smouldering galaxies at $5 < z < 8$, which we define to have weak emission lines, and thus, like the quiescent galaxies discussed above, generally have low sSFR and may even be in a (temporary) quiescent state. The deep public NIRCam imaging from the JWST Advanced Deep Extragalactic Survey \citep[JADES,][]{Eisenstein2023, Rieke2023, Bunker2024, Hainline2024} second data release \citep{Eisenstein2023b} is essential for this search, as these systems are faint, relatively quiescent dwarf galaxies in the EoR. Moreover, the F335M and F410M medium band imaging is key, enabling a clear identification of the lack of rest-frame optical emission lines in these smouldering galaxies, thus revealing their lack of active star formation. Indeed, unlike previous studies searching for extreme emission-line galaxies by demanding the offset in medium--wide-band photometry to be substantial \citep[e.g.\@][]{Withers2023, Simmonds2024}, we instead require the medium- and wide band measurements to be level, indicative of weak rest-frame optical emission lines in these systems. We conduct our smouldering galaxy search at $z \sim 6$ ($5.3 < z < 6.6$) and at $z \sim 7$ ($6.8 < z < 7.8)$, placing constraints on the star-formation properties and number abundances of these systems, which we compare to the predictions from various simulations. We further validate our smouldering galaxy colour selection procedure, investigating the nature of the sources selected in the FLARES simulations \citep{Lovell2021, Vijayan2021}, discussing methods to remove low-redshift interlopers, as well as alternate causes (beyond weak sSFR) for weak line-emission in smouldering galaxies.

This article is structured as follows. In Section~\ref{sec:data} we discuss the public JADES photometric catalog that underpins our analysis, as well as our procedure for using the {\tt Bagpipes} SED-fitting code \citep{Carnall2018} to determine the star formation properties of our sample. In Section~\ref{sec:selection} we discuss our colour selection procedure for identifying smouldering galaxy candidates, outlining the colour cuts and applicable redshift range, as well as introducing our sample of $5 < z < 8$ smouldering galaxy candidates. In Section~\ref{sec:properties} we discuss the star-formation properties inferred for our smouldering galaxy candidates, together with their number abundances. In Section~\ref{sec:discussion} we further validate our colour selection method for identifying smouldering galaxies. Finally, we conclude in Section~\ref{sec:conclusions}. We assume a flat $\Lambda$CDM cosmology with $H_0 = 70~\mathrm{km~s^{-1}~Mpc^{-1}}$, $\Omega_\mathrm{m} = 0.3$ and $\Omega_\Lambda = 0.7$ throughout and adopt the AB magnitude system \citep{Oke1983}. Throughout this work, when we refer to quiescent galaxies in the context of our smouldering selection procedure, we mean sources with a complete lack of rest-frame optical line emission (e.g.\@ [\ion{O}{III}], H$\alpha$), indicating no star formation over the past 10~Myr. Thus, by relatively quiescent galaxies we mean systems with emission line equivalent widths that are (well) below the median for their redshift, presumably indicating a relatively low sSFR.

\section{Data} \label{sec:data}

In this section we discuss the data that underpins our analysis. In Section~\ref{subsec:jades} we discuss the public JADES photometric catalog that our high-redshift galaxy sample is based on. In Section~\ref{subsec:bagpipes} we outline how we fit the observed photometry with the SED-fitting code {\tt Bagpipes} to place constraints on the star-formation properties of these systems. 

\subsection{JADES photometric catalog} \label{subsec:jades}

We make use of the second JADES public data release \citep{Eisenstein2023b}. This covers the GOODS-S region in the vicinity of the Hubble Ultra Deep Field (HUDF), with both NIRCam imaging \citep{Rieke2023} and NIRSpec spectroscopy \citep{Bunker2024}. We primarily make use of the photometric catalog \citep{Hainline2024}, restricting our analysis to the regions where F335M medium band imaging is available, as this is key for identifying smouldering galaxies via our colour selections. We therefore utilise 34~arcmin$^2$ of imaging, with 25~arcmin$^2$ and 9~arcmin$^2$ coming from the JADES Deep \citep{Eisenstein2023} and JADES Origins Field \citep{Eisenstein2023b}, respectively. This data is deep, with ${\sim}29$--$30$~AB~mag $5\sigma$ depth in 0.3~arcsec diameter apertures across the fields, covering the F090W, F115W, F150W, F200W, F277W, F335M, F410M and F444W filters. This imaging is also supplemented by medium band imaging from the JWST Extragalactic medium band Survey \citep[JEMS,][]{Williams2023} on the HUDF, providing F182M, F210M, F430M, F460M and F480M coverage at ${\sim}28.25$--$29.25$~AB~mag depth. Finally, F182M and F210M imaging at ${\sim}28.25$--$28.50$~AB~mag depth from the First Reionization Epoch Spectroscopically Complete Observations \citep[FRESCO,][]{Oesch2023} program is also available for a large fraction of the JADES footprint we analyse in this work. This NIRCam imaging is further supplemented by existing \emph{HST}/ACS imaging in the F435W, F606W, F775W, F814W and F850LP bands, together with \emph{HST}/WFC3 imaging in the F105W, F125W, F140W and F160W filters. 

We utilise the JADES DR2 public photometric catalog, which contains aperture and Kron (i.e.\@ ``total'') photometry for over 45,000 sources across the filters mentioned above. Specifically we use the circular aperture photometry adopting a 0.3~arcsec diameter aperture, providing a good balance between signal-to-noise (still being relatively high) and aperture correction (being relatively low). The reported flux densities stem from the native resolution images and have been aperture-corrected, include a local background subtraction, with the errors we use in our SED-fitting analysis having been estimated by placing a series of empty apertures in the vicinity of the source of interest. 

We further make use of the photometric redshifts reported in the public photometric catalog. These were estimated using the {\tt EAZY} SED-fitting code \citep{Brammer2008}, utilising the original {\tt EAZY} templates supplemented by seven additional templates that were created to better span the observed colour space of galaxies in the JAGUAR simulations \citep{Williams2018}. All available NIRCam photometry is included in the SED-fitting process, together with \emph{HST}/ACS photometry in the F435W, F606W, F775W, F814W and F850LP filters, adopting a minimum error of 5 per cent in each band. 

Utilising these photometric redshifts, we construct our high-redshift galaxy sample ($5.3 < z < 7.8$) by applying additional quality cuts to the JADES photometric catalog:

\begin{itemize}
\item We require the source to be ${>}5\sigma$ detected in all NIRCam bands redward (but not including) the Ly$\alpha$ break.
\item We require the source to be ${<}3\sigma$ detected (i.e.\@ non-detected) in all bands blueward of the Ly$\alpha$ break.
\item We require the reduced-$\chi^2 < 6$ for our {\tt Bagpipes} fits, so that our models provide a reasonable description of the data.
\item We visually inspect each source to remove objects likely affected by artefacts. We remove two sources which lie on stellar diffraction spikes in this manner. 
\end{itemize}

\subsection{SED-fitting with Bagpipes} \label{subsec:bagpipes}

We fit the \emph{HST}+NIRCam photometry with {\tt Bagpipes} \citep{Carnall2018} to place constraints on the star-formation properties of our high-redshift galaxy sample. Similar to the JADES photometric redshift catalog, we utilise all available NIRCam photometry together with the \emph{HST}/ACS filters. We adopt a minimum error of 5 per cent in each band, to account for photometric flux uncertainties and template mismatch between the models and the observations. 

In order for our {\tt Bagpipes}
models to have the flexibility to describe the star-formation histories of our smouldering galaxy candidates, some of which may be quiescent following a rapid decline in star formation, we adopt a double power law parameterisation for the SFH \citep[see e.g.\@][]{Carnall2018, Strait2023}. We fix the redshift to the photometric redshift (or spectroscopic redshift, where this is available) in the JADES catalog. We adopt priors on the falling $\alpha$ and rising $\beta$ power law slopes that are uniform in $\log_{10}$ between 0.01--1000. The time of peak star formation $\tau$ has a uniform prior between 0--$\mathrm{age_{univ}}(z)$. We further assume a uniform prior on the logarithm of the stellar mass $1 < \log (M_*/\mathrm{M}_\odot) < 15$, the logarithm of the ionisation parameter $-4 < \log U < -2$ and the V-band attenuation $0 < A_\mathrm{V} < 5$, adopting a \citet{Calzetti2000} dust attenuation curve. Finally, we adopt a prior on the metallicity that is uniform in $\log_{10}$ between 0.005--1~$\mathrm{Z}_\odot$. We utilise the default stellar (i.e.\@ \citealt{Bruzual2003} with a \citealt{Kroupa2001} initial mass function) and nebular {\tt Bagpipes} templates.

\begin{figure*}
\centering
\includegraphics[width=0.6\linewidth]{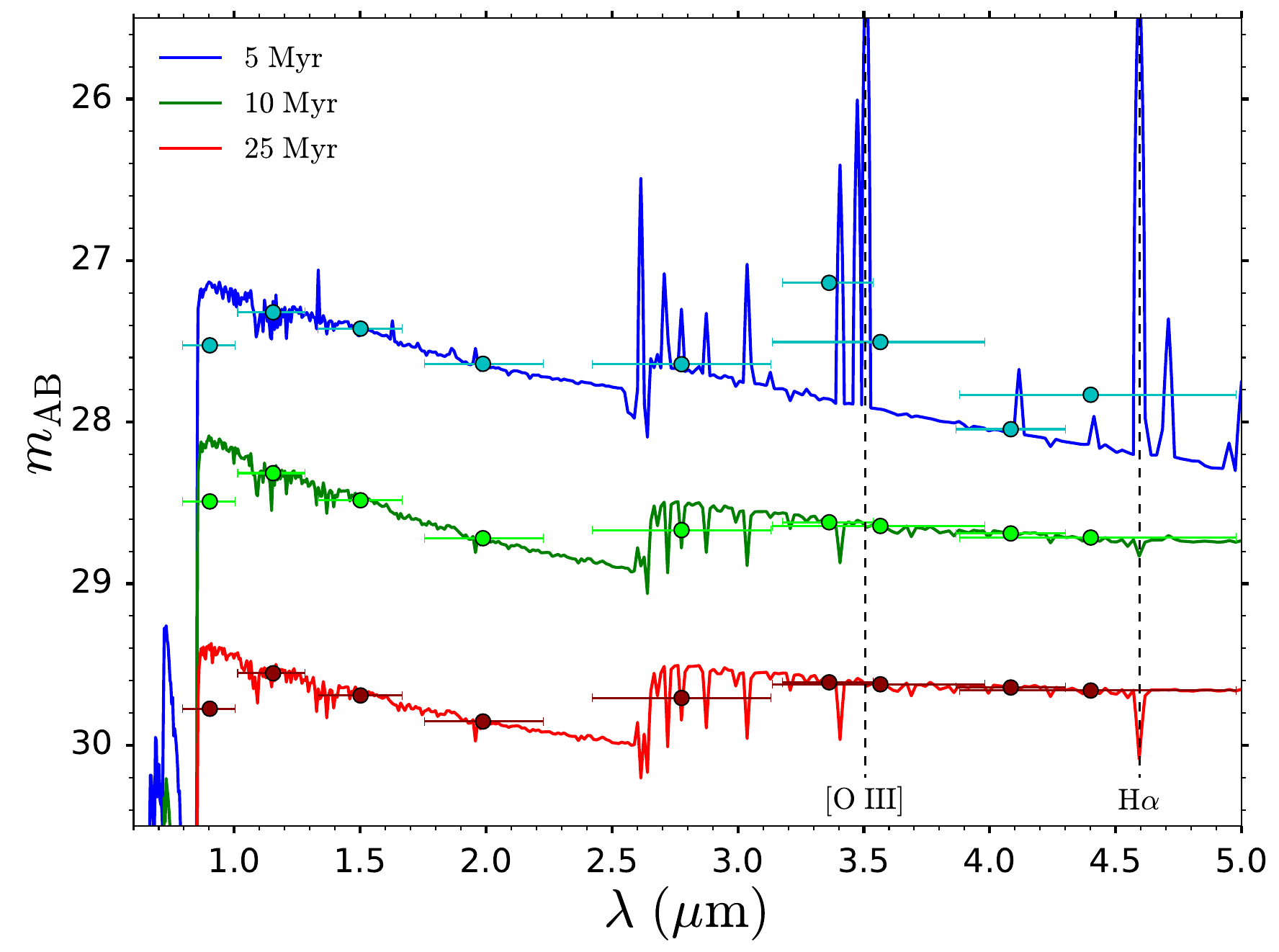}
\caption{The SEDs and photometry (circles, mirroring the main JADES bands in this study) of $z = 6$  {\tt Bagpipes} model galaxies seen 5~Myr (blue), 10~Myr and 25~Myr (red) after an instantaneous starburst, adopting $\log U = -2$ and $Z = 0.2~\mathrm{Z}_\odot$. The locations of the \OIII\ and \Ha\ lines are denoted by the dashed vertical lines. Owing to their greater emission-line-sensitivity and ability to better trace the continuum, the bandpass-averaged flux density in the medium bands is offset from the neighbouring wide band measurement in the presence of emission lines, such as for a young stellar population. It is only when the line emission is weak or absent, that the medium- and wide bands are approximately level, such as in more evolved stellar populations. It is this absence of emission lines, as imprinted on the medium band photometry, that serves as the basis of our identification of weak line-emitting smouldering galaxies.}
\label{fig:starburst_seds}
\end{figure*}

We slightly modify the {\tt Bagpipes} code to natively determine the equivalent width of the \Ha\ and \OIII\ + \Hb\ lines in the model fits. Then, following the Bayesian framework in {\tt Bagpipes}, we generate median values and 16--84 percentile errors by sampling the posterior for the following quantities: stellar mass $M_*$, specific star formation rate $\mathrm{sSFR}_{10}$ (adopting the average SFR over the past 10~Myr), and rest-frame equivalent widths for \OIII\ + \Hb\ ($\mathrm{EW_{[O\ III] + H\beta}}$) and \Ha\ ($\mathrm{EW_{H\alpha}}$). The stellar masses derived using {\tt Bagpipes} are based on the 0.3~arcsec diameter circular aperture photometry. We convert these to total stellar masses by scaling by the flux density ratio between the Kron and circular aperture measurement (imposing a minimum correction factor of 1) in the appropriate filter probing the rest-frame optical. We use the F277W and F356W filters for our $z \sim 6$ and $z \sim 7$ samples (discussed in the next section), respectively.

\section{Smouldering galaxy selection} \label{sec:selection}

In this section we outline the basis for identifying smouldering galaxy candidates. The key principle being that medium band photometry helps establish the weak emission lines in these systems, thus revealing their lack of active star formation. In Section~\ref{subsec:colour_selection} we discuss the colour cuts we apply to select smouldering galaxy candidates. In Section~\ref{subsec:redshift_range} we discuss the redshift range over which these colour cuts can be applied. In Section~\ref{subsec:flares_selection}, we vet our smouldering galaxy colour selection procedure by applying it to galaxies from the FLARES simulations \citep{Lovell2021, Vijayan2021}. Finally, in Section~\ref{subsec:candidates} we introduce our sample of smouldering galaxy candidates from the JADES data.

\subsection{Colour selection} \label{subsec:colour_selection}

\subsubsection{Main colour cuts}

Smouldering galaxies are defined to have weak or even non-existent emission lines, which is generally indicative of low levels of active star formation and perhaps even (temporary) quiescence. Therefore traditional quiescent galaxies comprise a subset of the smouldering population. Furthermore, some smouldering galaxies may be post-starburst systems, though they do not all have to be. That is, galaxies can enter a weak line-emitting smouldering state following a burst of star formation, or through a more gradual decline of their star formation rate. Hence smouldering galaxies do not necessarily have spectra/photometry dominated by A-type stars as with post-starburst galaxies. Nor do they necessarily exhibit prominent Balmer breaks and red rest-frame optical--NIR colours as traditional quiescent galaxies. Instead, medium band photometry is key in the accurate identification of weak line-emitting smouldering galaxies. As these filters span a narrower wavelength interval than the corresponding wide bands, their bandpass-averaged flux density is more greatly affected by emission lines, thus offering greater emission-line-sensitivity. Moreover, these same medium bands, owing to their narrower nature, are more likely to enable a clean (i.e.\@ unaffected by emission lines) measurement of the continuum level, unlike the wide band filters which usually have their bandpass-averaged flux density being affected by the one or more emission lines which typically reside with the filter range.

Thus for strong line-emitting systems we naturally expect there to be a differential in the photometry measured by a medium band and its accompanying wide band filter. As can be seen for the $z=6$, 5~Myr instantaneous starburst {\tt Bagpipes} model (shown in solid blue, with $\log U = -2$ and $Z = 0.2~\mathrm{Z}_\odot$) in Fig.~\ref{fig:starburst_seds}, when the same emission line (complex, in this case \OIII + \Hb) resides in both a medium band filter (F335M) and its neighbouring wide band filter (F356W), the bandpass-averaged flux density in the medium band filter is highly elevated compared to the wide band, owing to the greater emission-line-sensitivity in the narrower filter. Similarly, when a given emission line (complex, in this case just \Ha) is only present in the wide band filter (F444W), with the medium band (F410M) tracing the continuum level, the bandpass-averaged flux density in the wide band is elevated compared to the medium band. Thus the differential in the photometry between the medium- and accompanying wide band betrays the strong line emission in this system.

Now, as the stellar population ages, to 10~Myr (green, see Fig.~\ref{fig:starburst_seds}) and 25~Myr (red) after the instantaneous starburst, the strength of the line emission wanes and ultimately fades, its impact on the medium- and wide band photometry is reduced and ultimately is non-existent, with these bands now both purely tracing the same continuum level, with the medium- and wide band photometry thus being level. Hence it is level photometry, between medium and wide bands where we would otherwise expect emission lines to reside, that indicates a lack of line emission, and thus serving as the key principle behind selecting smouldering galaxy candidates. 

The JADES NIRCam photometry we use in this analysis has both the F335M and F410M medium bands, in addition to the 7 wide bands (F090W, F115W, F150W, F200W, F277W, F356W, F444W). Thus we are able to apply two key colour cuts that demand weak line emission in two distinct emission-line complexes, which places stronger constraints on the weak line-emitting nature of these systems, also removing more spurious sources and contaminants than from a single colour cut alone (which would otherwise often be the case as the F410M filter has usually been the only medium band filter utilised in other datasets thus far). 

Our two colour cuts for selecting smouldering galaxy candidates require the photometry to be level in the medium--wide-band filter pairs. These cuts are:

\begin{equation} \label{eq:MW1}
-0.1 < m_\mathrm{F335M} - m_\mathrm{F356W} < 0.1
\end{equation}
\begin{equation} \label{eq:MW2}
-0.1 < m_\mathrm{F410M} - m_\mathrm{F444W} < 0.1
\end{equation}

Now, noise on the observed photometry can in principle cause both emission-line-free systems with intrinsically level photometry to fail this criterion (being scattered out of the range), and also cause intermediate-strength line emitters to spuriously satisfy this criterion (being scattered into the range). To minimise this, we require the photometry to be ${>}10\sigma$ detected in the bands comprising the colour selection, so that we are very confident in the level nature of the photometry. 

\begin{figure}
\centering
\includegraphics[width=.9\linewidth]{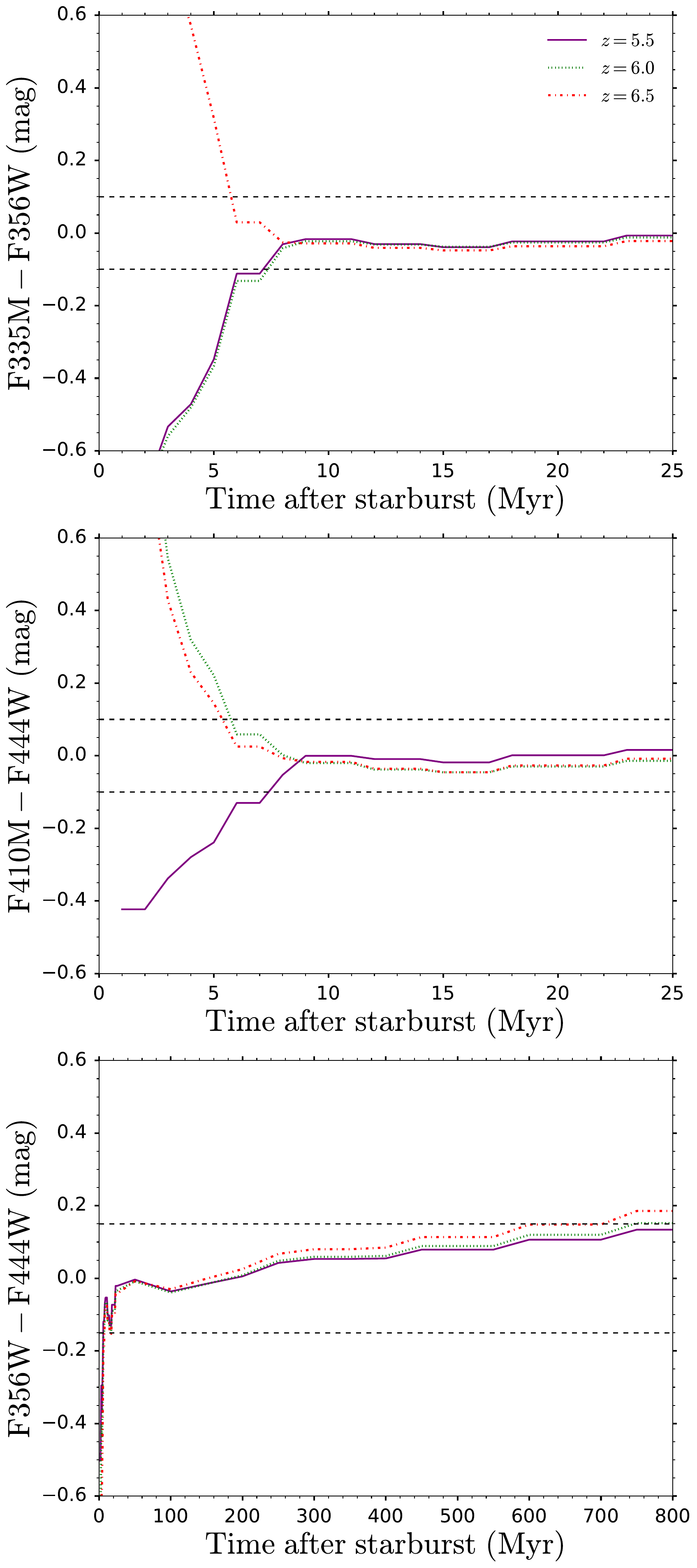}
\caption{The evolution of colours with time after an instantaneous starburst, for {\tt Bagpipes} model galaxies at $z = 5.5$ (solid purple), $z = 6.0$ (dotted green) and $z = 6.5$ (dash-dotted red). Both the $m_\mathrm{F335M}-m_\mathrm{F356W}$ colour (top panel) tracing \OIII\ + \Hb\ and the $m_\mathrm{F410M}-m_\mathrm{F444W}$ colour (middle panel) tracing \Ha\ at these redshifts become level (within the $\pm0.1$~mag dashed lines) within ${\sim}10$~Myr following an instantaneous starburst. Thus the line emission fades and these galaxies are selected via our smouldering colour cut procedure before the Balmer break starts to substantially develop (${\sim}40$~Myr). Note the colours are either initially blue or red depending on whether the emission line (complex) is in both the medium and wide band filters, or just the wide band filter, respectively. Bottom panel: As the stellar population ages, the rest-frame optical slope gradually becomes redder, and our level $m_\mathrm{F356W}-m_\mathrm{F444W}$ colour criterion ($\pm 0.15$~mag, Equation~\ref{eq:supplementary}) fails to be satisfied after ${\sim}700$~Myr has elapsed, with the exact timescale depending on e.g.\@ the metallicity of the stellar population, and being reduced if there is additional reddening via dust attenuation. Hence our selection procedure will likely not select very old (at these redshifts) or particularly dusty quiescent systems.}
\label{fig:starburst_colours}
\end{figure}

From our own investigations (generating mock spectra with a range of emission line equivalent widths) into the impact of emission lines on photometry, as well as using the approximate formula \citep[see e.g.\@][]{Marmol-Queralto2016} $\Delta m = -2.5 \log _{10} (1 + \mathrm{EW_{rest}}(1+z)/\Delta \lambda)$ which relates the magnitude boost $\Delta m$ that emission lines have on the bandpass-averaged flux density to their rest-frame equivalent width $\mathrm{EW_{rest}}$ (where $\Delta \lambda$ is the width of the filter), we find that the above colour cuts imply the following constraints on the line equivalent widths. A relative shift of 0.1~mag between a medium band and its accompanying wide band filter constrains the rest-frame line equivalent width to ${\sim}100$~\AA. This constraint becomes even stronger (${\sim}50$~\AA) if neighbouring pairs of medium band filters are used, such as [F335M, F360M] or [F410M, F430M], where one filter is affected by the emission line and the other traces the continuum level. This approach is not adopted in this work as the additional medium band filters required (e.g.\@ F360M, F430M) are generally not available in the JADES dataset. Since these line equivalent widths are relatively small compared to the large line equivalent widths \citep[${\sim}1000$~\AA\ in \Ha\ and \OIII+\Hb, e.g.\@][]{Matthee2023, Matthee2024, Endsley2024} generally seen in EoR galaxies, the high-redshift galaxies we select in this manner can be considered to be relatively weak line-emitters, i.e.\@ smouldering galaxies.

As can be seen from Fig.~\ref{fig:starburst_colours}, the medium and wide band photometry tracing the \OIII+\Hb\ line complex (F335M, F356W) and \Ha\ (F410M, F444W) at $z\sim6$ become level approximately ${\sim}10$~Myr after an instantaneous starburst, with the $m_\mathrm{F356W}-m_\mathrm{F444W}$ colour remaining level for several hundred Myr, depending on the metallicity and amount of dust, which both affect the rest-frame optical spectral slope. Here the medium--wide colours are initially blue or red immediately after a starburst depending on whether the emission lines are in both the medium- and wide bands, or solely in the wide band, respectively, which is redshift-dependent.

\begin{figure}
\centering
\includegraphics[width=\linewidth]{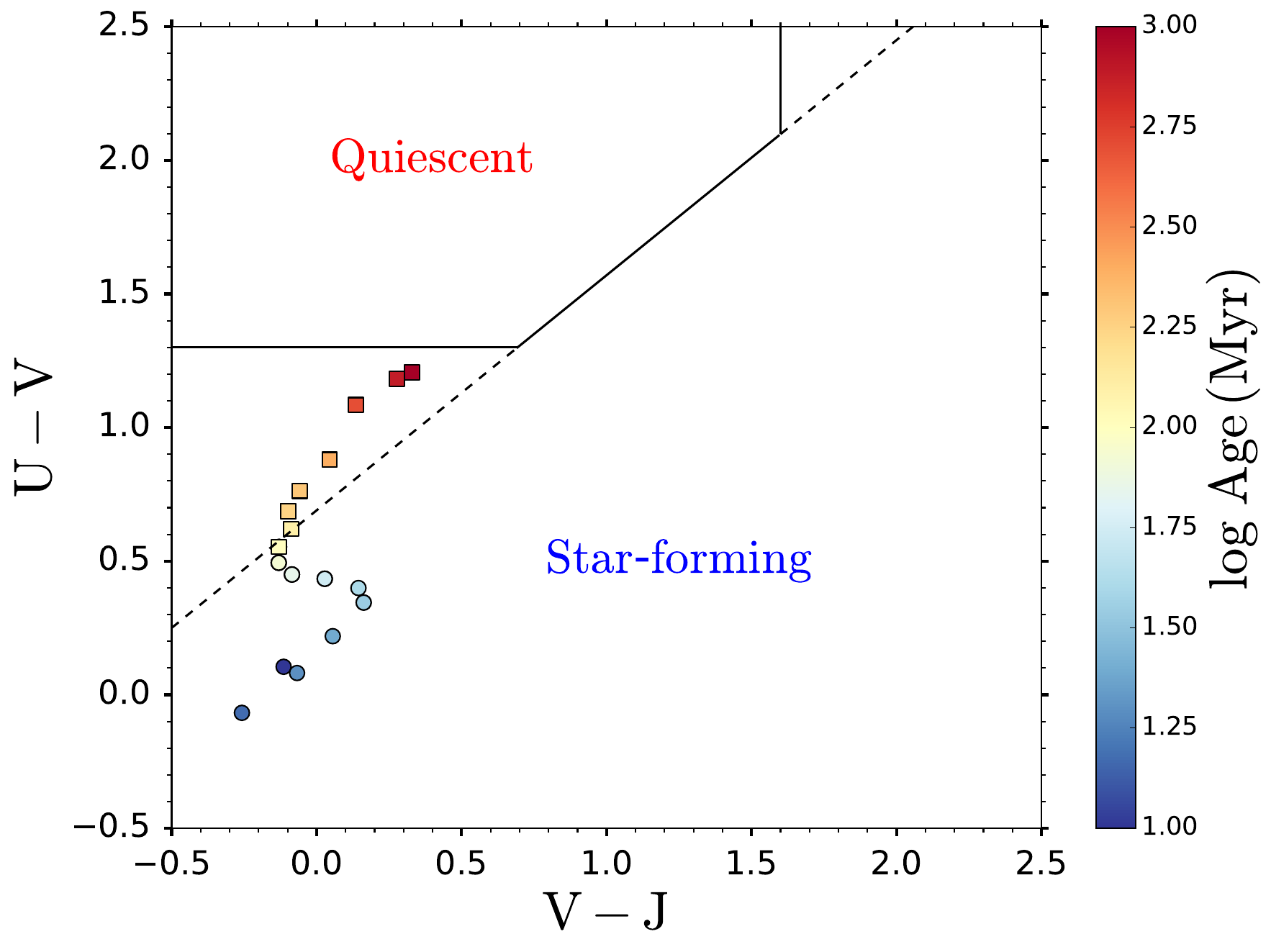}
\caption{The positions of {\tt Bagpipes} model galaxies in the \citet{Williams2009} rest-frame UVJ colour plane, colour-coded by their instantaneous starburst age. Circle and square symbols denote systems that are younger and older than 100~Myr, respectively. In the absence of dust attenuation (shifting colours towards the top right), it takes ${\sim}1$~Gyr of passive evolution for galaxies to enter the quiescent region of the UVJ diagram, and ${\sim}100$~Myr to enter the linear extension (dashed line) of this region. Given the young age of the Universe in the EoR, together with the fact that the SEDs of these high-redshift galaxies are often likely dominated by their recent star formation activity, these traditional methods for selecting quiescent systems may be missing a large fraction of the passive or smouldering galaxy population.}
\label{fig:smouldering_uvj}
\end{figure}

Thus, our colour cuts can select weak emission-line and quiescent systems before the Balmer break substantially develops \citep[e.g.\@ ${\sim}40$~Myr for a $\sim$0.5~mag Balmer break,][]{Trussler2024} and well before such systems would be identified as quiescent using the traditional rest-frame UVJ colour diagram \citep{Williams2009}. Indeed, using instantaneous starburst {\tt Bagpipes} models, we see in Fig.~\ref{fig:smouldering_uvj} that it takes ${\sim}1$~Gyr for a galaxy to be classified as UVJ-passive, barring any dust. Furthermore, it also still takes ${\sim}100$~Myr for a galaxy to enter the linear extension \citep[dashed lines, see e.g.\@][]{Belli2019, Carnall2023} of the traditional UVJ-passive region. Given the young age of the Universe during the EoR, together with the likely bursty star formation in the first galaxies \citep[e.g.\@][]{Looser2023b, Trussler2024}, where the SEDs of systems are dominated by their recent star formation activity (or lack thereof), these traditional methods for selecting quiescent systems may simply be missing a large fraction of the passive or smouldering galaxy population (see also \citealt{Lovell2023} who introduce revised UVJ-selections to account for the young age of passive galaxies emerging from the EoR). Hence our selection method, which is instead based on the lack of strong emission lines, may aid in the photometric identification of such systems. Though we note that our application of an additional $m_\mathrm{F356W}-m_\mathrm{F444W}$ colour cut (outlined in the next subsection) aiming to remove spurious sources also likely results in the removal of old ($\gtrsim 500~$Myr) or particularly dusty quiescent systems from our smouldering galaxy sample.

\subsubsection{Further colour cuts}

As the aim of this initial work on smouldering galaxies is to be as confident as possible in their relatively emission-line-free nature, we apply two further colour cuts to remove spurious line-emitting sources and low-redshift interlopers. These additional cuts likely come at the cost of completeness which we deem acceptable for this study.

As can be seen from Fig.~\ref{fig:starburst_seds}, we expect the F356W and F444W wide band photometry to also be approximately level for systems with weak or non-existent emission lines. Thus as a consistency check, we also apply the following colour cut:

\begin{equation} \label{eq:supplementary}
-0.15 < m_\mathrm{F356W} - m_\mathrm{F444W} < 0.15
\end{equation}

Here we adopt a greater tolerance (0.15~mag) than in the medium--wide band comparison (0.1~mag), owing to the greater wavelength spacing between the two wide bands, which can result in a non-zero $m_\mathrm{F356W}-m_\mathrm{F444W}$ colour even in the absence of emission lines if there is a notable red or blue slope in the rest-frame optical. Admittedly, this cut therefore does remove particularly red (e.g.\@ dusty or old) and blue (e.g.\@ metal-poor and young) smouldering galaxy candidates. However, it also reduces contamination, which is the primary concern in this initial study. After applying the further colour cuts described below, Equation~\ref{eq:supplementary} removes 9/21 sources from our $z\sim6$ smouldering sample. Based off of the {\tt Bagpipes} median fits, we deem that 5 sources could possibly have rather blue slopes, 1 source has a rather red slope, and 3 appear to be possible contaminants. Equation~\ref{eq:supplementary} also removes 1/5 sources from our $z\sim7$ sample, which we infer to have $\mathrm{EW_{[O\ III] + H\beta}} = 127$~\AA.

Low-redshift passive galaxies can somewhat mimic the SED profile of a high-redshift smouldering galaxy. Here the Balmer break of the former is confused with the Lyman break of the latter. Moreover, the red rest-frame optical slope of the passive galaxy is potentially confused as being a red rest-frame ultraviolet slope for the smouldering galaxy. As can be seen from Fig.~\ref{fig:starburst_seds}, the rest-frame ultraviolet slope of the 25~Myr population is still blue, while the Balmer break is already starting to develop. Thus, in the absence of dust, by the time (${>}100$~Myr) the ultraviolet slope is substantially red, the Balmer break must be substantial (${\sim}1$~mag). Hence we require that any smouldering candidates with a red rest-frame ultraviolet slope must have a significant Balmer break.  For our $z \sim 6$ sample of smouldering galaxies (which we will shortly introduce), this corresponds to the following additional colour criterion:

\begin{equation}
    \begin{aligned} \label{eq:cut1}
    \mathrm{if}\ m_\mathrm{F115W} - m_\mathrm{F200W} > 0.3\\
    \mathrm{then}\ m_\mathrm{F200W} - m_\mathrm{F356W} > 0.6 
\end{aligned}
\end{equation}

For our $z \sim 7$ sample of smouldering galaxies this instead corresponds to:

\begin{equation}
    \begin{aligned} \label{eq:cut2}
    \mathrm{if}\ m_\mathrm{F150W} - m_\mathrm{F277W} > 0.3\\
    \mathrm{then}\ m_\mathrm{F277W} - m_\mathrm{F444W} > 0.6 
\end{aligned}
\end{equation}

This cut therefore successfully removes low-redshift passive galaxies which satisfy the first criterion due to their red rest-frame optical slopes, but do not satisfy the second criterion as their SEDs begin to plateau in the rest-frame NIR. Thus, this procedure complements our redshift cuts (described in this next section), which may fail to otherwise properly discriminate between these two distinct systems. However, this cut also removes young dusty smouldering galaxies which have red UV slopes but no prominent Balmer break. Again, we deem this tradeoff between completeness and contamination acceptable for this initial study on smouldering galaxies.

\begin{figure}
\centering
\includegraphics[width=\linewidth]{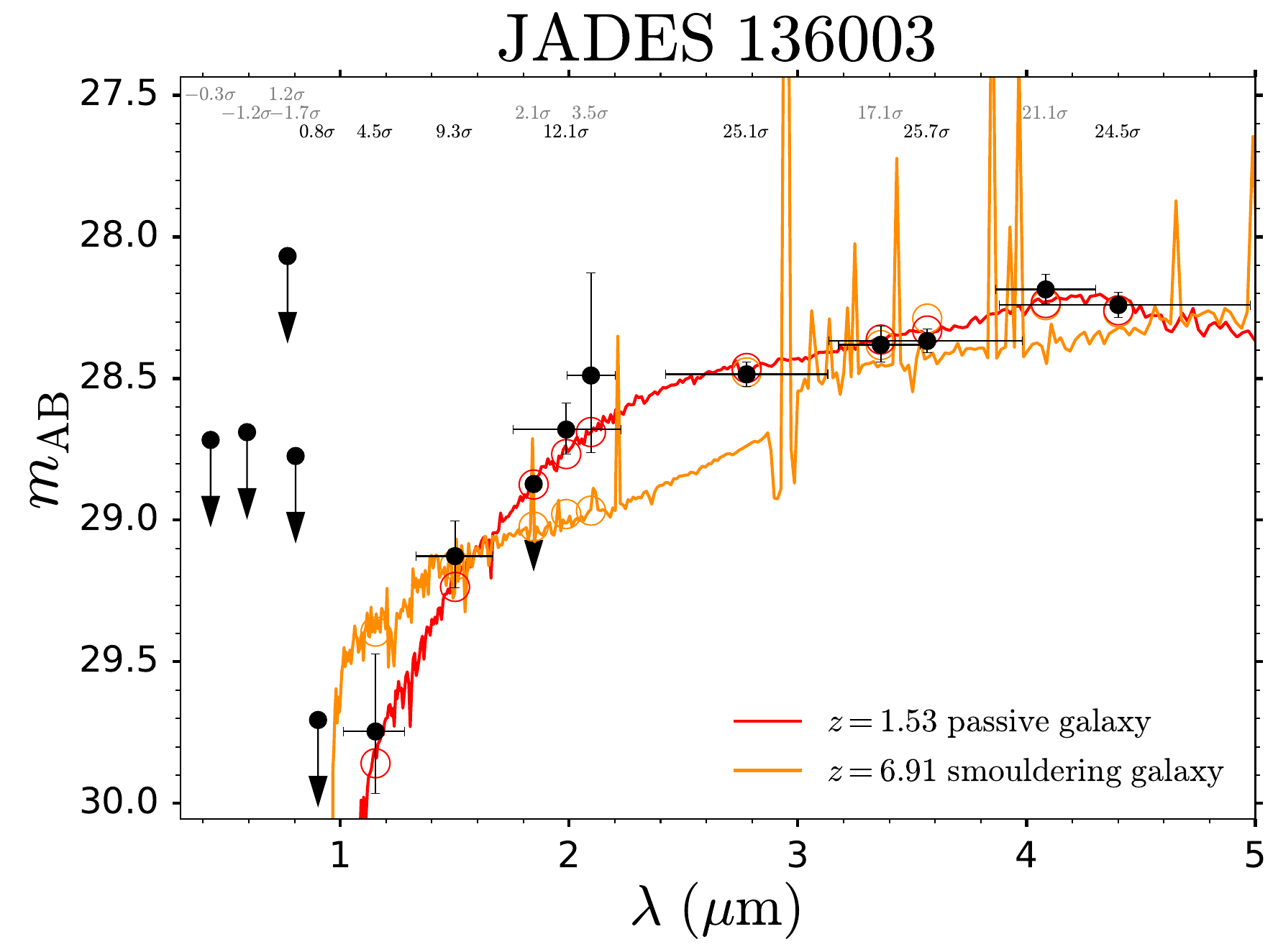}
\caption{The HST+NIRCam photometry (black) of a low-redshift passive galaxy (red) can in principle also resemble that of a high-redshift smouldering galaxy (orange). This comes about because of confusion between the Lyman and Balmer breaks, with the red rest-frame optical continuum of the passive source potentially being misinterpreted as the red rest-frame UV of a high-redshift smouldering galaxy. Thus we apply additional colour cuts (Equations \ref{eq:cut1}/\ref{eq:cut2}) to remove these ambiguous, possibly contaminating cases (such as JADES 136003 which has $z = 6.91$ in the public JADES catalog) from our sample.}
\label{fig:passive_vs_smouldering}
\end{figure}

We show an example of a $z=6.91$ galaxy (orange) in the JADES catalog that would otherwise be selected as a smouldering candidate in Fig.~\ref{fig:passive_vs_smouldering}. As the prominent red rest-frame UV continuum is inferred to be attributable to dust ($A_\mathrm{V} = 0.87$), rather than an old stellar population, the Balmer break is too weak to satisfy our additional colour cut. From Fig.~\ref{fig:passive_vs_smouldering}, we further see that this source's photometry is well-described by a $z=1.53$ passive galaxy solution, motivating this additional quality cut to remove possible low-redshift interlopers.

\subsection{Redshift range} \label{subsec:redshift_range}

As our main colour cuts for selecting smouldering galaxy candidates are based on the imprint of emission lines on photometry (or lack thereof), we require our sources to be in a given redshift range such that the relevant emission lines driving the selection are in the appropriate wide band filters. 

For our $z\sim6$ sample, our colour cuts target the \OIII\ $\lambda5007$ and \Ha\ emission lines, demanding that these are relatively weak. Therefore we restrict the redshift range for this sample to $5.3 < z < 6.6$, with the lower limit corresponding to when the \OIII\ $\lambda5007$ emission line begins to enter the spectral range of the F356W filter (i.e.\@ when the filter throughput reaches 50 per cent of its maximum value), and the upper limit corresponding to when the \Ha\ line gets redshifted out of the F444W filter.

For our $z \sim 7$ sample, our colour cuts target \OIII\ $\lambda5007$ and the complex of weaker rest-frame optical lines (e.g.\@ \Hg, \NeIII\ $\lambda 3869$ and the \OII\ doublet). It is primarily the weak \OIII\ $\lambda5007$ emission in the F444W filter that is the driving factor for this selection, with the weak cumulative emission in the F356W filter by the inherently weaker rest-frame optical lines serving more as a consistency check that there is little-to-no line emission in the system of interest. Thus the redshift range for this sample is $6.8 < z < 7.8$, with the lower limit corresponding to when the \OIII\ $\lambda5007$ line begins to enter the F444W filter. The upper redshift limit is in principle more flexible, here being set by when the Balmer break is 10 per cent through the F356W filter. As we will see, many of our smouldering galaxy candidates exhibit a Balmer break, which could cause such systems to begin to fail our supplementary colour cut in Equation~\ref{eq:supplementary} if the redshift is too large, as the F356W filter then increasingly traces the lower-lying continuum level blueward of the break. As we will see, owing to the lack of direct \Ha\ constraints (as the line is redshifted out of the NIRCam wavelength range) for our $z \sim 7$ sample of smouldering galaxies, the sSFRs inferred for this sample are generally higher than for the $z \sim 6$ sample. 

\subsection{Application of colour selections to the FLARES simulations} \label{subsec:flares_selection}

\begin{figure*}
\centering
\includegraphics[width=.475\linewidth] {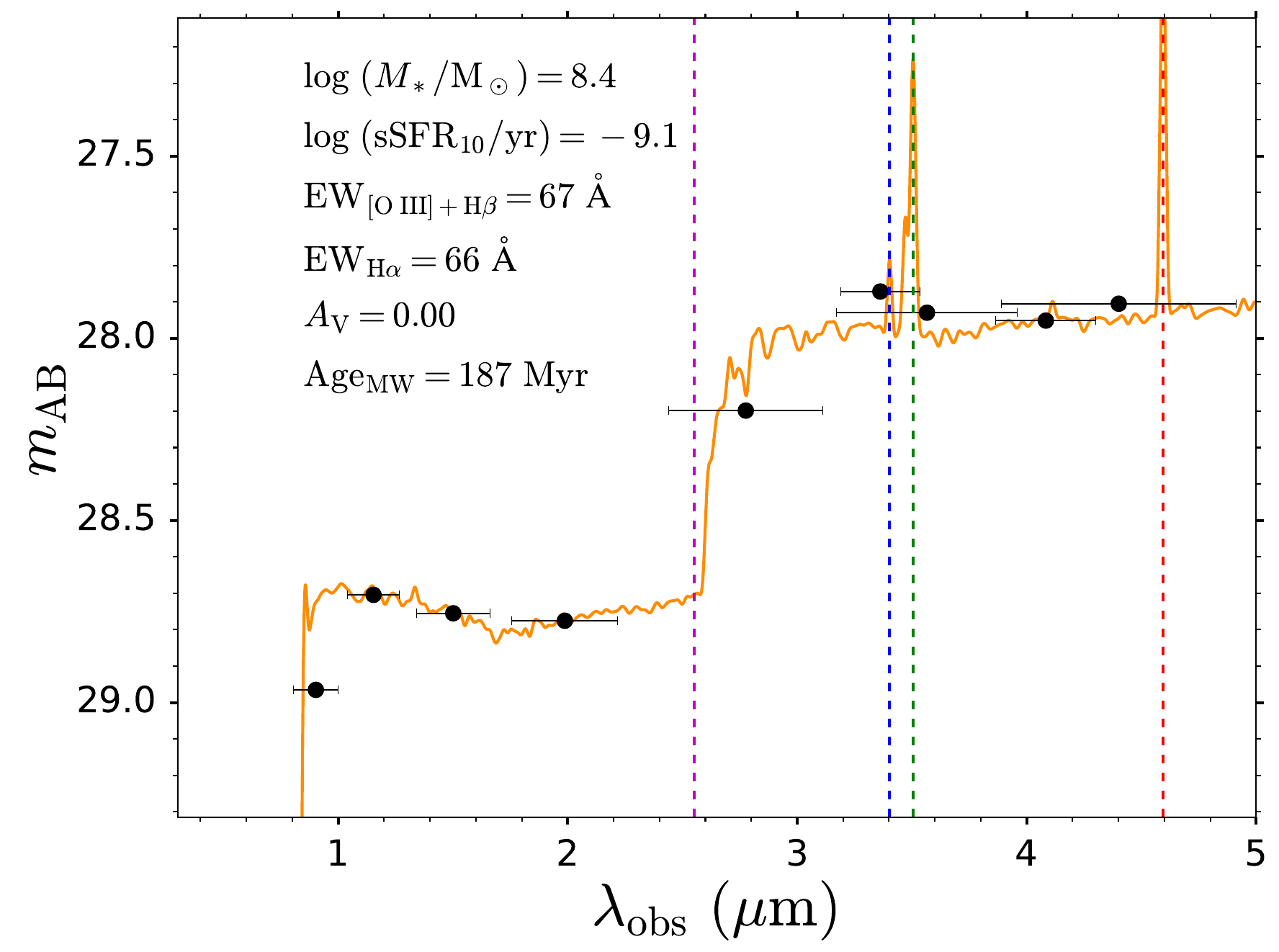} \hfill
\includegraphics[width=.475\linewidth]{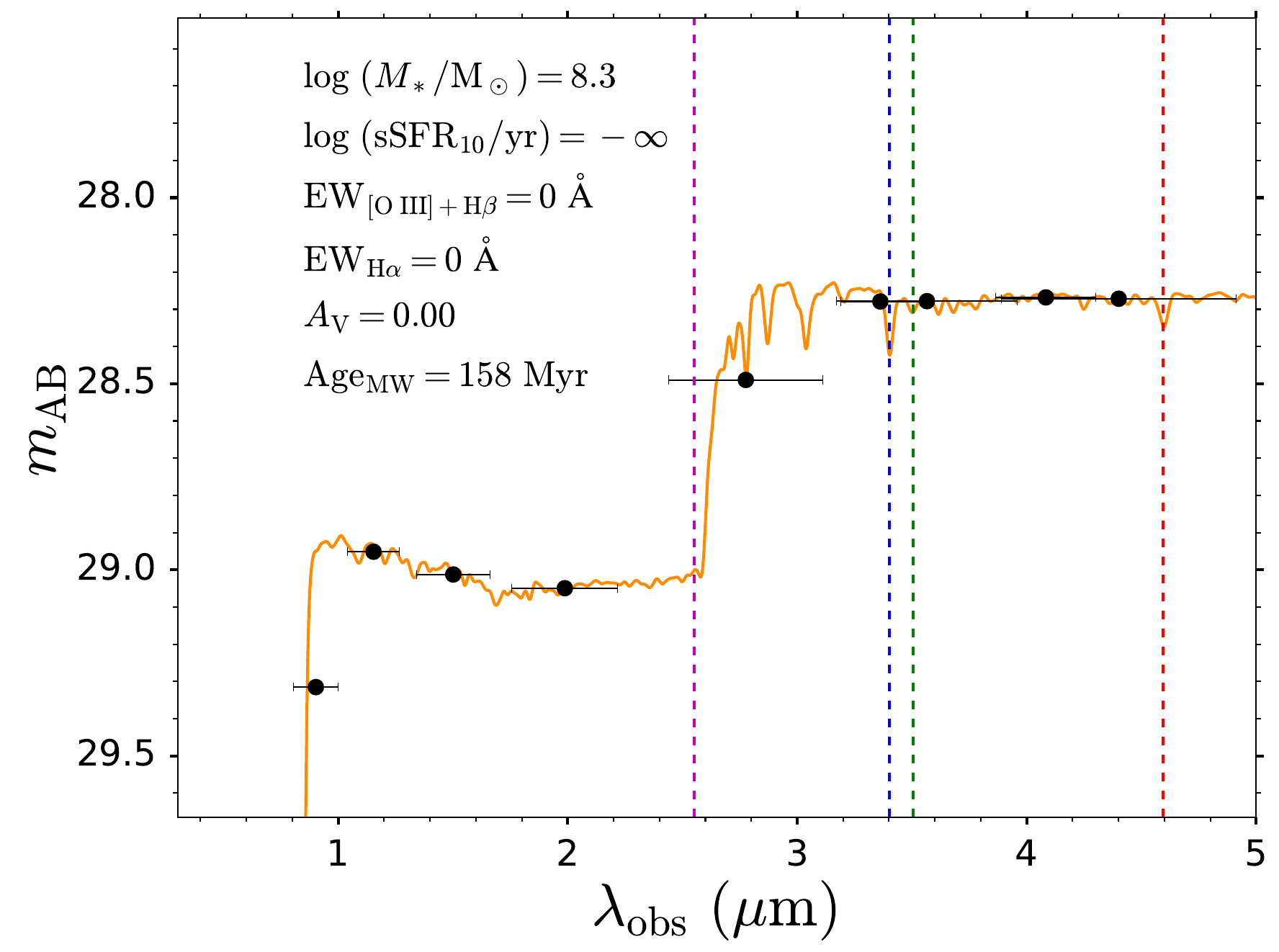} \\[4.5ex]
\includegraphics[width=.475\linewidth] {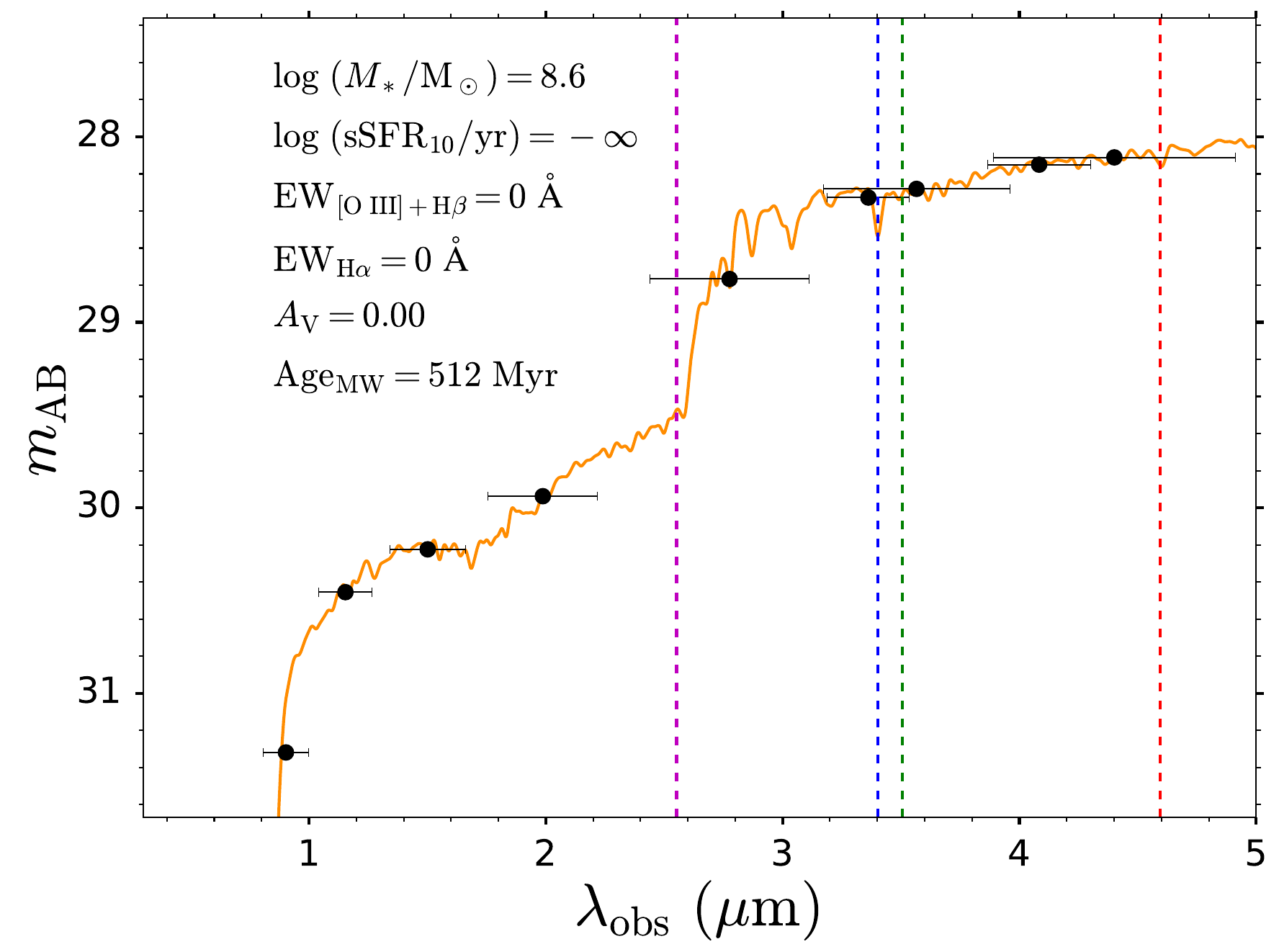} \hfill
\includegraphics[width=.475\linewidth]{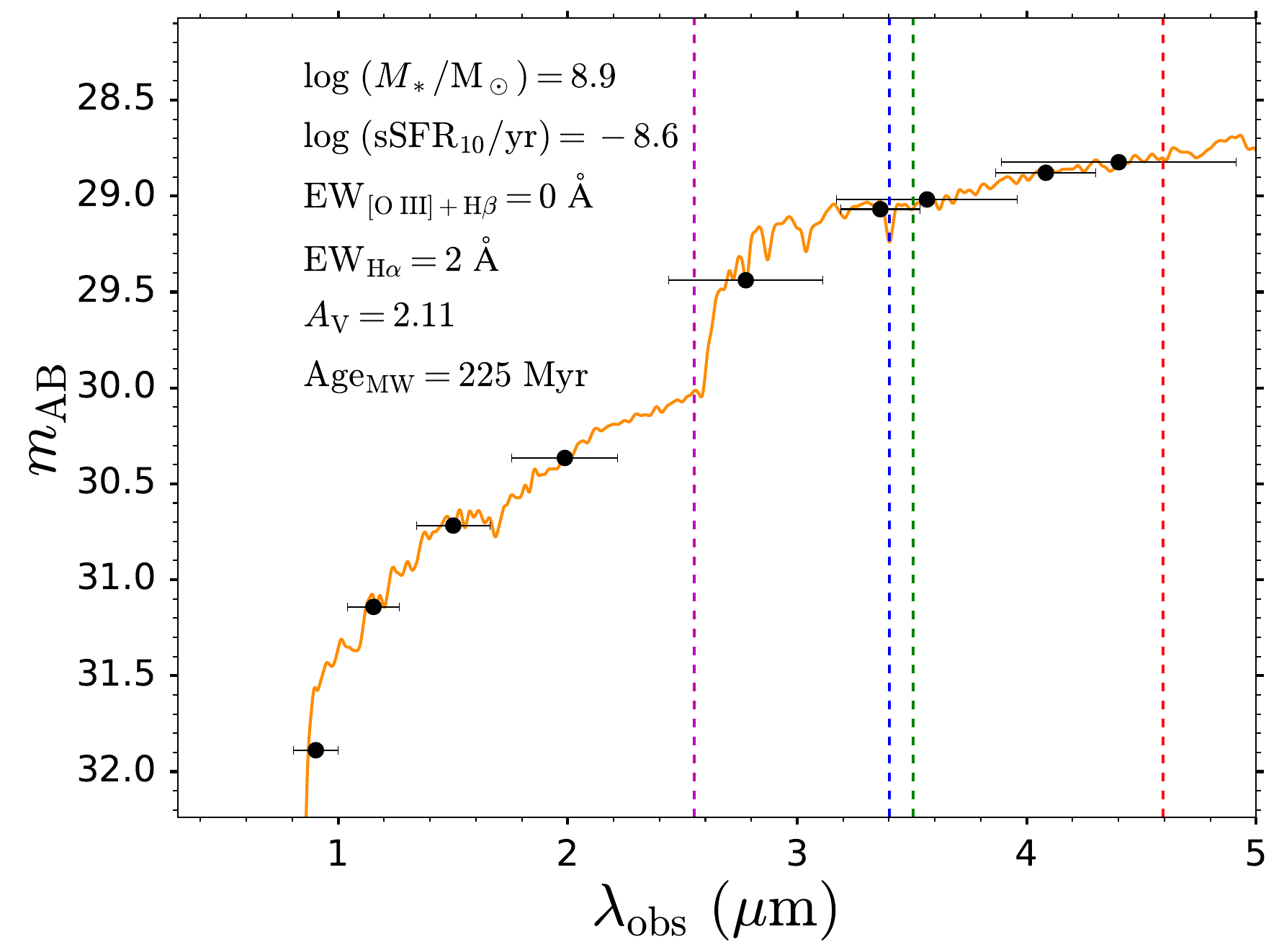}
\caption{SEDs (orange) and NIRCam photometry (black) of $z=6$ FLARES galaxies. Stellar masses $M_*$, specific star formation rates $\mathrm{sSFR}_{10}$, \Ha\ and \OIII\ + \Hb\ rest-frame equivalent widths, V-band attenuation $A_\mathrm{V}$ and mass-weighted stellar ages are also displayed. Top panels: Example FLARES galaxies selected via our smouldering galaxy colour selection procedure (Equations \ref{eq:MW1}, \ref{eq:MW2} and \ref{eq:supplementary}). These systems either have weak (top-left) or non-existent emission lines (top-right), consistent with a relative/complete lack of ongoing star formation. Bottom panels: Example FLARES galaxies with weak emission lines that are not selected via our procedure, due to their particularly red rest-frame optical colours failing to satisfy our level $m_\mathrm{F356W}-m_\mathrm{F444W}$ colour criterion (Equation \ref{eq:supplementary}). These (rare) red systems either have substantially old stellar populations (e.g.\@ 512~Myr, bottom-left) or are rather dusty (e.g.\@  $A_\mathrm{V} = 2.11$, bottom-right).} 
\label{fig:flares_examples}
\end{figure*}

\begin{figure*}
\centering
\includegraphics[width=\linewidth]{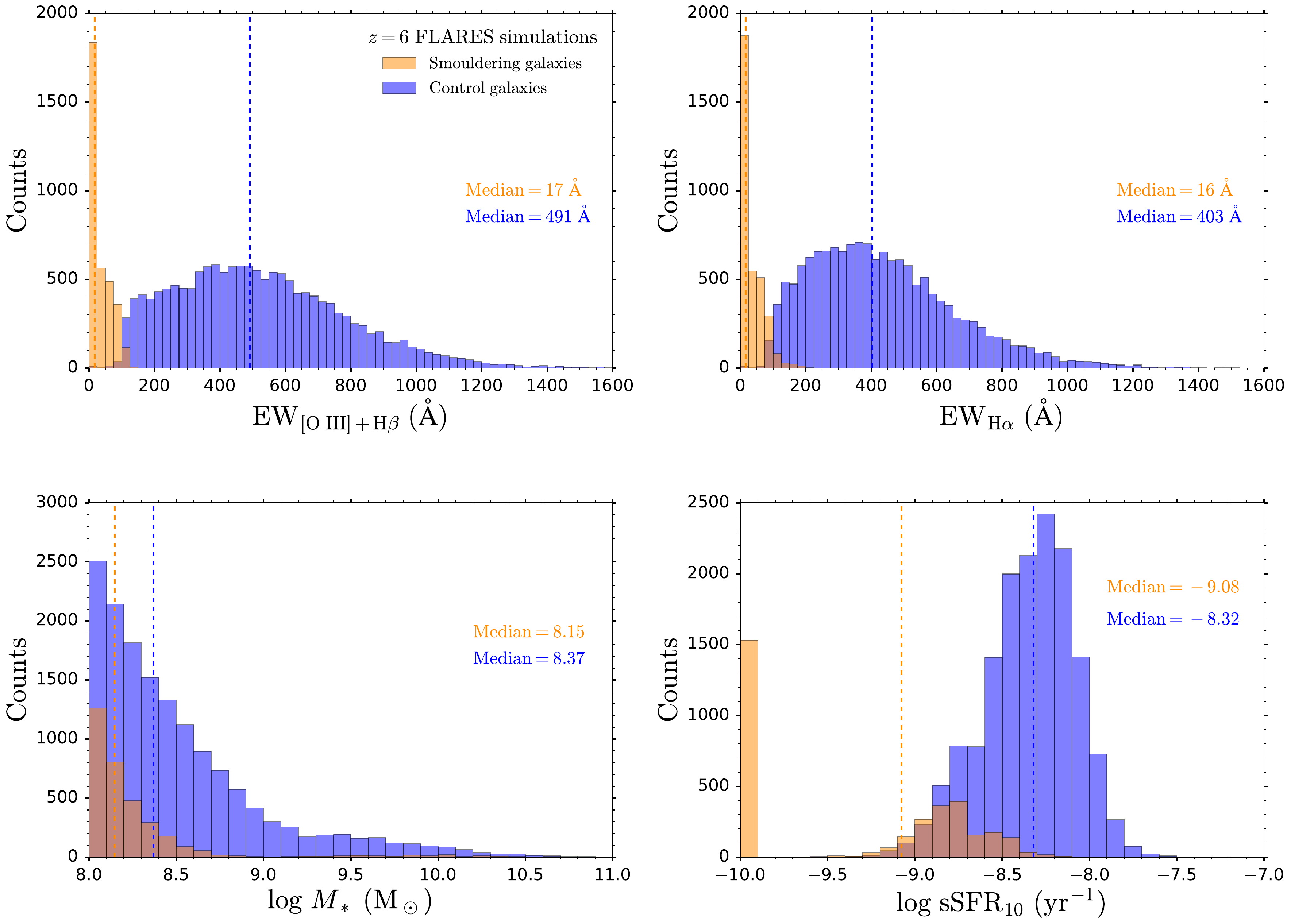}
\caption{Histograms for the rest-frame equivalent widths of \OIII\ + \Hb\ (top-left panel) and \Ha\ (top-right), as well as the stellar mass (in a 30~kpc aperture, bottom-left) and $\mathrm{sSFR}_{10}$ (bottom-right) for smouldering (orange) and control galaxies (blue) at $z = 6$ in the FLARES simulations. Smouldering galaxies identified through our colour selections typically have much smaller line equivalent widths (vertical dashed lines indicate the median), lower stellar masses and lower specific star formation rates than the control sample. The relative number counts for the smouldering and control galaxies as a function of stellar mass highlight how smouldering activity increasingly dominates at lower stellar masses.}
\label{fig:flares_histograms}
\end{figure*}

Here we vet our smouldering galaxy colour selection procedure by applying it to galaxies from the FLARES simulations \citep{Lovell2021, Vijayan2021}, which are a suite of hydrodynamic simulations of galaxy formation and evolution, based on the physics of EAGLE, with simulation outputs provided at $z = $ 5, 6, 7, 8, 9, 10. We apply our colour selections (Equations \ref{eq:MW1}, \ref{eq:MW2} and \ref{eq:supplementary}) to the $z=6$ galaxies in these simulations, where both the strength of the \OIII\ + \Hb\ and \Ha\ emission can be constrained via NIRCam photometry. We wish to establish the nature of the sources selected in the case of ideal photometry that is noise-free and without data quality issues, and where the properties of the sources such as redshift, stellar mass and sSFR are perfectly known. 

We show example SEDs of $z = 6$ FLARES galaxies selected by applying our smouldering galaxy colour cuts in the top panels of Fig.~\ref{fig:flares_examples}. These systems clearly have weak (left) or non-existent (right) emission lines and a notable Balmer break, consistent with a relative/complete lack of ongoing star formation and an older (underlying) stellar population. However, there are also cases (though rare in the FLARES simulations) where our procedure fails to select systems with weak emission lines, because these have particularly red rest-frame optical colours, thus not satisfying our level $m_\mathrm{F356W}-m_\mathrm{F444W}$ colour criterion (Equation~\ref{eq:supplementary}). We show example SEDs of these rare red systems in the bottom panels of Fig.~\ref{fig:flares_examples}, corresponding to a galaxy with substantially old stellar populations (512~Myr) and a galaxy with substantial dust attenuation ($A_\mathrm{V}=2.11$) in the left and right panels, respectively.

We show histograms of various star-formation-related quantities for the $z = 6$, $\log\, (M_*/\mathrm{M}_\odot) \geq 8.0$ FLARES galaxies in Fig.~\ref{fig:flares_histograms}. We find that the smouldering galaxies (orange) selected via our colour cuts tend to have much weaker \OIII\ + \Hb\ (top-left panel) and \Ha\ (top-right panel) emission than the control sample of galaxies (blue), with a substantial fraction of the smouldering galaxies having an equivalent width of zero, indicative of no ongoing star formation (i.e.\@ quiescence). Indeed the median \OIII\ + \Hb\ and \Ha\ equivalent widths for smouldering galaxies (orange vertical dashed lines) of 17~\AA\ and 16~\AA\ are an order of magnitude lower than for the control galaxies (491~\AA\ and 403~\AA). As the \Ha\ equivalent width can serve as a rough proxy for the $\mathrm{sSFR}$, we would expect smouldering galaxies to tend to have lower $\mathrm{sSFRs}$ than their control galaxy counterparts, which is confirmed in the bottom-right panel, with the median sSFR of the smouldering and control galaxies being $\mathrm{sSFR}_{10} = 10^{-9.08}~\mathrm{yr}^{-1},\ 10^{-8.32}$~yr$^{-1}$, respectively. Note that galaxies with $\mathrm{sSFR}_{10} \leq 10^{-10}$~yr$^{-1}$ (such as the quiescent galaxies) are plotted at $10^{-10}$~yr$^{-1}$, for clarity. In contrast to the observational results (which are affected by incompleteness, to be discussed in Section~\ref{subsec:sf_properties}), smouldering galaxies in FLARES tend to be less massive (median $\log (M_*/\mathrm{M}_\odot) = 8.15$) than the control galaxies ($\log (M_*/\mathrm{M}_\odot) = 8.37$), indicating that smouldering activity preferentially takes place in low-mass systems (bottom-left panel). This is consistent with the increasing smouldering fraction with decreasing stellar mass that will be shown in Section~\ref{subsec:number_abundances}, which can also be deduced by comparing the relative counts of smouldering and control galaxies as a function of stellar mass in Fig.~\ref{fig:flares_histograms}.

\begin{figure}
\centering
\includegraphics[width=.85\linewidth]{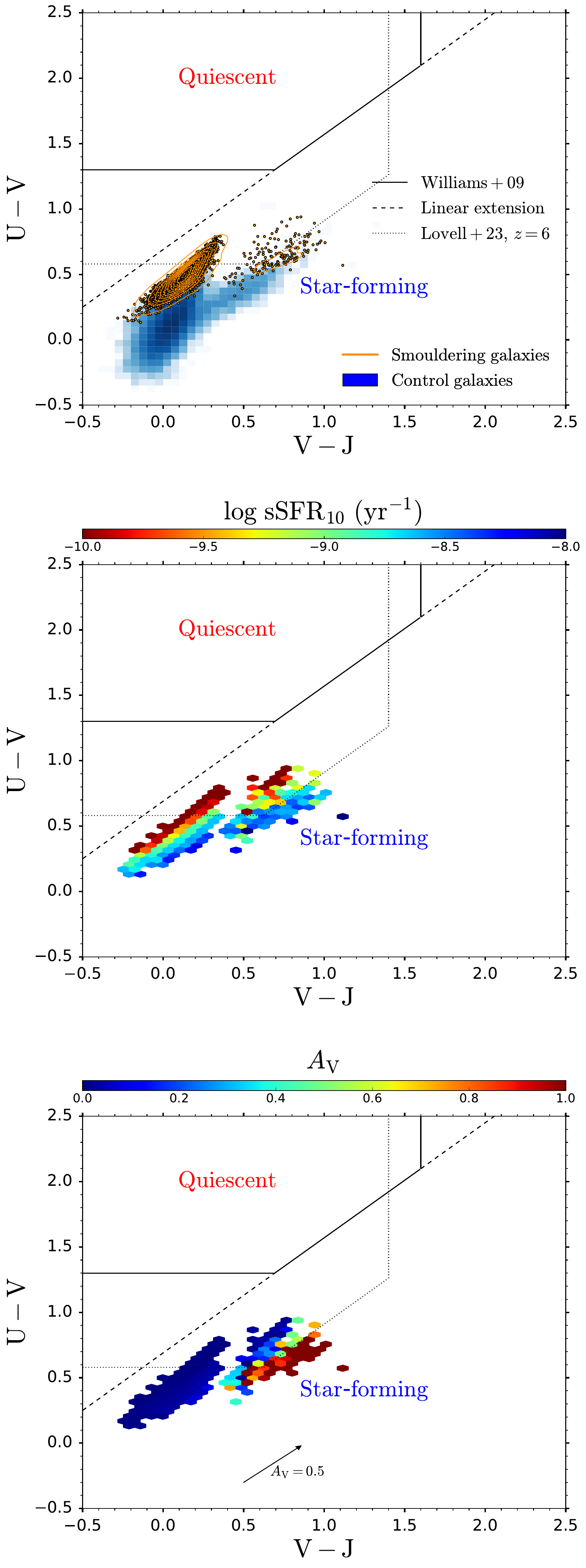}
\caption{The locations of $z = 6$ FLARES smouldering (orange) and control galaxies (blue) in the rest-frame UVJ colour plane. Smouldering galaxies exhibit (subtly) different colours to the control sample. Crucially, these systems are generally located in the star-forming region of the UVJ diagram, rather than being in the quiescent region \citep[solid line,][]{Williams2009} or linear extension thereof (dashed line). Though a subset of the oldest smouldering galaxies do reside in the quiescent region of the \citet{Lovell2023} revised UVJ selection (dotted), extrapolated to $z=6$. Hence traditional methods for selecting quiescent galaxies are likely missing a substantial fraction of the passive and smouldering population at these redshifts. The two distinct smouldering galaxy populations in the UVJ diagram exhibit a comparable spread and sequence in $\mathrm{sSFR}_{10}$ (middle panel), though have differing amounts of dust attenuation $A_\mathrm{V}$ (bottom panel), with the bluer and redder populations consisting of relatively dust-free and dusty smouldering galaxies, respectively.}
\label{fig:flares_uvj}
\end{figure}

We show the positions of the $z = 6$ FLARES galaxies in the rest-frame UVJ colour plane in Fig.~\ref{fig:flares_uvj}. While smouldering galaxies (orange) exhibit (subtly) distinct colours compared to control galaxies (blue), crucially they still reside within the star-forming region of the diagram, rather than being in the quiescent region of the traditional UVJ diagram \citep[solid line,][]{Williams2009} or linear extension thereof (dashed line). Thus these weak line-emitting and potentially even young quiescent systems would be missed using conventional methodologies for selecting passive galaxies. Indeed, \citet{Lovell2023}, who studied the emergence of passive galaxies in the FLARES simulations, also recognised this. Hence they introduced new redshift-dependent UVJ selections that can be used to identify the younger quiescent galaxies emerging from the EoR. Extrapolating their $z \leq 5$ selection criterion to $z=6$, we find that a subset of the oldest smouldering galaxies can be selected in this way, thus certainly offering considerable improvement over the traditional \citet{Williams2009} selection, but not capturing the entirety of the smouldering population. Moreover, direct constraints on the rest-frame J-band (rather than inaccurate extrapolations) for $z = 6$ and $z = 7$ galaxies would require MIRI F770W and F1000W imaging, which would be exceedingly expensive to acquire over the JADES area for these generally faint high-redshift galaxies. Though \citet{Alberts2023} demonstrate that the need for direct J-band constraints is alleviated if dense wavelength coverage with NIRCam (i.e.\@ including several medium bands in addition to the wide bands) is available.

Interestingly, we note that smouldering galaxies in the FLARES simulations are located in two distinct regions within the UVJ plane, thus perhaps being comprised of two differing populations of galaxies. From the middle panel of Fig.~\ref{fig:flares_uvj}, where we colour code by $\mathrm{sSFR}_{10}$, we see that galaxies in these two regions exhibit a comparable spread and sequence in sSFR, so their star formation activities are roughly the same and thus not the cause for the two distinct smouldering regions within the UVJ diagram. In contrast, from the bottom panel, we can see that galaxies in these two regions have differing amounts of dust attenuation $A_\mathrm{V}$, with the increased dust reddening shifting the dustier smouldering galaxies further towards the top-right in the UVJ diagram. Indeed, owing to the correlation between stellar mass and dust attenuation $A_\mathrm{V}$ in the FLARES simulations, the bluer and redder regions in the UVJ diagram correspond to relatively dust-free low-mass ($M_* = 10^{8\textrm{--}9}~\mathrm{M}_\odot$), and dusty massive ($M_* > 10^{9}~\mathrm{M}_\odot$) smouldering galaxies, respectively.

\subsection{Smouldering galaxy candidates} \label{subsec:candidates}

\begin{figure*}
\centering
\includegraphics[width=\linewidth]{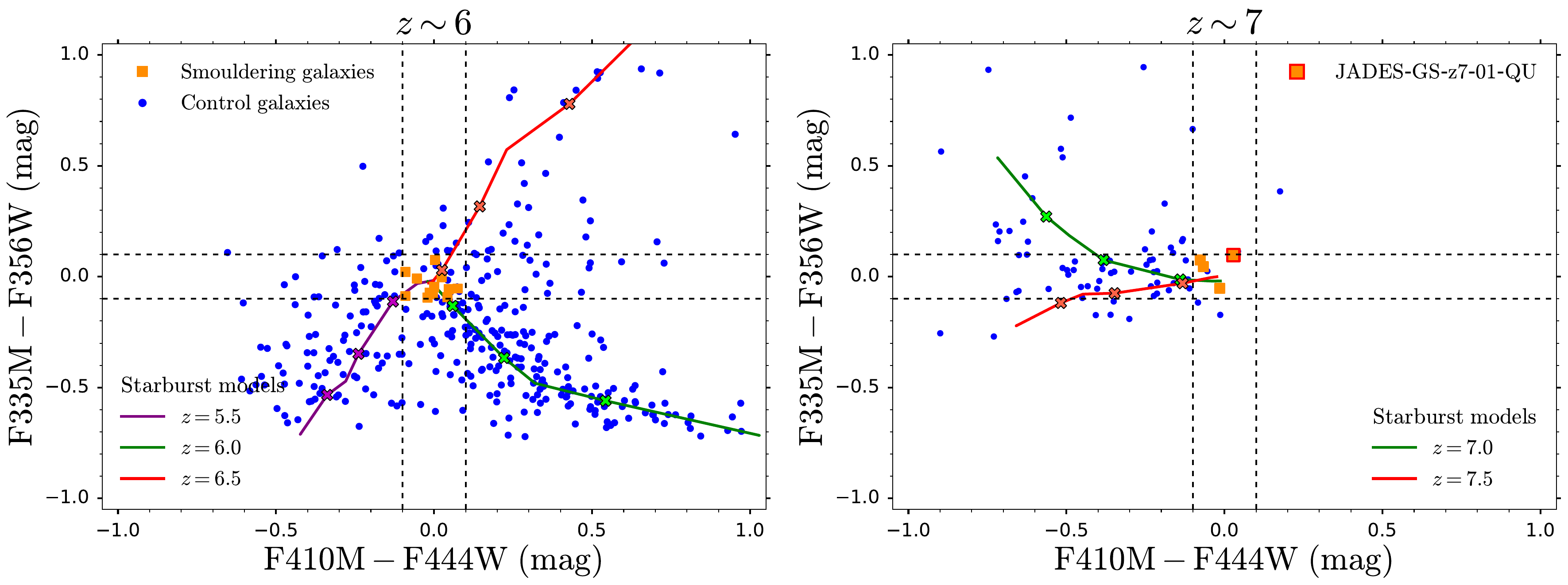}
\caption{Our main colour selection procedure for identifying smouldering galaxy candidates. Smouldering galaxies (orange squares) are defined to have weak emission lines, as indicated by the level photometry (within the $\pm 0.1$~mag dashed lines) in the pairs of medium--wide-bands. They are further required to satisfy additional colour cuts, aiming to minimise contamination. The $z=7.29$ recently-quenched galaxy JADES-GS-z7-01-QU \citep{Looser2024} selected via our methodology is shown with a red border. The remaining systems that do not satisfy these cuts, and therefore likely have appreciable emission lines, constitute our control sample (blue circles). At $z \sim 6$ (left panel), the $m_\mathrm{F335M} - m_\mathrm{F356W}$ and $m_\mathrm{F410M} - m_\mathrm{F444W}$ colour cuts demand the \OIII\ + \Hb\ and \Ha\ lines to be weak, respectively. At $z \sim 7$, these cuts instead demand the complex of weaker rest-frame optical lines (e.g.\@ \Hg, \NeIII\ $\lambda 3869$, and the \OII\ doublet) and \OIII\ + \Hb\ to be weak, respectively. The colours of {\tt Bagpipes} model galaxies between 1--10~Myr after an instantaneous starburst are also shown (solid lines), with the crosses denoting ages of [3, 5, 7]~Myr. The models assume a metallicity $Z = 0.2~\mathrm{Z}_\odot$ and ionisation parameter $\log U = -2$, at (left panel) $z=5.5,\ 6.0,\ 6.5$ (purple, green and red, respectively) and at (right panel) $z=7.0, 7.5$ (green and red, respectively).}
\label{fig:colour_selection}
\end{figure*}

Having vetted our selection procedure using FLARES, we now show the two main colour cuts used to select smouldering galaxy candidates from the JADES data in Fig.~\ref{fig:colour_selection}. Our $z \sim 6$ and $z \sim 7$ samples are shown in the left and right panels, respectively. Smouldering galaxies, which have level photometry in the medium and wide bands (i.e.\@ within the dashed lines in the diagram), that also satisfy our further quality colour cuts (Equation~\ref{eq:supplementary} and \ref{eq:cut1}/\ref{eq:cut2}, if relevant) are shown as orange squares. The remaining galaxies in the same redshift interval, which seemingly have stronger emission lines, constitute our control sample and are shown as blue circles. We select 12 and 4 smouldering galaxy candidates at $z \sim 6$ and $z \sim 7$, respectively. The control sample constitutes 338 and 71 galaxies at these redshifts, respectively. 

\begin{figure*}
\centering
\includegraphics[width=.425\linewidth] {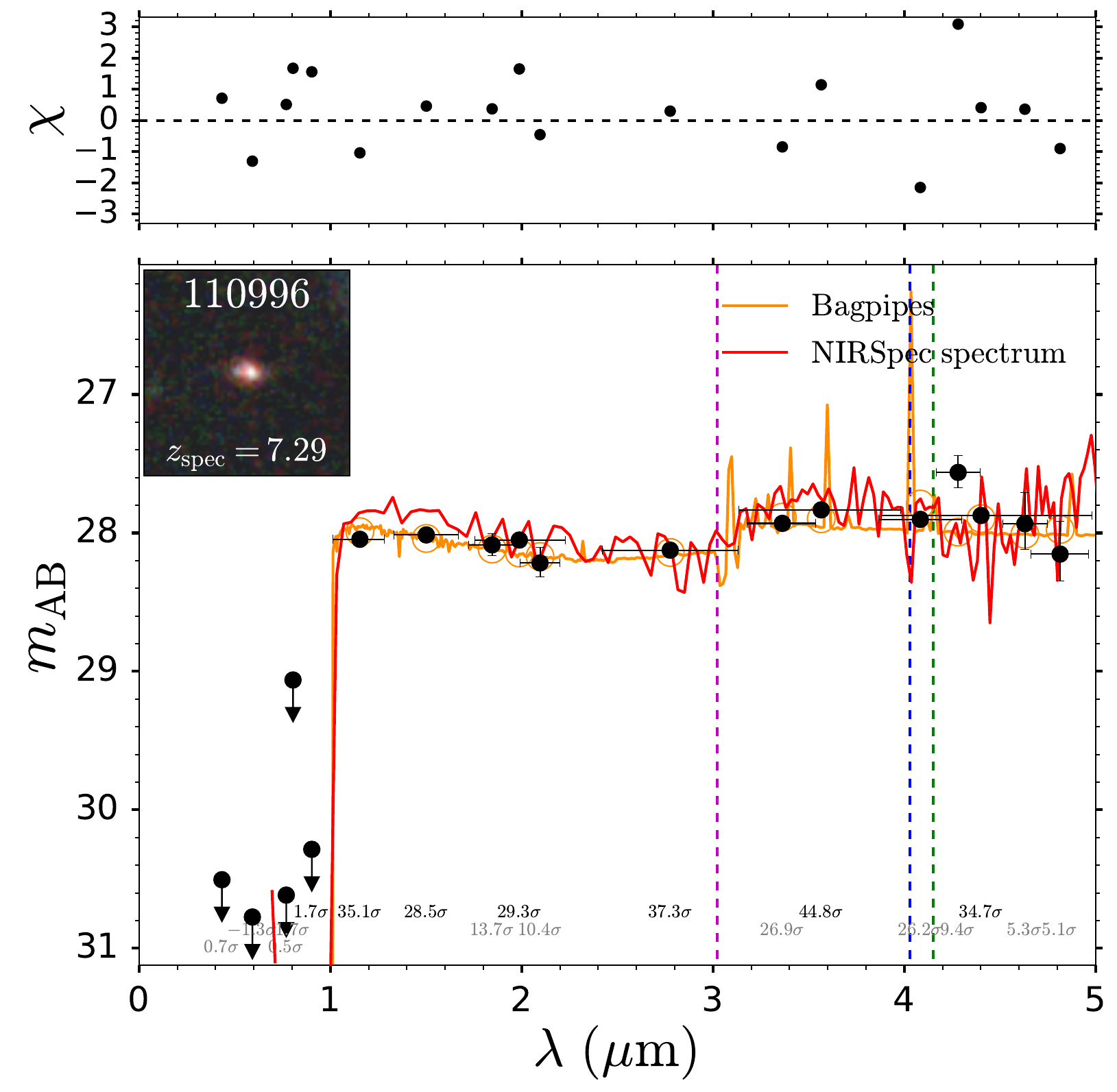} \hfill
\includegraphics[width=.425\linewidth] {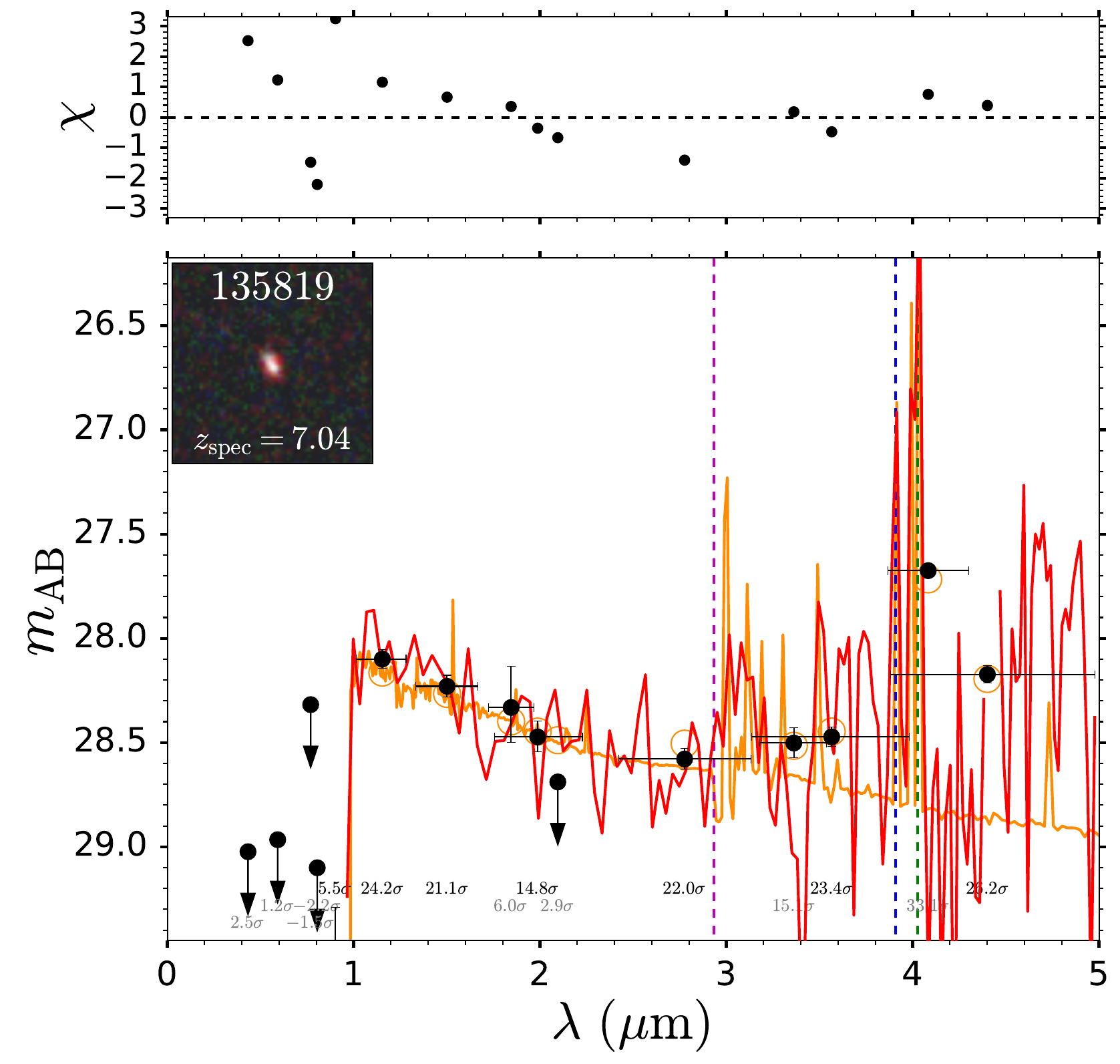} \\[4.5ex]
\includegraphics[width=.425\linewidth]{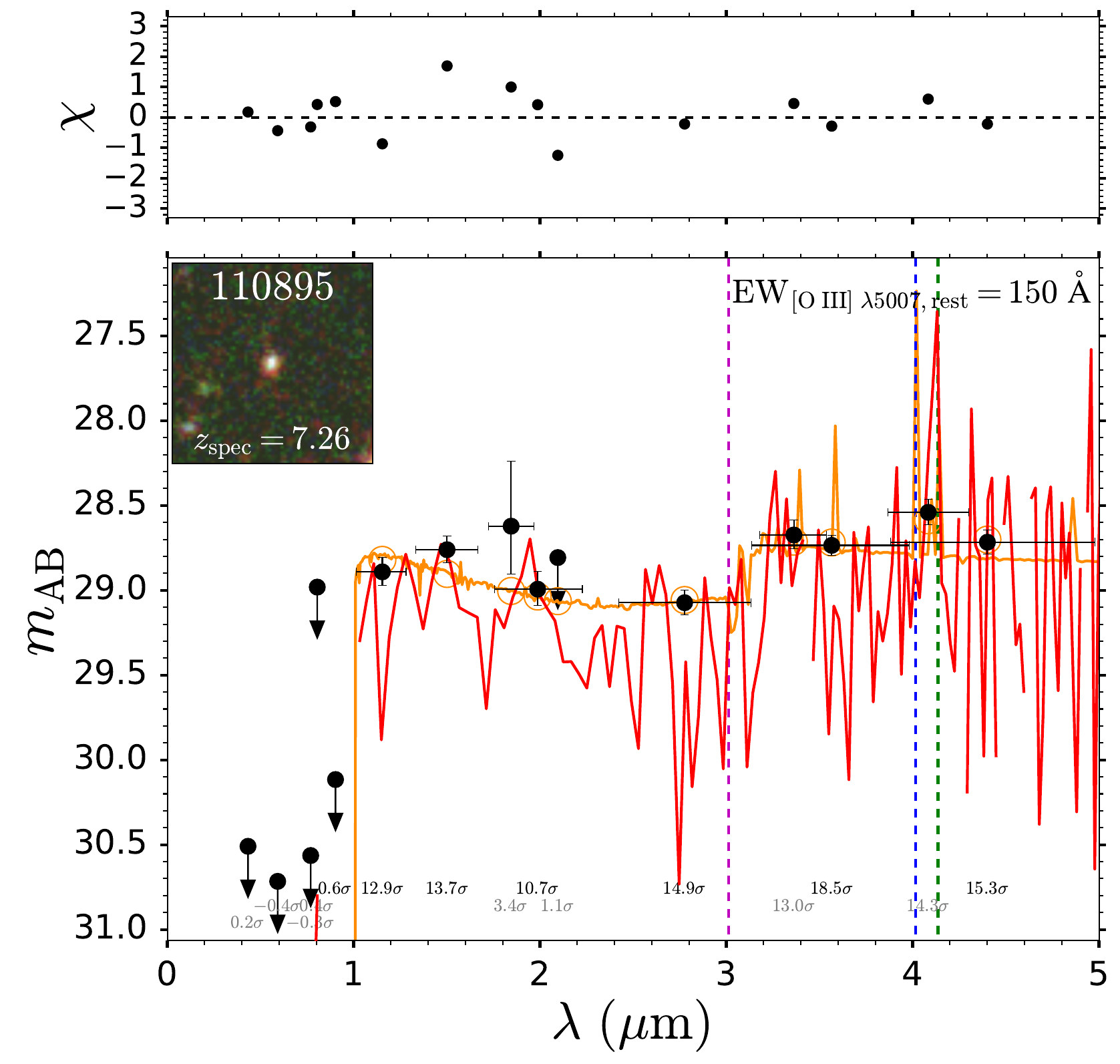} \hfill 
\includegraphics[width=.425\linewidth] {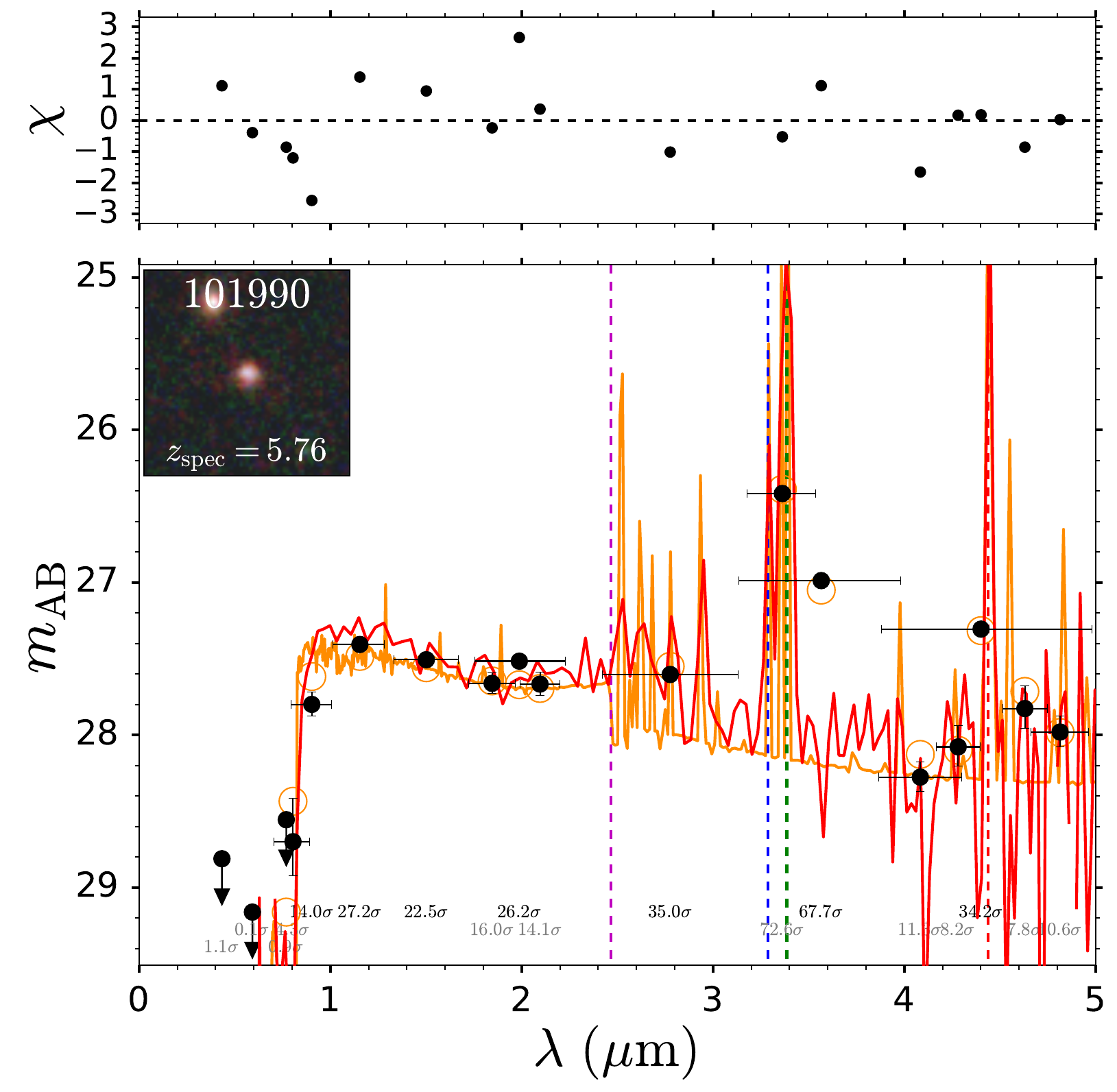} \\[4.5ex]
\caption{Top-left panel: \emph{HST} + NIRCam photometry (black), median {\tt Bagpipes} fit to this photometry (orange) and NIRSpec spectrum (red) of the $z = 7.29$ quiescent galaxy JADES-GS-z7-01-QU reported by \citet{Looser2024}, that is selected through our smouldering galaxy colour cut procedure. The JADES catalog ID and spectroscopic (or for other sources the photometric) redshift is noted in the RGB (R: F444W; G: F200W; B: F115W, $2\times2$~arcsec) cutout at the top-left, with the SNR on the photometry shown at the bottom, with non-detections (${<}3\sigma$) displayed as downward arrows at the $3\sigma$ depth. $\chi$ values for the photometry are presented in the upper subpanel. Vertical dashed lines in the main panel note the location of the Balmer break (purple), \Hb\ (blue), \OIII\ $\lambda5007$ (green) and \Ha\ (red, if applicable). The medium and wide band photometry of this source is level redward of the small Balmer break, indicative of weak, or as established by the spectroscopy, non-existent emission lines. Bottom-left panel: A spectroscopically-confirmed $z = 7.26$ source that is not selected as a smouldering galaxy candidate following our procedure, owing to the non-negligible photometric excess in the F410M filter relative to F444W, indicating moderately strong line emission, which is confirmed to be \OIII\ $\lambda5007$ with $\mathrm{EW_{rest}} = 150$~\AA\ by NIRSpec. Right panels: Emission-line galaxies at $z=7.04$ (top) and $z=5.76$ (bottom) that form part of our control sample, due to the substantial offset between F410M and F444W (top and bottom), and F335M and F356W (bottom) indicating very strong line emission.}
\label{fig:spectroscopic_examples}
\end{figure*}

We note that following our selection procedure, we do indeed select the spectroscopically-confirmed quiescent galaxy JADES-GS-z7-01-QU reported in \citet{Looser2024}. The NIRCam + \emph{HST} photometry (black), publicly-released \citep{Bunker2024} JADES NIRSpec spectrum ($4\times$ binned for clarity of the continuum level, red) and {\tt Bagpipes} fit to the photometry (orange) are shown in the top-left panel of Fig.~\ref{fig:spectroscopic_examples}. As with the other SED examples shown in this paper, the spectroscopic (if available) or JADES catalog photometric redshift is displayed within the RGB (R: F444W; G: F200W; B: F115W, $2\times2$~arcsec) cutout in the top-left, with the SNR in each photometric band reported at the bottom below each respective datapoint. Non-detections (${<}3\sigma$) are displayed as downward arrows. $\chi$ values for the photometry are presented in the upper subpanel. The purple, blue, green and (in future figures) red vertical dashed lines in the main panel correspond to the locations of the Balmer break, \Hb, \OIII\ $\lambda 5007$ and \Ha\ emission lines, respectively. As can be seen from the figure, the photometry redward of the Balmer break is relatively level, with the (F335M, F356W) and (F410M, F444W) flux densities being very comparable. We do note the discrepant F430M flux density (not seen in the JADES DR1 photometric catalog), indicating potential outstanding difficulties with the photometric flux calibration. As discussed extensively in \citet{Looser2024}, the deep spectrum of this source reveals no emission lines. Though we note that the {\tt Bagpipes} fit to the photometric data exhibits weak but non-zero emission lines ($\mathrm{EW}_{\mathrm{[O\ III\ + H\beta]}} \approx 100~$\AA), which highlights the level of inference on weak emission-line strengths that can reliably be made given the error budget (here adopting a 5 per cent error floor) on photometric data. Moreover, both the photometry and the spectrum exhibit a minor Balmer break (${\sim}0.3$~mag). As noted by \citet{Looser2024}, this Balmer break would be much too weak to identify this system as quiescent using the conventional rest-frame UVJ colour diagram. However, our approach, which instead selects based on a lack of emission lines, can identify such young quiescent systems.

In the bottom-left panel of Fig.~\ref{fig:spectroscopic_examples}, we also show the photometry, {\tt Bagpipes} fit and public-released JADES spectrum of a spectroscopically-confirmed source not selected following our procedure. Here the F410M measurement is elevated ${\sim}0.15$~mag above the F444W reading, indicating the presence of \OIII\ + \Hb\ line emission of weak-to-moderate strength, thus not satisfying our smouldering galaxy colour selection criterion. Indeed, from the NIRSpec spectrum we see the \OIII\ $\lambda5007$ emission line (detected at 8.5$\sigma$), which from our own investigations has a rest-frame equivalent width of 150~\AA, exactly as would be expected from the photometric excess in the F410M filter. These two examples therefore help to validate our colour selection methodology for identifying smouldering galaxy candidates with weak rest-frame optical emission lines. As constrast, we further show two emission-line galaxies at $z = 7.04$ and $z=5.76$ in the top-right and bottom-right panels of Fig.~\ref{fig:spectroscopic_examples}, respectively. These two galaxies form part of our control sample, exhibiting a substantial offset between F410M and F444W (top and bottom), and F335M and F356W (bottom), indicating very strong line emission, which is clearly evident from the NIRSpec spectra.

We briefly note that the two sources, JADES 110996 and JADES 110895, are closely separated on the sky, by $\Delta z = 0.03$ ($=1200$~kpc physical along the line-of-sight) and 6~arcsec ($=30$~kpc physical in projection). Given the quiescent nature of the former, and the relatively weak line emission in the latter, this example may indicate that there is a connection between the smouldering nature of some sources and their wider environment, at least for some galaxies. We defer a detailed study of the environment and morphologies (possibly indicative of mergers) of smouldering galaxies to future work. 

\begin{figure*}
\centering
\includegraphics[width=.425\linewidth] {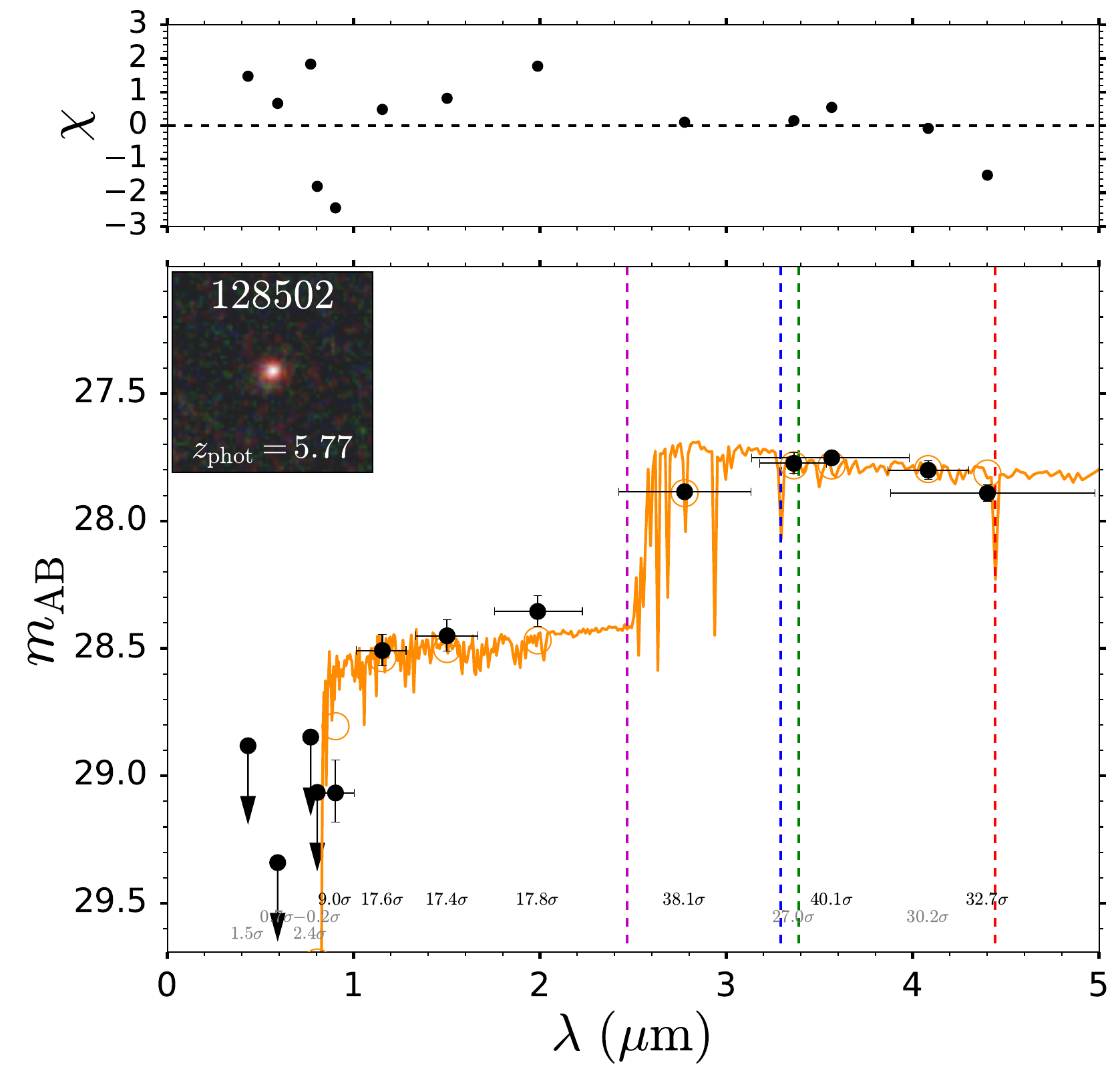} \hfill
\includegraphics[width=.425\linewidth]{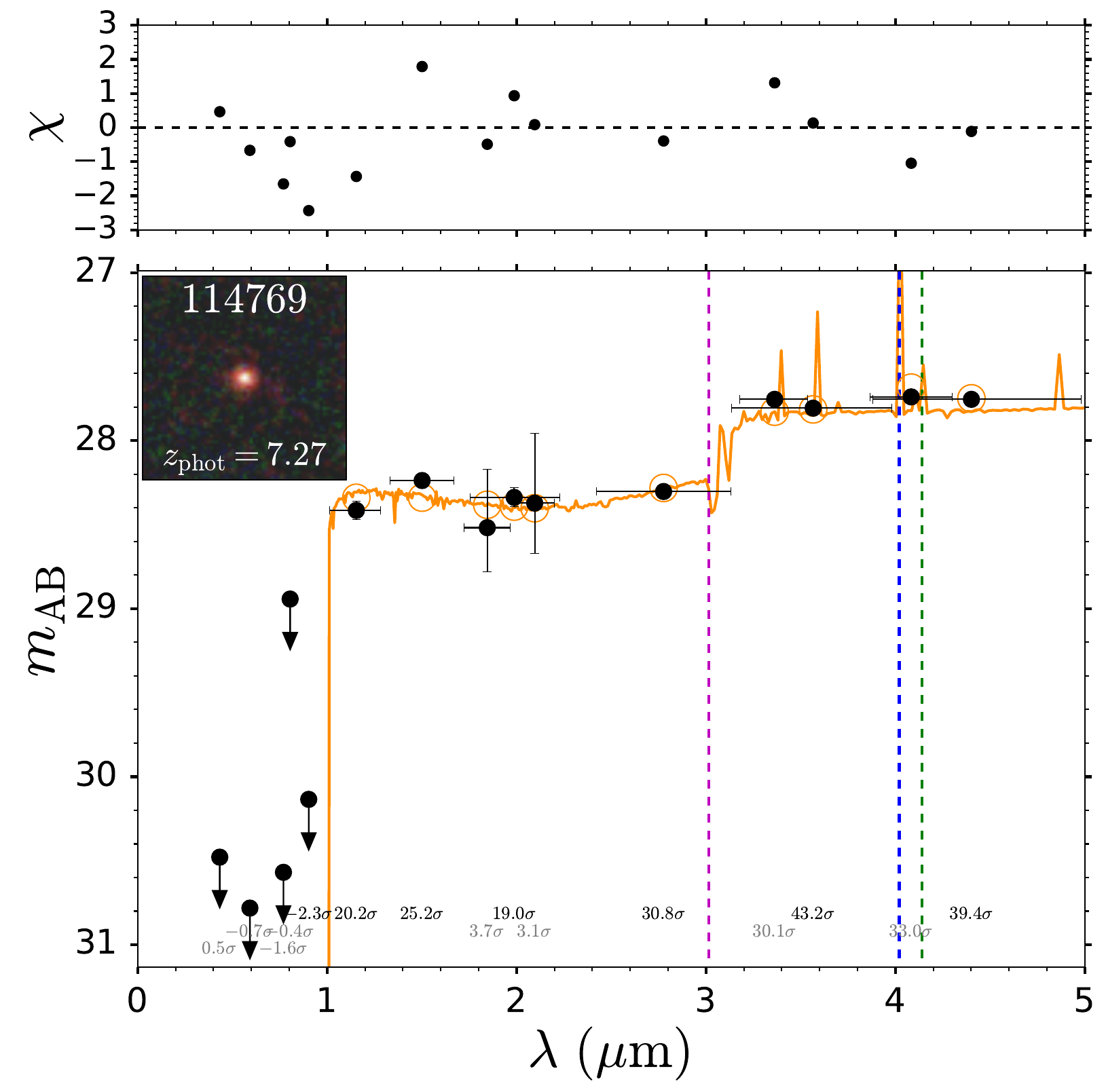} \\[4.5ex]
\includegraphics[width=.425\linewidth] {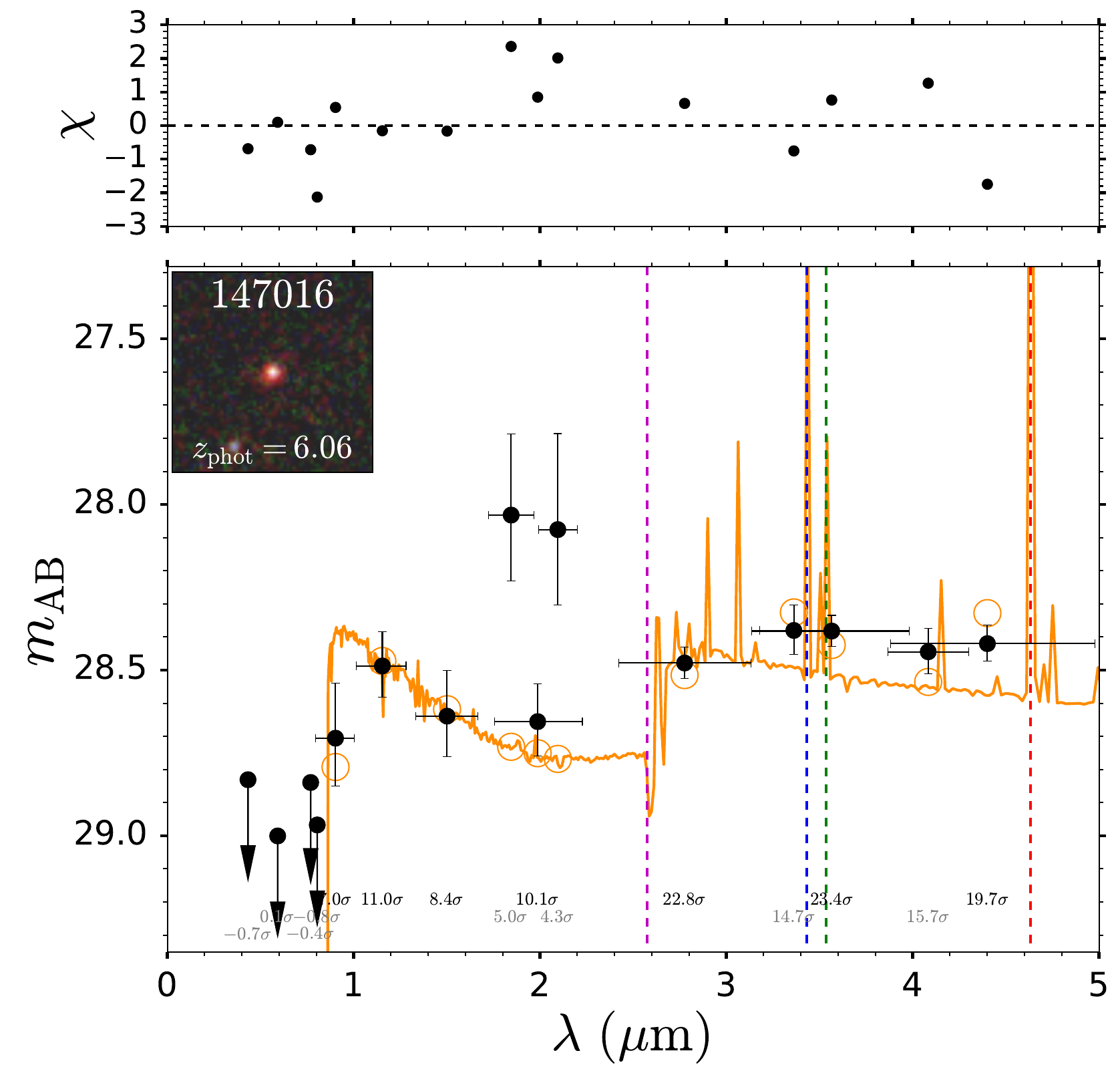} \hfill
\includegraphics[width=.425\linewidth]{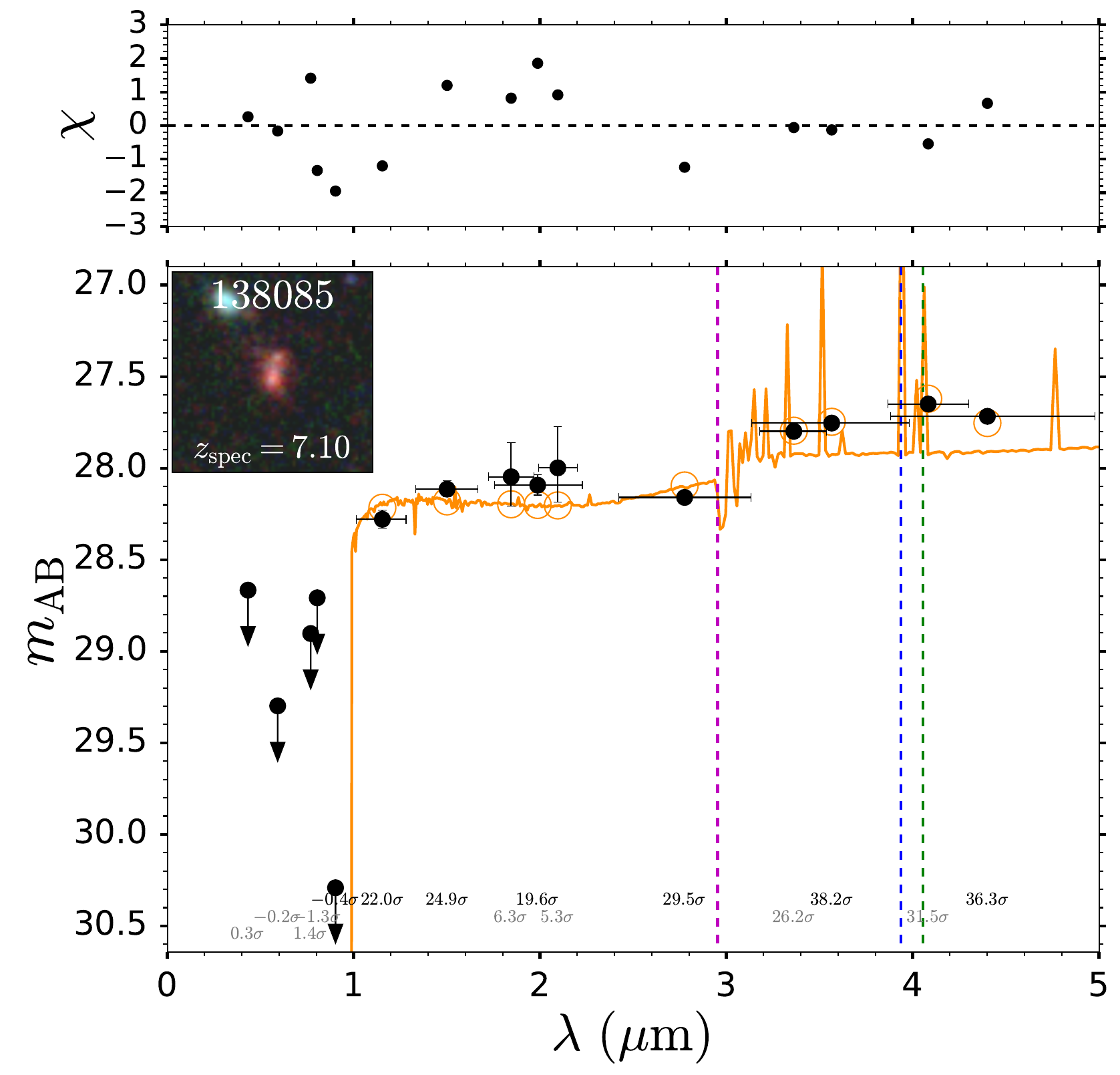} \\[4.5ex]
\includegraphics[width=.425\linewidth] {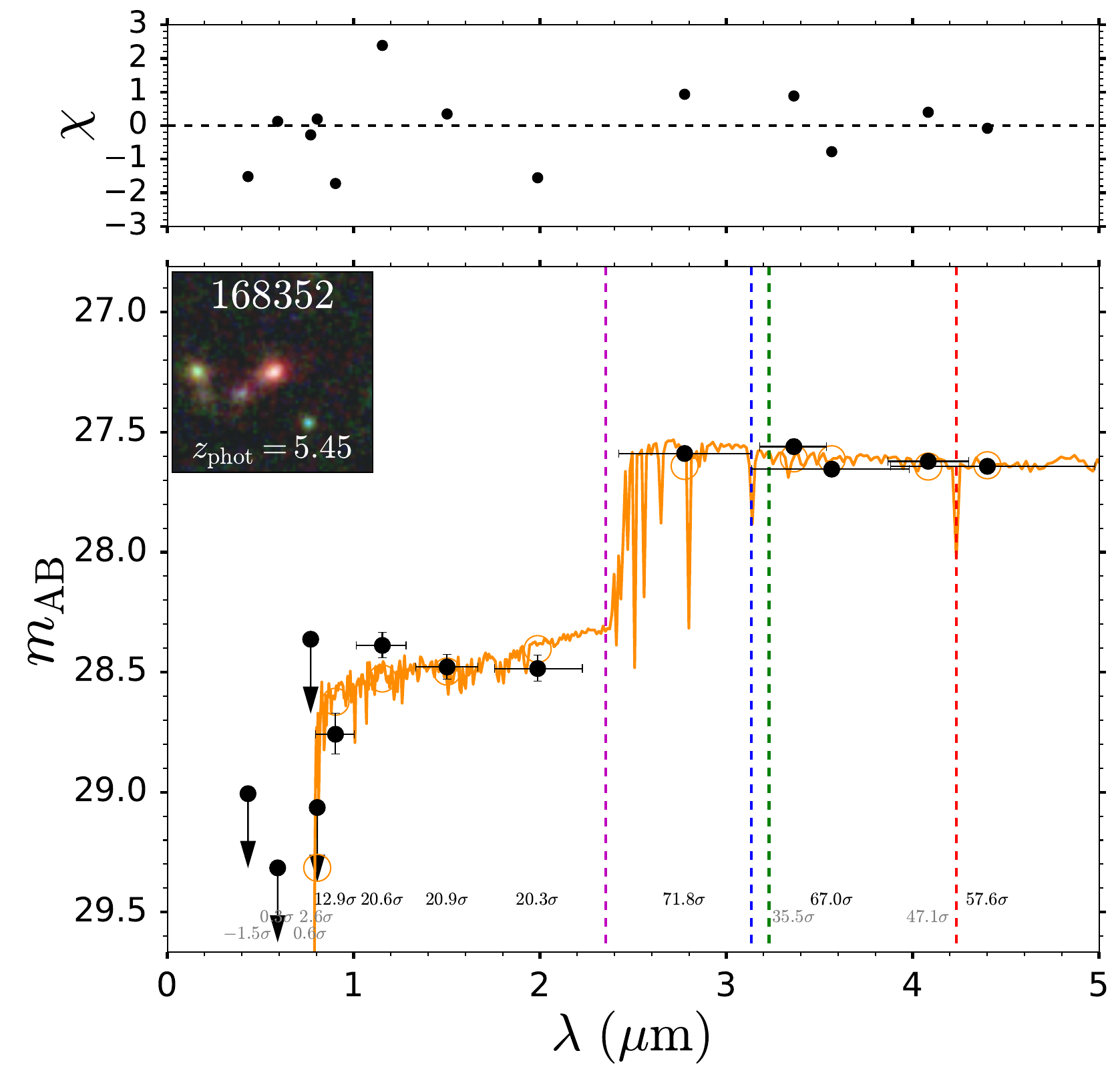} \hfill
\includegraphics[width=.425\linewidth]{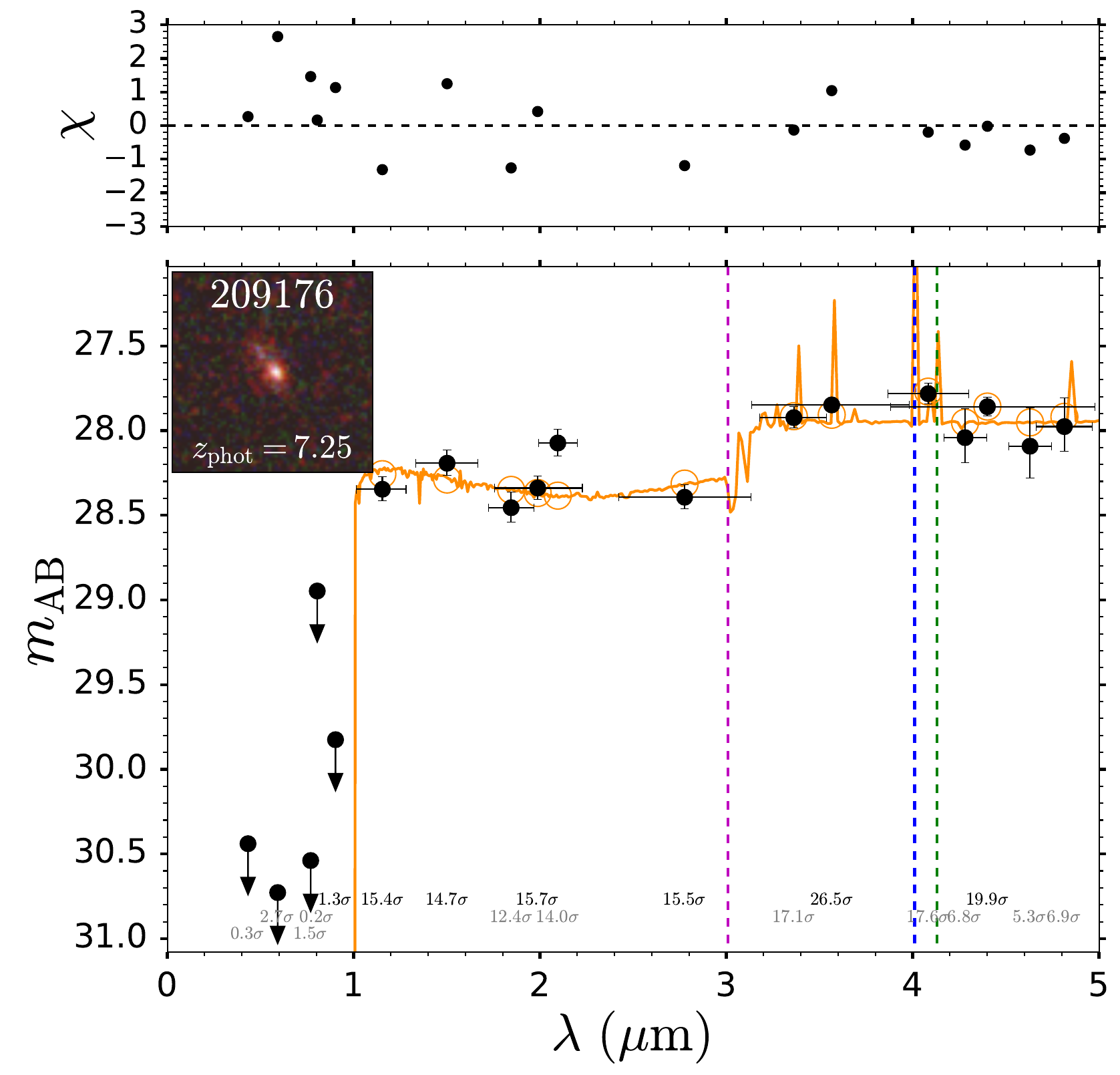} 
\caption{\emph{HST} + NIRCam photometry (black) and median {\tt Bagpipes} fits (orange) for example smouldering galaxy candidates at $z \sim 6$ ($5.3 < z < 6.6$, left panels) and the remainder (i.e.\@ in addition to JADES-GS-z7-01-QU in Fig.~\ref{fig:spectroscopic_examples}) of the $z \sim 7$ ($6.8 < z < 7.8$) smouldering sample (right). The format is similar to Fig.~\ref{fig:spectroscopic_examples}, with the vertical red dashed line indicating the location of \Ha\ for the $z \sim 6$ sample. The smouldering galaxy candidates are colour-selected to have weak emission lines, which is generally indicative of relatively low levels of ongoing star formation and perhaps even (temporary) quiescence. Whilst we do not explicitly select these systems to have Balmer breaks, a non-negligible Balmer break is a common consequence of demanding the line emission to be weak in these systems.} 
\label{fig:candidates}
\end{figure*}

We show additional examples of smouldering galaxy candidates selected using our methodology in Fig.~\ref{fig:candidates}. As with the other two column figures in this work, the $z \sim 6$ and $z \sim 7$ samples are shown in the left and right panels, respectively. As can be seen from the photometry and SEDs of these different sources, although we do not explicitly select for a Balmer break in our methodology, having a Balmer break is a common consequence of mandating that the rest-frame optical emission lines in these systems are weak. Such smouldering systems likely have little-to-no active star formation, allowing the older and fainter stellar populations, with their (moderately-sized) Balmer breaks, to dominate the spectrum. 

We report the coordinates, redshifts, colours, stellar masses, $\mathrm{sSFRs}$, inferred emission line equivalent widths, rest-frame UV colours, and proxies for the Balmer break strength for our $z \sim 6$ and $z \sim 7$ smouldering candidates in Tables~\ref{tab:z6_candidates} and \ref{tab:z7_candidates}, respectively. From our investigations in stellar mass recovery using mock photometry outlined in Appendix~\ref{app:masses}, we anticipate that the stellar masses derived for our $M_* \sim 10^{8\text{--}9}~\mathrm{M}_\odot$ smouldering galaxy candidates may be systematically overestimated by ${\sim}0.2$~dex.

\begin{table*}
\begin{center}
\begin{tabular}{|c|c|c|c|c|c|c|c|c|c|} 
\hline
ID & R.\@A.\@ & Dec.\@ & $z$ & $\log M_*$ & $\log \mathrm{sSFR}_{10}$ & \OIII\ + \Hb\ EW & \Ha\ EW & F115W$-$F200W & F200W$-$F356W \\
 &  & & & ($\mathrm{M}_\odot$) & ($\mathrm{yr}^{-1}$) & (\AA) & (\AA) & (mag) & (mag) \\
\hline
8810 & 53.05199 & -27.89676 & $6.08^{+0.17}_{-0.73}$ & $8.20^{+0.07}_{-0.06}$ & $-8.11^{+0.12}_{-0.13}$ & $201.8^{+50.7}_{-113.2}$ & $267.3^{+95.5}_{-119.3}$ & $0.00 \pm 0.07$ & $0.15 \pm 0.06$ \\
42312 & 53.03911 & -27.86860 & $5.63^{+0.15}_{-1.56}$ & $8.36^{+0.20}_{-0.23}$ & $-8.81^{+0.48}_{-0.67}$ & $5.8^{+15.8}_{-5.2}$ & $21.2^{+97.3}_{-19.4}$ & $0.12 \pm 0.06$ & $-0.11 \pm 0.05$ \\
57559 & 53.11022 & -27.85913 & $6.13^{+0.08}_{-0.39}$ & $8.39^{+0.04}_{-0.06}$ & $-10.17^{+1.57}_{-8.62}$ & $1.6^{+102.5}_{-1.6}$ & $7.4^{+321.2}_{-7.4}$ & $-0.33 \pm 0.13$ & $0.63 \pm 0.11$ \\
128502 & 53.19546 & -27.77683 & $5.77^{+0.18}_{-3.99}$ & $8.53^{+0.02}_{-0.02}$ & $-13.32^{+3.19}_{-7.25}$ & $0.0^{+1.8}_{-0.0}$ & $0.0^{+2.9}_{-0.0}$ & $0.15 \pm 0.09$ & $0.60 \pm 0.07$  \\
147016 & 53.17817 & -27.75437 & $6.06^{+0.28}_{-0.47}$ & $8.25^{+0.09}_{-0.13}$ & $-8.43^{+0.17}_{-0.20}$ & $106.0^{+63.0}_{-50.1}$ & $426.8^{+71.7}_{-206.6}$ & $-0.17 \pm 0.15$ & $0.27 \pm 0.12$ \\
166576 & 53.04712 & -27.88186 & $5.58^{+0.03}_{-1.19}$ & $8.72^{+0.05}_{-0.05}$ & $-10.27^{+1.44}_{-3.73}$ & $1.3^{+57.5}_{-1.3}$ & $6.3^{+258.5}_{-6.3}$ & $-0.07 \pm 0.17$ & $0.63 \pm 0.11$ \\
168352 & 53.07772 & -27.87958 & $5.45^{+0.11}_{-0.36}$ & $8.63^{+0.03}_{-0.02}$ & $-11.68^{+2.32}_{-7.53}$ & $0.1^{+24.4}_{-0.1}$ & $0.1^{+27.6}_{-0.1}$ & $-0.10 \pm 0.08$ & $0.83 \pm 0.06$ \\
200687 & 53.13262 & -27.80595 & $5.35^{+0.08}_{-0.39}$ & $8.44^{+0.09}_{-0.07}$ & $-8.76^{+0.14}_{-0.47}$ & $171.4^{+67.3}_{-153.8}$ & $133.1^{+49.9}_{-89.3}$ & $0.04 \pm 0.12$ & $0.49 \pm 0.09$ \\
200992 & 53.16401 & -27.80504 & $5.83^{+0.18}_{-0.10}$ & $8.67^{+0.04}_{-0.06}$ & $-8.79^{+0.09}_{-0.14}$ & $204.5^{+30.2}_{-43.2}$ & $146.2^{+35.5}_{-33.1}$ & $-0.14 \pm 0.18$ & $1.10 \pm 0.15$ \\
208681 & 53.13047 & -27.77825 & $5.86^{+0.17}_{-0.20}$ & $9.12^{+0.07}_{-0.05}$ & $-8.79^{+0.06}_{-0.09}$ & $108.3^{+34.2}_{-44.4}$ & $189.7^{+40.3}_{-22.2}$ & $-0.08 \pm 0.07$ & $0.80 \pm 0.05$ \\
211492 & 53.18694 & -27.76988 & $6.05^{+0.04}_{-0.22}$ & $8.29^{+0.06}_{-0.08}$ & $-8.87^{+0.40}_{-1.69}$ & $19.6^{+38.4}_{-19.3}$ & $88.5^{+148.0}_{-87.2}$ & $-0.37 \pm 0.09$ & $0.15 \pm 0.09$ \\
296750 & 53.08069 & -27.83523 & $5.50^{+0.19}_{-0.28}$ & $8.51^{+0.09}_{-0.10}$ & $-8.61^{+0.11}_{-0.11}$ & $158.4^{+41.1}_{-30.8}$ & $208.9^{+49.9}_{-66.1}$ & $-0.06 \pm 0.07$ & $0.56 \pm 0.06$ \\
\hline
\end{tabular}
\caption{JADES DR2 public catalog ID, coordinates, photometric redshift, total stellar mass $\log M_*$, specific star formation rate $\mathrm{sSFR}_{10}$ (using the average SFR over the past 10~Myr), combined \OIII\ $\lambda\lambda4959, 5007$ + \Hb\ rest-frame equivalent width inferred using {\tt Bagpipes}, \Ha\ equivalent width, rest-frame UV colour F115W$-$F200W and proxy for the Balmer break strength F200W$-$F356W for our $z \sim 6$ ($5.3 < z < 6.6$) sample of smouldering galaxy candidates.}
\label{tab:z6_candidates}
\end{center}
\end{table*}

\begin{table*}
\begin{center}
\begin{tabular}{|c|c|c|c|c|c|c|c|c|c|} 
\hline
ID & R.\@A.\@ & Dec.\@ & $z$ & $\log M_*$ & $\log \mathrm{sSFR}_{10}$ & \OIII\ + \Hb\ EW & \Ha\ EW & F150W$-$F277W & F277W$-$F444W \\
 &  & & & ($\mathrm{M}_\odot$) & ($\mathrm{yr}^{-1}$) & (\AA) & (\AA, extrapolated) & (mag) & (mag) \\
\hline
110996$^*$ & 53.15508 & -27.80178 & $7.29$ (spec)$^\dagger$ & $8.54^{+0.07}_{-0.08}$ & $-8.19^{+0.13}_{-0.15}$ & $112.0^{+24.0}_{-29.0}$ & $588.1^{+132.3}_{-199.9}$ & $-0.11 \pm 0.05$ & $0.25 \pm 0.04$ \\
114769 & 53.18632 & -27.79561 & $7.27^{+0.28}_{-0.23}$ & $8.90^{+0.05}_{-0.07}$ & $-8.60^{+0.15}_{-0.07}$ & $81.6^{+23.7}_{-13.7}$ & $389.7^{+75.4}_{-55.1}$ & $-0.07 \pm 0.06$ & $0.55 \pm 0.04$ \\
138085 & 53.19208 & -27.76662 & $7.10$ (spec)$^\ddagger$ & $8.78^{+0.09}_{-0.10}$ & $-8.01^{+0.19}_{-0.19}$ & $173.5^{+50.6}_{-38.0}$ & $769.1^{+183.0}_{-144.3}$ & $-0.05 \pm 0.06$ & $0.44 \pm 0.05$ \\
209176 & 53.17867 & -27.77633 & $7.25^{+0.62}_{-0.20}$ & $8.87^{+0.06}_{-0.07}$ & $-8.47^{+0.17}_{-0.17}$ & $104.4^{+32.0}_{-24.2}$ & $458.1^{+98.2}_{-78.0}$ & $-0.20 \pm 0.10$ & $0.53 \pm 0.09$ \\
\hline
\end{tabular}
\caption{Similar to Table~\ref{tab:z6_candidates}, but now for our $z \sim 7$ ($6.8 < z < 7.8$) sample of smouldering galaxy candidates. Due to lack of direct \Ha\ constraints, the inferred EW is now based on extrapolation. The F150W$-$F277W and F277W$-$F444W colours now trace the rest-frame UV slope and Balmer break strength, respectively. $^*$The $z=7.29$ recently-quenched galaxy JADES-GS-z7-01-QU reported by \citet{Looser2024}. $^\dagger$Spectroscopic redshift from JADES NIRSpec observations. $^\ddagger$Spectroscopic redshift from FRESCO NIRCam slitless spectroscopic observations reported in the JADES DR1 release.}
\label{tab:z7_candidates}
\end{center}
\end{table*}

\section{Smouldering galaxy properties} \label{sec:properties}

In this section we discuss the properties inferred for our smouldering galaxy candidates, comparing and contrasting these against the control sample of galaxies. In Section \ref{subsec:sf_properties} we discuss the star-formation properties of the smouldering galaxies, as inferred from SED-fitting the \emph{HST}+NIRCam photometry. In Section \ref{subsec:number_abundances}, we discuss the number abundances of smouldering galaxies, also comparing against the predictions from various cosmological simulations. 

\subsection{Star-formation properties} \label{subsec:sf_properties}

Since the basis for the identification of smouldering galaxy candidates is their weak line emission as imprinted on the medium and wide band photometry, we now verify whether these systems are indeed inferred to generally be weak line-emitters once the full photometry and observational errors are taken into account. 

\begin{figure*}
\centering
\includegraphics[width=\linewidth]{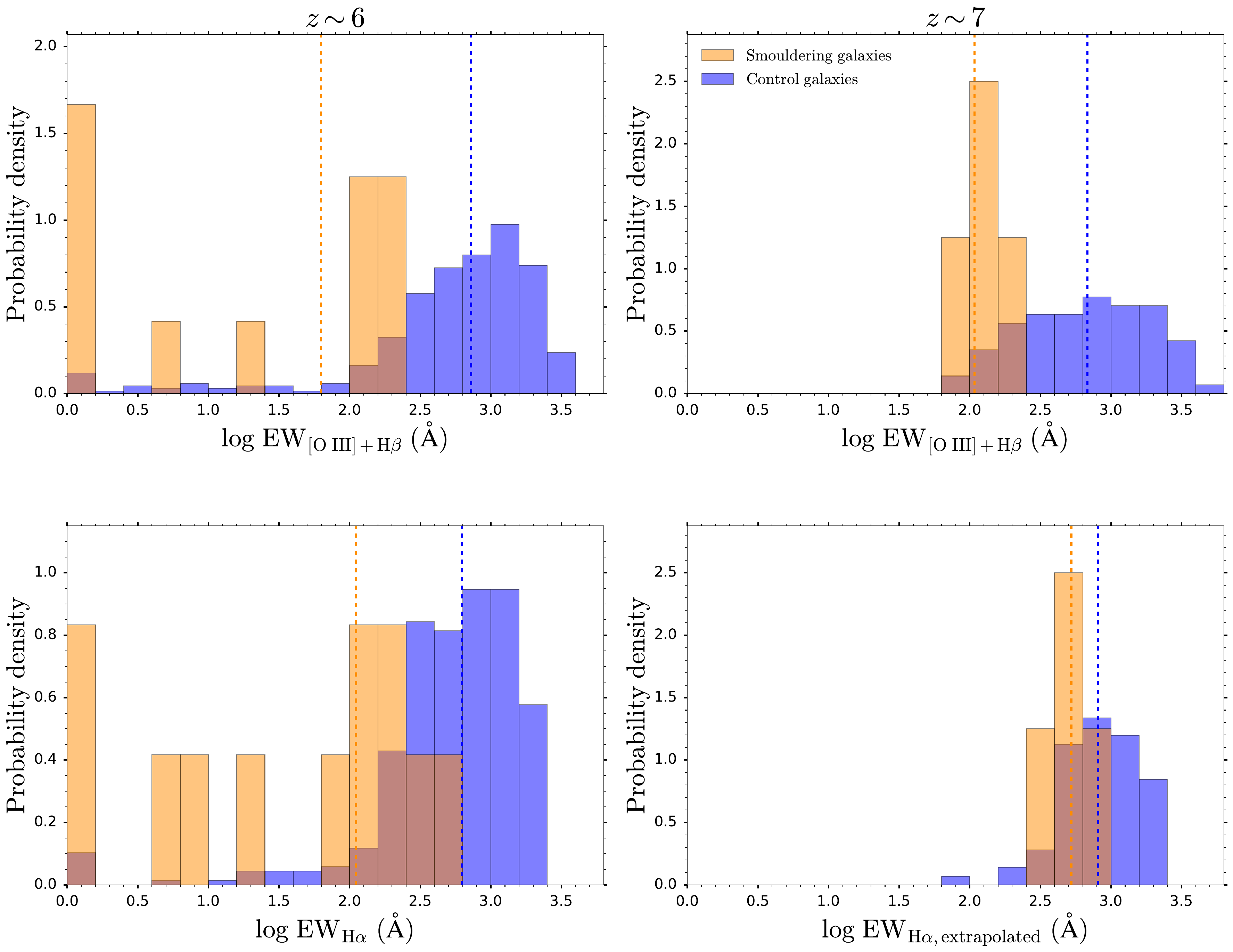}
\caption{Normalised histograms of the rest-frame emission line equivalent widths inferred by {\tt Bagpipes} for \OIII\ + \Hb\ (top panels) and \Ha\ (bottom) for the $z\sim 6$ sample (left) and $z \sim 7$ sample (right). As expected based on the colour selection criterion, smouldering galaxies (orange) generally have lower inferred equivalent widths than the control galaxies (blue), with the vertical dashed lines denoting the median value for each type. As \Ha\ is redshifted out of the NIRCam range by $z = 6.6$, we have no direct constraints on the \Ha\ EW for the $z \sim 7$ sample, the values reported being based on extrapolation from the weak \OIII\ + \Hb\ emission. For this reason, our colour selections are likely somewhat less effective at only selecting weak \Ha\ emitters in this redshift range, this requiring MIRI F560W imaging. Direct NIRSpec follow-up on these smouldering candidates to further constrain their \OIII\ + \Hb\ emission would therefore perhaps be more efficient.}
\label{fig:ew_histograms}
\end{figure*}

As can be seen from Fig.~\ref{fig:ew_histograms}, our smouldering galaxy candidates (orange) typically are inferred to have much lower emission line rest-frame equivalent widths (i.e.\@ weaker emission lines) than the control sample (blue). As expected from our earlier discussion of the selection method, the median rest-frame equivalent width inferred is ${\sim}100$~\AA, consistent with the ${<}0.1$~mag offset between medium and wide band photometry that we impose. This is substantially smaller than the median equivalent width inferred (${\sim}600$~\AA) for the control sample, consistent with the notion that galaxies in the EoR are typically very strong line-emitters \citep[e.g.\@][]{Matthee2023, Matthee2024, Endsley2024}. Thus our smouldering galaxy colour selection method is successful at preferentially selecting weak line-emitting systems. 

We do note that there are some (though few) control galaxies that are inferred to have weak emission lines. These are often systems that either narrowly miss our medium--wide-band ($\vert \Delta m \vert < 0.1$~mag) selection criteria (by a few hundredths of a mag), systems that fail the wide--wide ($\vert \Delta m \vert < 0.15$~mag) selection criterion due to a particularly blue or red rest-frame optical slope, or are systems that are only ${\sim}10\sigma$-detected and thus likely fail our selection criterion due to noise in the data shifting the observed photometry by more than $1\sigma$ (= 0.1~mag) from the model-predicted values.

Furthermore, the \Ha\ equivalent widths inferred for the $z \sim 7$ smouldering galaxies are relatively large (${\sim}500$~\AA) and thus only slightly smaller than those for the control sample (${\sim}800$~\AA). As pointed out earlier, this comes about because we have no direct \Ha\ constraints for these $z > 6.6$ systems, the line emission being redshifted out of the NIRCam wavelength range. Hence the \Ha\ equivalent widths inferred are based on extrapolation. Still, the weak \OIII\ + \Hb\ emission demanded by our colour cut does result in a generally smaller inferred \Ha\ equivalent width for the smouldering galaxies, as the \OIII\ + \Hb\ and \Ha\ line strengths are related. Indeed, assuming the \Hb\ emission dominates over \OIII, then our colour cut implies that $\mathrm{EW}_{\mathrm{H}\beta} < 100$~\AA. Assuming the standard ratio between the \Ha\ and \Hb\ flux of 2.86, combined with the fact that the continuum level between \Ha\ and \Hb\ (in $f_\nu$) for smouldering galaxies is approximately flat (thus $f_\lambda$ decreases with increasing wavelength), this implies that $\mathrm{EW}_{\mathrm{H}\alpha} < 500$~\AA. Hence, even if the \Hb\ emission is constrained to be weak, the \Ha\ emission can still be relatively strong. As we will see, this implies that the $z \sim 7$ smouldering galaxies we select can still (be inferred to) have relatively high sSFRs. 

Direct constraints on the \Ha\ emission can be placed, and thus an improved identification of true smouldering systems at $z \sim 7$ can be achieved, through MIRI F560W imaging. However, owing to the lower sensitivity (${\sim}2$~mag) and imaging footprint ($4\times$) of MIRI over NIRCam, this will be exceedingly expensive. Thus perhaps a more efficient approach would be to directly follow-up with NIRSpec spectroscopy on the sources colour-selected to have weak \OIII\ +  \Hb\ line emission. Similar to JADES-GS-z7-01-QU, these observations would provide the definitive constraints on the line emission (or lack thereof), and in the case of PRISM continuum spectroscopy, would further constrain the stellar population parameters, thus giving the best insights on the star-formation histories of these smouldering systems.

\begin{figure*}
\centering
\includegraphics[width=\linewidth]{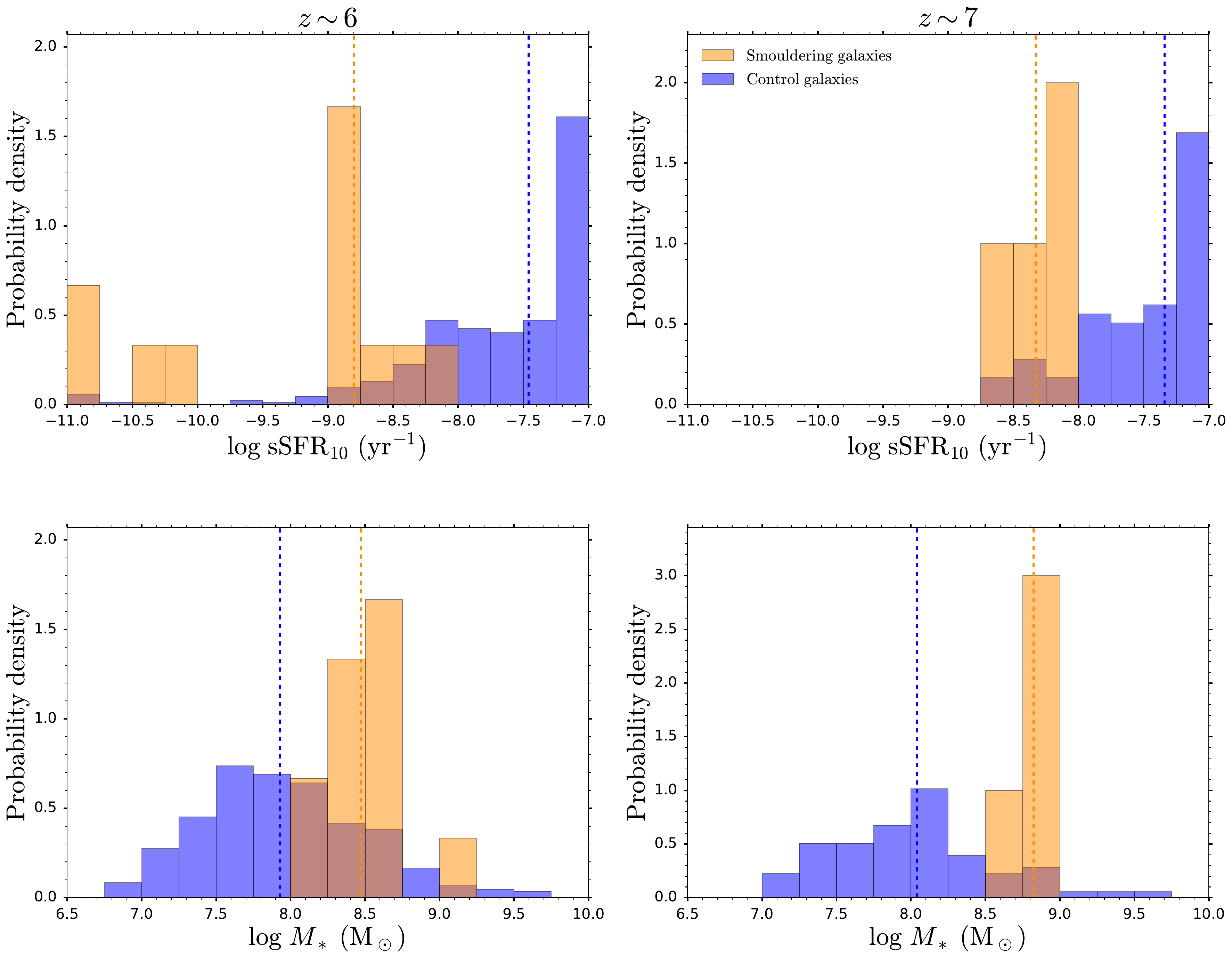}
\caption{Similar to Fig.~\ref{fig:ew_histograms}, but now showing the inferred specific star formation rates $\mathrm{sSFR}_{10}$ (using the average SFR over the past 10~Myr, top panels) and the total stellar mass (bottom panels). Following on from their weaker emission lines, smouldering galaxies are generally inferred to have lower sSFR than the control sample, though this is not as prominent at $z \sim 7$ due to the lack of direct constraints on the \Ha\ EW. Smouldering galaxies are also generally inferred to be more massive than the control sample. This is likely due to incompleteness, with the less massive smouldering galaxies, owing to their lower light-to-mass ratios than the control sample, likely being too faint to be detected and/or pass our SNR cuts, hence biasing the smouldering galaxy sample towards higher masses.}
\label{fig:ssfr_mass_histograms}
\end{figure*}

We show the inferred sSFRs (top panels) and stellar masses (bottom panels) for the smouldering and control galaxies in Fig.~\ref{fig:ssfr_mass_histograms}. Here the specific star formation rates consider the average SFR over the past 10~Myr, i.e.\@ sSFR$_{10}$. For clarity, sSFRs below $10^{-11}$~yr$^{-1}$ are plotted at $10^{-11}$~yr$^{-1}$. Furthermore, the stellar masses incorporate the aperture correction going from 0.3~arcsec diameter apertures (used in the SED-fitting) to the total Kron elliptical apertures, utilising the F277W and F356W corrections for the $z \sim 6$ and $z \sim 7$ samples, respectively. 

As expected from the selection procedure, smouldering galaxies, with their weaker emission lines, tend to be inferred to have lower specific star formation rates. Indeed, for the $z \sim 6$ sample, the median inferred $\mathrm{sSFR}_{10}$ for smouldering galaxies ($10^{-8.80}$~yr$^{-1}$) is an order of magnitude lower than for the control sample ($10^{-7.50}$~yr$^{-1}$). This is also true for the $z \sim 7$ sample ($10^{-8.30}$~yr$^{-1}$ vs.\@ $10^{-7.35}$~yr$^{-1}$), though as outlined before the sSFRs for the $z \sim 7$ smouldering candidates cannot be constrained to be as low as for the $z \sim 6$ sample, due to the lack of direct \Ha\ constraints. Indeed, following our SED-fitting methodology on the photometric data, we infer the $\mathrm{sSFR}_{10}$ of JADES-GS-z7-01-QU to be $10^{-8.19}$~yr$^{-1}$. This is 2--3 orders of magnitude larger than the sSFR obtained by \citet{Looser2024} ($10^{-11.3}$--$10^{-9.5}$~yr$^{-1}$) using the joint spectroscopic--photometric constraints on the emission lines and SED profile. Hence accurately recovering the sSFR of these $z \sim 7$ smouldering galaxies using photometric data alone can be challenging. 

For the inferred total stellar masses in the bottom panels of Fig.~\ref{fig:ssfr_mass_histograms}, we find that the smouldering galaxies are generally more massive than the control sample, with a median stellar mass of $\log\, (M_*/\mathrm{M}_\odot) = 8.5$ (vs.\@ 7.9) and $\log\, (M_*/\mathrm{M}_\odot) = 8.8$ (vs.\@ 8.0) for the $z \sim 6$ and $z \sim 7$ samples, respectively. This can be interpreted in two ways. First, that the stellar mass threshold for triggering smouldering activity is higher than the typical stellar mass of high-redshift galaxies. Or second, that the less massive smouldering galaxies, owing to their generally lower light-to-mass ratios than the control sample, are simply too faint to be detected and/or pass our SNR cuts, i.e.\@ our observational sample of smouldering galaxies is rather incomplete at the low-mass end. Indeed, there are theoretical grounds on which to expect smouldering activity to dominate at lower masses, which we will discuss in more detail in Section~\ref{subsec:number_abundances}. We only remark that from our own investigations, generating mock SEDs using {\tt Bagpipes}, that smouldering galaxies with intermediate starburst ages (25--50~Myr) and relatively little dust ($A_\mathrm{V} = 0.10${--}$0.25$) start to fail our SNR requirements at $\log\, (M_*/\mathrm{M}_\odot) \lesssim 8.5$. Hence there is possibly a significant population of low-mass ($\log\, (M_*/\mathrm{M}_\odot) \leq 8.0$) smouldering galaxies that even the deep JADES data lacks the sensitivity to adequately detect \citep[but gravitational lensing may be able to uncover, see e.g.\@][]{Strait2023}.

\subsection{Number abundances} \label{subsec:number_abundances}

The quiescent fraction of galaxies has traditionally served as a valuable statistic, greatly informing and constraining cosmological models of galaxy formation which aim to reproduce both the properties and number abundances of these extreme systems. Indeed, the mass- and environment-dependence of the quiescent fraction provides valuable insights into the nature of the physical processes primarily driving the cessation of star formation, and the parameter regimes within which they begin to dominate \citep[e.g.\@][]{Peng2010}. Here we extend such studies to very high redshift by considering the smouldering fraction, defined to be the number of smouldering galaxies in a particular mass bin, divided by the total (i.e.\@ smouldering plus control) galaxies in that bin. 

\begin{figure*}
\centering
\includegraphics[width=\linewidth]{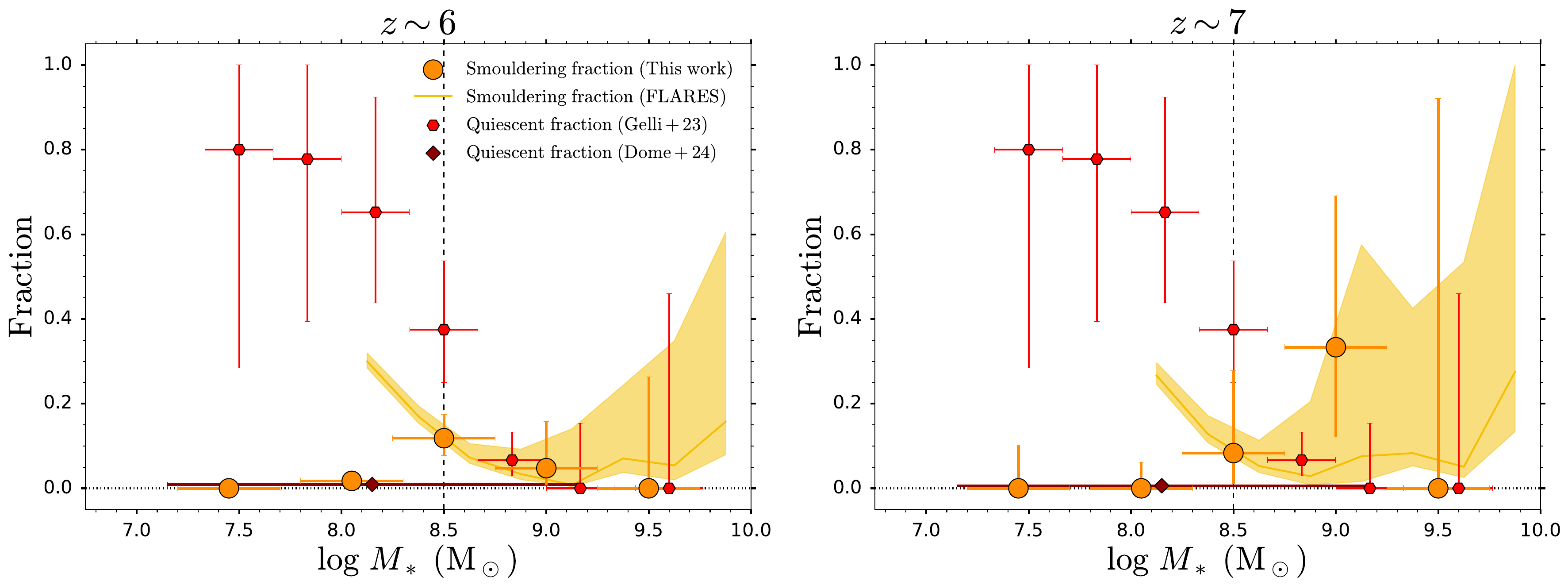}
\caption{Smouldering fractions (orange) from this work (orange circles) and from the FLARES simulations (light orange line), as well as quiescent fractions (red) from the \citet{Gelli2023} SERRA zoom-in simulations (red hexagons) and the analysis of the IllustrisTNG simulations by \citet{Dome2024} (dark red diamond), at $z \sim 6$ (left panel) and $z \sim 7$ (right). The plotted masses and associated error bars for the  $\log (M_*/\mathrm{M}_\odot) = 7.5,\ 8.0$ mass bins from this work, the $\log (M_*/\mathrm{M}_\odot) = 9.5$ mass bin from \citet{Gelli2023} and the single \citet{Dome2024} mass bin are shifted slightly ($\mp$ 0.05, +0.10, +0.15~dex respectively) for visual clarity. As predicted by various simulations, the observed smouldering fraction (initially) increases with decreasing stellar mass, reflecting the increasingly fragile nature of these first dwarf galaxies, being susceptible to feedback-driven winds in their shallow potential wells. The observed smouldering fraction then falls to zero, likely due to incompleteness (which we do not correct for) at $\log\, (M_*/\mathrm{M}_\odot) \lesssim 8.5$ (indicated by the dashed vertical line), thus complicating the observational--theoretical comparison. Nevertheless, smouldering galaxies constitute a considerable fraction of the EoR dwarf galaxy population, with these observational constraints thus aiding our understanding of the primordial baryon cycle, as current simulations greatly disagree on whether these systems are rare (${\sim}1\%$, IllustrisTNG) or common (${\sim}50\%$, FLARES and SERRA) in the EoR. }
\label{fig:smouldering_fraction}
\end{figure*}

We show the smouldering fraction for our $z \sim 6$ and $z \sim 7$ samples as orange circles in Fig.~\ref{fig:smouldering_fraction}. The error bars on the smouldering fraction are obtained by propagating the errors on the source counts (described by Poisson statistics) in quadrature. The error bars on the mass bins display the mass range spanned by that bin. We also show the smouldering fraction inferred using the FLARES simulations \citep{Lovell2021, Vijayan2021} (light orange line), by applying our smouldering galaxy colour selection procedure (Equations \ref{eq:MW1}, \ref{eq:MW2} and \ref{eq:supplementary}) to the FLARES galaxies, considering $z = 6$ and $z = 7$ FLARES galaxies in the left and right panels, respectively.  We limit our study to FLARES galaxies with $\log\, (M_\mathrm{*}/\mathrm{M}_\odot) = 8$ so as to not enter the gas mass resolution regime of the simulation ($\log\, (M_\mathrm{g}/\mathrm{M}_\odot) = 6$).

We also show the traditional quiescent fraction inferred from the SERRA cosmological zoom-in simulations by \citet{Gelli2023} (red hexagons), and from the analysis of the IllustrisTNG simulations by \citet{Dome2024} (dark red diamond). Quiescent galaxies, being well below the star-forming main sequence and thus having essentially non-existent emission lines (assuming these are solely powered by star formation and not e.g.\@ AGN activity), can generally be considered to be a subset of the smouldering population which are defined to have weak (but not necessarily no) emission lines. Hence the quiescent fraction will in general be lower than the smouldering fraction, which should be taken into account when comparing the quiescent and smouldering fractions in Fig.~\ref{fig:smouldering_fraction}. However, as discussed in Section~\ref{subsec:colour_selection}, our smouldering galaxy colour selections also potentially suffer from incompleteness, removing dusty and particularly old quiescent systems that have very red rest-frame optical slopes. 

\begin{figure*}
\centering
\includegraphics[width=\linewidth]{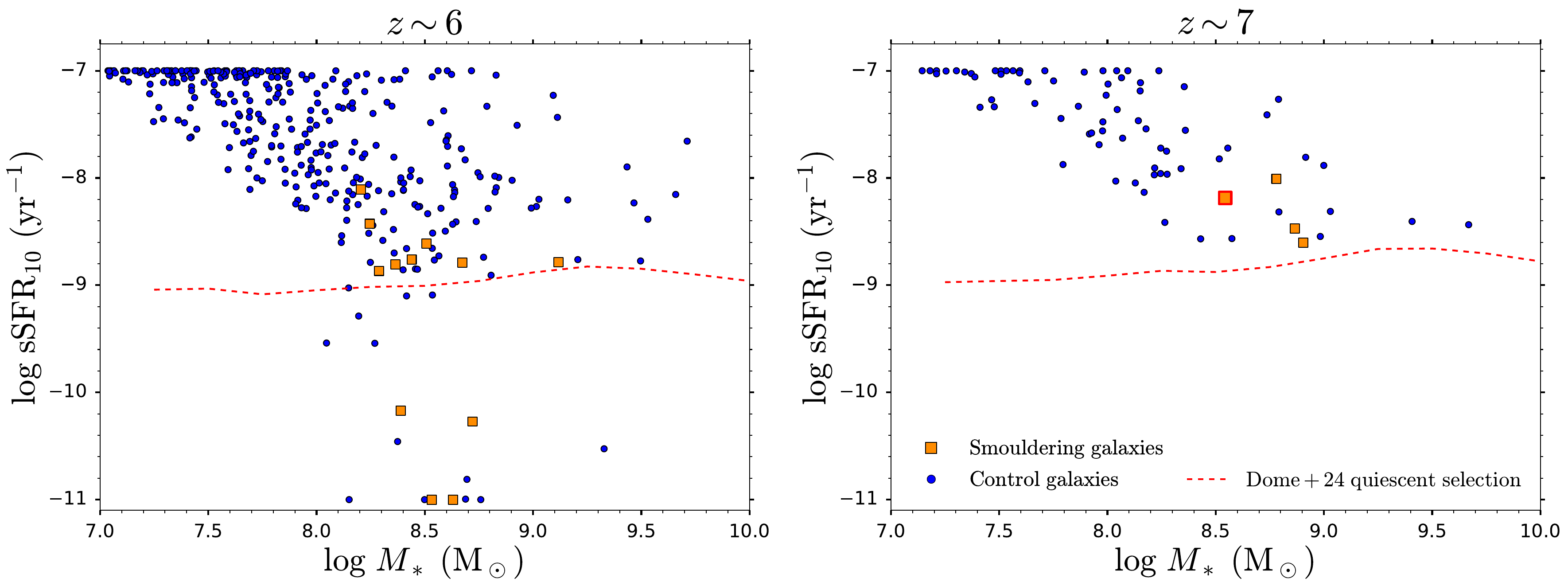}
\caption{Positions of smouldering galaxies (orange squares, JADES-GS-z7-01-QU is shown with a red border) and our control sample (blue circles) in the $\mathrm{sSFR}_{10}$--$M_*$ plane, for $z\sim6$ (left panel) and $z \sim 7$ (right). The \citet{Dome2024} quiescent criterion, corresponding to a SFR 0.75~dex below their star-forming main sequence is shown (dashed red). Sources with $\mathrm{sSFR}_{10} < 10^{-11}$~yr$^{-1}$ are plotted at $\mathrm{sSFR}_{10}= 10^{-11}$~yr$^{-1}$ for display purposes.}
\label{fig:ssfr_m}
\end{figure*}

To highlight these differences in selection, we show the positions of smouldering (orange squares) and control galaxies (blue circles) in the $\mathrm{sSFR}$--$M_*$ diagram in Fig.~\ref{fig:ssfr_m}. The \citet{Dome2024} quiescent selection (10~Myr-averaged SFRs being 0.75 dex below the star-forming main sequence) is shown in dashed red. \citet{Gelli2023} defines quiescent galaxies as those with no (or negligible) star formation activity at the epoch of observation.  Only a subset of our smouldering candidates satisfy the \citet{Dome2024} quiescent criterion. Partly, this could stem from the limited $\mathrm{sSFR}$ constraints from photometry, such as for JADES-GS-z7-01-QU. Moreover, there are numerous control galaxies that satisfy the \citet{Dome2024} quiescent criterion. We find that these either narrowly miss our medium--wide band colour selections (presumably scattered out via noise), have relatively blue/red rest-frame optical slopes, or their inferred weak line-emitting nature is not evident from the medium--wide bands comprising our colour selections. These added complications from different selection procedures can be lifted in future analyses by mimicking our smouldering galaxy colour selection procedure across different cosmological simulations (as was done for FLARES).

From Fig.~\ref{fig:smouldering_fraction}, we see that the observed smouldering fraction starts off small (in this case at zero) at high masses ($\log\, (M_*/\mathrm{M}_\odot) {\sim} 9.5$), initially increases and plateaus at intermediate masses ($\log\, (M_*/\mathrm{M}_\odot) {\sim} 8.5$, $9.0$ at $z \sim 6$, $7$, respectively) before decreasing as we approach the lowest stellar masses ($\log\, (M_*/\mathrm{M}_\odot) \lesssim 8.0$). As discussed earlier in Section~\ref{subsec:sf_properties}, this either indicates that smouldering activity dominates at intermediate stellar masses, or that our observational study lacks the sensitivity to detect the lowest mass smouldering galaxies, which may in fact constitute a substantial fraction of the low-mass galaxy population. From this discussion, and as displayed by the vertical dashed line on Fig.~\ref{fig:smouldering_fraction}, our $\log\, (M_*/\mathrm{M}_\odot) \lesssim 8.5$ bins are likely affected by incompleteness.

Indeed, the three cosmological simulations discussed in this work all agree that the smouldering fraction (FLARES) and quiescent fraction \citep{Gelli2023, Dome2024} should increase with decreasing stellar mass below $M_* = 10^{9}~\mathrm{M}_\odot$. The increasing smouldering fraction with increasing stellar mass in the FLARES simulations at $M_* > 10^{9}~\mathrm{M}_\odot$ may indicate the emergence of the traditional massive quiescent population quenched by AGN feedback \citep[for more details on the quenching of massive galaxies in FLARES, see][]{Lovell2023}. 

Collectively, \citet{Looser2024}, \citet{Gelli2023} and \citet{Dome2024} attribute the quiescent nature of these high-redshift dwarf galaxies to stellar feedback, AGN feedback, galaxy--galaxy interactions, and a lack of gas accretion. \citet{Looser2023b} observationally investigate the SFHs of high-redshift galaxies using JADES NIRSpec spectroscopy, finding that low-mass galaxies have particularly bursty SFHs, while the more massive systems evolve more smoothly. Indeed, \citet{Gelli2023} find from their zoom-in simulations that the SFHs of high-redshift dwarf galaxies are punctuated by bursts, drops and even complete halts of star formation, driven by stellar feedback decreasing or even suppressing star formation in these low-mass systems. Furthermore, the fraction of time spent in an active star-forming phase decreases with decreasing stellar mass. Moreover, \citet{Dome2024} find from IllustrisTNG that tidal interactions during close galaxy--galaxy interactions can play an important role in shaping the SFRs of low-mass galaxies, with such merger-related quiescence accounting for 30 and 67 per cent of all quiescent galaxies in IllustrisTNG50 at $z=6$ and $z=7$, respectively. We further speculate that dwarf galaxies, owing to their lower mass, are more likely to undergo major mergers when they do interact with other systems, as the bulk of the galaxy population is low mass. Thus low-mass dwarf galaxies have a propensity to enter a bursty state, either driven by secular- or merger-driven processes. This, combined with their shallower gravitational potential wells, which in general makes them more susceptible to feedback processes, whether it be from regular or bursty star formation, likely results in a considerable population of quiescent dwarf galaxies in the EoR. 

From the observations we can establish that smouldering galaxies constitute a considerable fraction of the high-redshift dwarf galaxy population, ranging from 0.05--0.35 at intermediate stellar masses (not affected substantially by incompleteness). Thus the observed smouldering galaxy fraction can in principle provide valuable constraints on the primordial dwarf galaxy baryon cycle, helping to inform the next-generation of high-redshift galaxy simulations, improving our understanding of star formation, merging and feedback processes in the first dwarf galaxies in the Universe. Indeed, while the current simulations shown all agree that the smouldering (or quiescent) fraction should increase with decreasing stellar mass, the normalisation of this fraction varies substantially between the simulations. At $\log\, (M_*/\mathrm{M}_\odot) = 8$ and $z = 6$, \citet{Dome2024} expect a quiescent fraction of ${\sim}$0.01, FLARES predicts a smouldering fraction of ${\sim}0.3$ and \citet{Gelli2023} have a quiescent fraction of ${\sim}$0.7. These differences in predicted fractions are likely due to differences in feedback implementations in the various simulations. Though we also note that the results of \citet{Gelli2023} are based on a zoom-in simulation on an overdense region, and may therefore perhaps not be completely representative of the average cosmic volume, with the resulting quiescent fractions therefore possibly being biased high. 

At $\log\, (M_*/\mathrm{M}_\odot) \geq 8.5$ we find that the observed smouldering fraction generally closely matches the FLARES predictions, suggesting that the simulation provides a reasonable description of the relative roles of starbursts, mergers and feedback in driving the baryon cycle of high-redshift dwarf galaxies.  The disagreement between the observations and the FLARES predictions is most pronounced for the $\log\, (M_*/\mathrm{M}_\odot) = 9.5$ mass bin at $z \sim 7$. Certainly our observational study of smouldering galaxies would benefit from greater statistics across a multitude of widely-spaced sightlines (rather than the single JADES region) to better mitigate cosmic variance. medium band imaging in additional filters would further constrain and clarify the strength of the line emission in these systems, also helping to remove possibly outstanding low-redshift interlopers, thus solidifying the observational--theoretical comparison. Follow-up NIRSpec PRISM spectroscopy would provide the definitive account on the (potentially bursty) star-formation history of these galaxies, serving as a valuable tool to confront against models, which are required to reproduce the inferred histories within a physically motivated framework describing the gas flows in these systems.

At $\log\, (M_*/\mathrm{M}_\odot) \leq 8.5$ our observational results are likely affected by incompleteness, complicating the comparison against the simulation predictions shown which have not been corrected for this effect. We do note that at $z \sim 6$, the quiescent fraction from \citet{Dome2024} is comparable to the observed smouldering fraction at $\log\, (M_*/\mathrm{M}_\odot) = 8$, despite the latter likely greatly underestimating the true smouldering fraction due to incompleteness. Hence it is plausible that the IllustrisTNG simulations underpredict the number of low-mass smouldering galaxies. Further work, accounting for incompleteness and comparing the observed smouldering fraction against the smouldering fraction (rather than quiescent fraction) across a number of simulations will enable more conclusive statements about the baryon cycle in low-mass, high-redshift dwarf galaxies to be made. 

\begin{figure*}
\centering
\includegraphics[width=\linewidth]{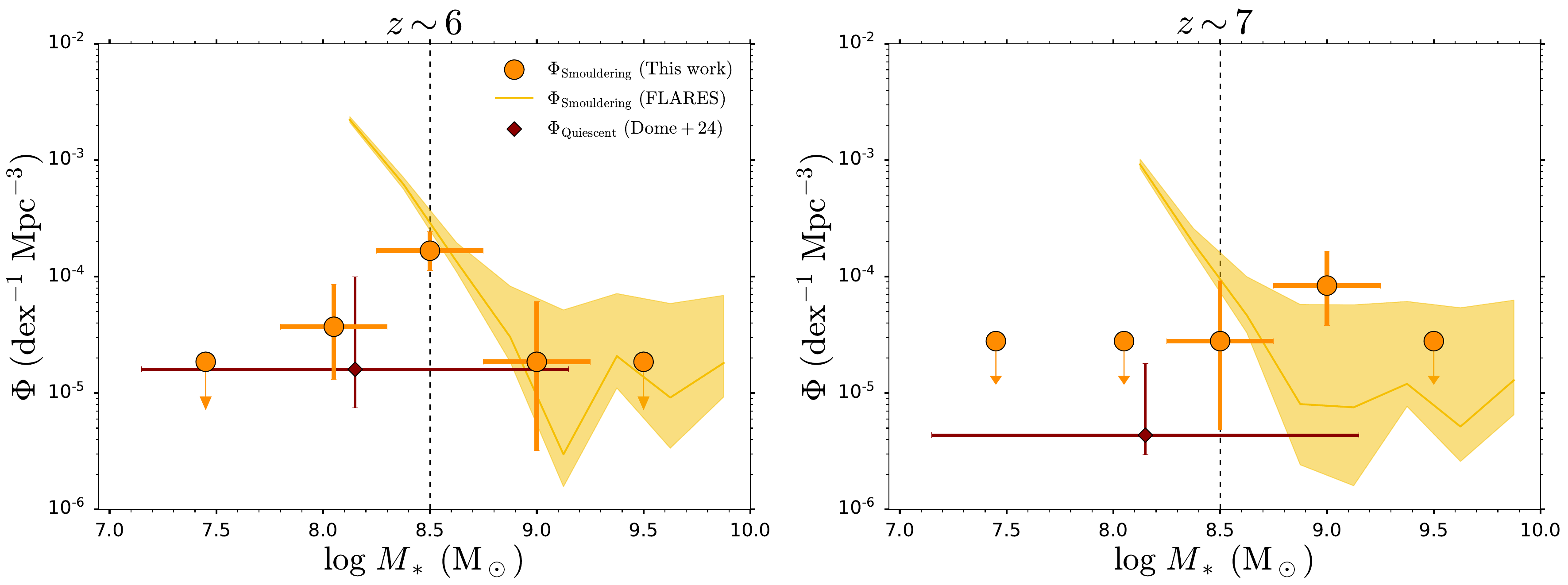}
\caption{Comoving number densities $\Phi$ of smouldering and quiescent galaxies at $z\sim6$ (left) and $z \sim 7$ (right). Symbols and colour-coding are similar to Fig.~\ref{fig:smouldering_fraction}. Vertical error bars on the observational data points and the FLARES simulations are from Poisson statistics only, and do not account for cosmic variance. Horizontal error bars indicate the width of the mass bins. Observational mass bins with no smouldering galaxies ($N_\mathrm{smouldering}=0$) are plotted with downward arrows at the comoving number density corresponding to $N_\mathrm{smouldering}=1$. These are not necessarily upper limits as we do not correct the observational results for incompleteness. The lower limit, median and upper limit for the \citet{Dome2024} data point corresponding to the results from the TNG50, joint TNG50+TNG100 and TNG100 simulations, respectively.}
\label{fig:smouldering_number_density}
\end{figure*}

We further show the comoving number density of smouldering and quiescent galaxies as a function of stellar mass in Fig.~\ref{fig:smouldering_number_density}. The symbols and colour coding are the same as Fig.~\ref{fig:smouldering_fraction}, and the observational results have not been corrected for completeness. Note that the error bars on the observational and FLARES results are due to Poisson statistics, and do not take the additional contribution from cosmic variance into account. As before, the observed smouldering number density starts off small at the high-mass end (with no systems being detected), before rising to intermediate masses and then plummeting (with no detections) at the low-mass end. From FLARES we expect the smouldering number density to continue to rise with decreasing stellar mass, indicating that such systems are common in the high-redshift Universe. Again, at $\log\, (M_*/\mathrm{M}_\odot) \geq 8.5$ the FLARES predictions are generally consistent with the observational results, though the error bars are substantial due to the low-number statistics at the high-mass end. Again, the observational results are generally consistent with \citet{Dome2024} at $\log\, (M_*/\mathrm{M}_\odot) = 8$, with the upper limit, median and lower limit of the \citet{Dome2024} datapoint corresponding to the results from the TNG50, joint TNG50+TNG100 and TNG100 simulations, respectively. However, as the observational results have not been corrected for incompleteness and the \citet{Dome2024} results are based on quiescent galaxies, it is difficult to draw firm conclusions. 

Owing to their faint nature, the incompleteness of low-mass smouldering galaxies will often be a concern, their perceived low statistics being driven by a lack of depth rather than area. This is in firm contrast to traditional studies of ultra-massive quiescent galaxies, which instead greatly benefit from wide-area surveys to find these relatively rare systems. This incompleteness can be combated in the following manners, though each has its limitations. First, by simply surveying even deeper than the (already very deep) JADES program. However, every added AB magnitude of depth demands a six-fold increase in exposure time, enabling 0.4~dex less massive systems (i.e.\@ one mass bin lower in Figs.~\ref{fig:smouldering_fraction} and \ref{fig:smouldering_number_density}) to be probed but at the cost of exceedingly long exposure times. Second, the $10\sigma$ detection threshold in the bands driving the colour selection can be lowered to $5\sigma$, enabling 0.75~AB~mag fainter (0.3~dex less massive) systems to be studied, at the cost of more contamination. Third, comparably deep exposures to the JADES program (a blank field) can instead be conducted on a cluster field, with the added magnification from the strong gravitational lensing enabling much intrinsically fainter systems to be studied. However, the very same magnification that enables the detection of low-mass smouldering galaxies also complicates the determination of the resulting smouldering fractions and number densities, and their variation with stellar mass. Regularly star-forming control galaxies, with their higher light-to-mass ratios, will require smaller magnifications, and thus be detectable over a larger area than the fainter smouldering galaxies of the same stellar mass, skewing the statistics. Moreover, the comoving number densities derived would have to account for magnification reducing the effective cosmic volume behind the cluster. Finally, the derived stellar masses would also have to be corrected for magnification, introducing further uncertainty into the observational results.

\section{Further discussion of selection method} \label{sec:discussion}

In this section we further discuss the effectiveness of our colour selections in identifying weak line-emitting smouldering galaxies. In Section~\ref{subsec:flares} we examine the connection between line emission and sSFR in the FLARES simulations. In Section~\ref{subsec:interlopers} we outline how additional medium band photometry can aid in the removal of low-redshift dusty interlopers. Finally in Section~\ref{subsec:weak_lines}, we discuss how sources with high ionising photon escape fractions and/or heavily dust-obscured star formation can also exhibit weak emission lines, outlining how to potentially distinguish these from low sSFR galaxies.

\subsection{Weak line-emission as a proxy for low sSFR} \label{subsec:flares}

Our smouldering colour selections aim to identify weak line-emitting systems with low emission-line equivalent widths. Loosely speaking, the line luminosity is indicative of the galaxy SFR and the continuum level is indicative of the galaxy stellar mass. Hence smouldering galaxies, with their low EWs, can generally be expected to have low sSFRs, thus being preferentially comprised of weakly star-forming and quiescent systems. Here we investigate using a low \Ha\ EW as a proxy for low sSFR in more detail using the $z = 6$ FLARES galaxies. 

\begin{figure}
\centering
\includegraphics[width=.85\linewidth]{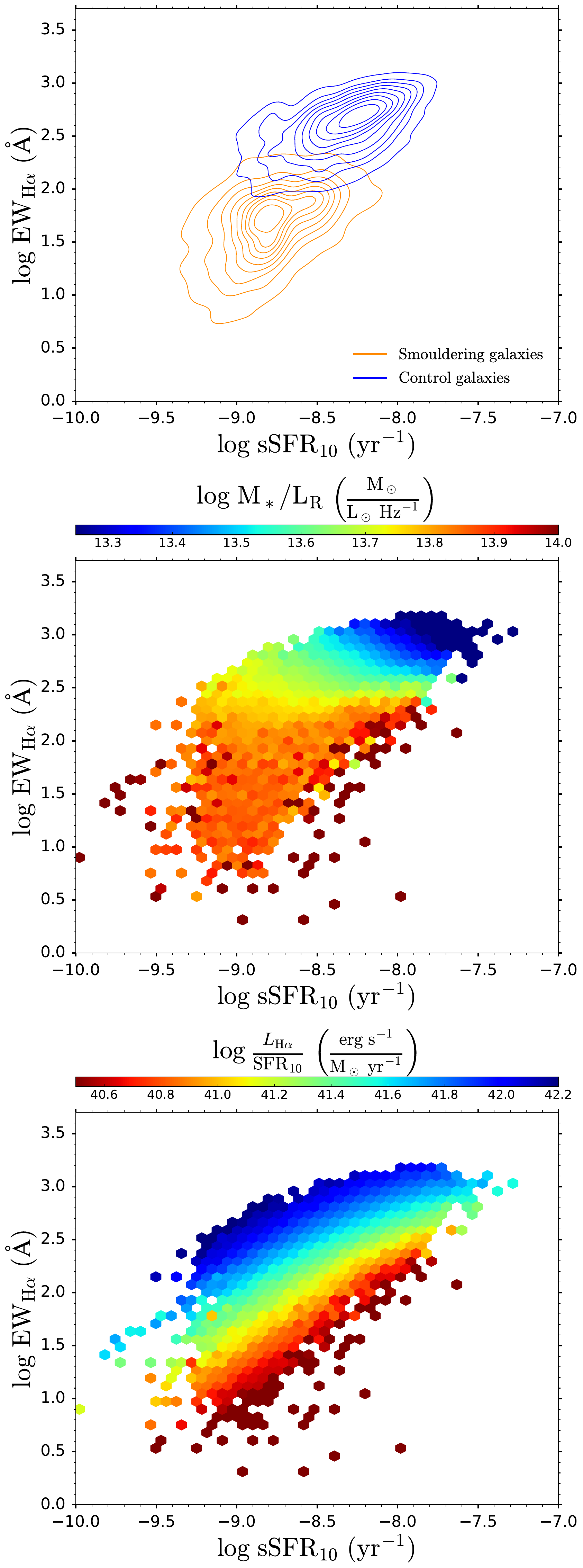}
\caption{The connection between $\mathrm{EW_{H\alpha}}$ and $\mathrm{sSFR}_{10}$ for $z = 6$ FLARES galaxies. Top panel: The two quantities are generally well correlated, with e.g.\@ low $\mathrm{EW_{H\alpha}}$ usually implying a low $\mathrm{sSFR}_{10}$, as can be seen from the contours for smouldering (orange) and control galaxies (blue). However, this is not a guarantee, with some smouldering galaxies still having a relatively high $\mathrm{sSFR}_{10}$ given their low $\mathrm{EW_{H\alpha}}$. Middle panel: The variation in $\mathrm{sSFR}_{10}$ at a given low $\mathrm{EW_{H\alpha}}$ is not driven by variations in the mass-to-light ratio $M_*/L_\mathrm{R}$ (colour-coding), this quantity essentially being relatively fixed. Bottom panel: The variation in $\mathrm{sSFR}_{10}$ is in fact anti-correlated with the \Ha\ brightness per unit SFR $L_{\mathrm{H}\alpha} / \mathrm{SFR}_{10}$ (colour-coding). Hence some smouldering galaxies can still have relatively high sSFR, despite their weak emission lines, because they are inefficient producers of \Ha\ emission. This can come about because the youngest stellar population is relatively old (closer to 10~Myr), the ionising photon escape fraction is high, or the star formation takes place in very dust-obscured environments.}
\label{fig:flares_ssfr_ew}
\end{figure}

As can be seen from the top panel of Fig.~\ref{fig:flares_ssfr_ew}, there is a general correlation between \Ha\ EW and sSFR. Thus, smouldering galaxies (orange) with their weak \Ha\ line-emission (by selection) generally do have low sSFRs, though crucially this is not a guarantee. It is possible for galaxies with weak emission-lines to still have $\mathrm{sSFRs}$ that are comparable to what is typically seen in the stronger line-emitting control galaxies (blue). As discussed in \citet{Belfiore2018}, $\mathrm{EW}_{\mathrm{H}\alpha} \sim L_{\mathrm{H}\alpha}/L_\mathrm{R}$ and $\mathrm{sSFR} = \mathrm{SFR}/M_*$, where $L_{\mathrm{H}\alpha}$ and $L_\mathrm{R}$ are the \Ha\ luminosity and the typical luminosity density in the R-band, respectively. Hence to better understand the connection and scatter between $\mathrm{EW}_{\mathrm{H}\alpha}$ and $\mathrm{sSFR}$, we can write:

\begin{equation}
\frac{\mathrm{EW}_{\mathrm{H}\alpha}}{\mathrm{sSFR}} \sim \frac{M_*}{L_\mathrm{R}}\frac{L_{\mathrm{H}\alpha}}{\mathrm{SFR}}
\end{equation}

That is, variations in the mass-to-light ratio ($M_* / L_\mathrm{R}$) or the \Ha\ brightness per unit SFR ($L_{\mathrm{H}\alpha} / \mathrm{SFR}$) can cause scatter in the correlation between $\mathrm{EW}_{\mathrm{H}\alpha}$ and $\mathrm{sSFR}$, resulting in some smouldering systems still having relatively high sSFR. To establish whether variations in the mass-to-light ratio are primarily responsible for this effect we now turn to the middle panel of Fig.~\ref{fig:flares_ssfr_ew}. We can see that at a given low $\mathrm{EW}_{\mathrm{H}\alpha}$ (relevant for smouldering galaxies), the sSFR can vary by up to one order of magnitude with little-to-no variation in $M_* / L_\mathrm{R}$, i.e.\@ the scatter in the sSFRs seen for smouldering galaxies is not driven by variations in $M_* / L_\mathrm{R}$. On the other hand, we see from the bottom panel that the \Ha\ brightness per unit $\mathrm{SFR}_{10}$ can vary by an order of magnitude at fixed $\mathrm{EW}_{\mathrm{H}\alpha}$, thus being strongly anti-correlated with the $\mathrm{sSFR}_{10}$. Hence the primary reason why some smouldering systems still have high sSFR, despite their weak emission lines, is because their $L_{\mathrm{H}\alpha} / \mathrm{SFR}$ is relatively low, i.e.\@ these systems are inefficient producers of \Ha\ emission given their SFR. 

This weak \Ha\ emission for a given $\mathrm{SFR}_{10}$ can come about for a number of reasons. First, because the most recent starburst population is relatively old, being closer to 10~Myr (when the \ion{H}{II} region and thus line emission has completely faded) than 0~Myr (when the line emission is at its strongest). In which case this scatter is more of an artefact of choosing to use $\mathrm{SFR}_{10}$ in our analysis, rather than a shorter timescale such as $\mathrm{SFR}_{5}$. Second, because the system has a high escape fraction of ionising photons, so the ongoing star formation translates into a weaker \Ha\ flux. Third, because the bulk of the star formation is taking place in very dust-obscured environments, with the \Ha\ emission thus being heavily attenuated, resulting in low EWs despite the high sSFR. We will discuss the latter two scenarios \citep[also raised in][]{Looser2024} in more detail in Section~\ref{subsec:weak_lines}.

\subsection{Removal of low-redshift dusty interlopers} \label{subsec:interlopers}

\begin{figure}
\centering
\includegraphics[width=\linewidth]{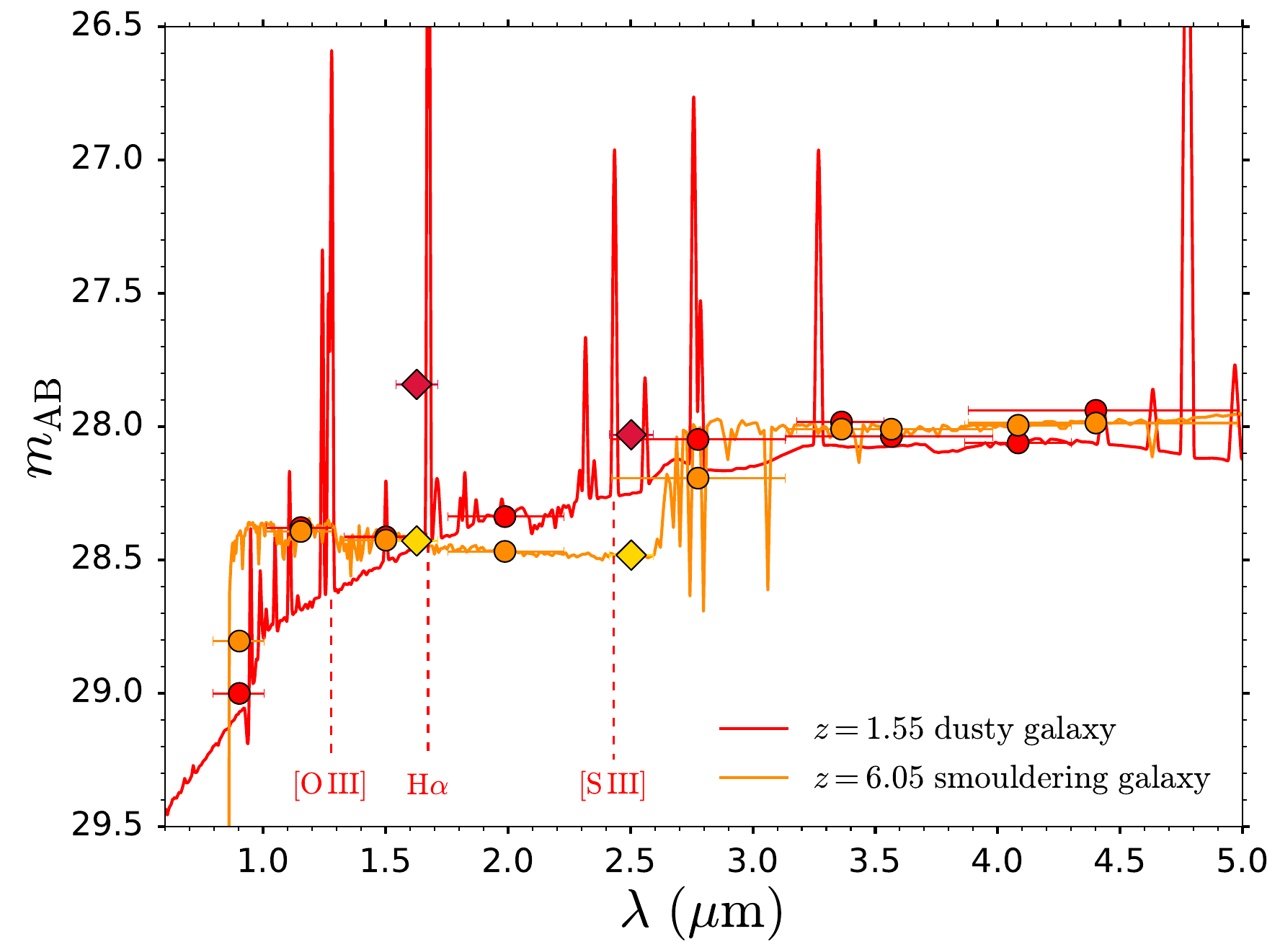}
\caption{The photometry of a low-redshift dusty galaxy (red) can in principle closely resemble that of a high-redshift smouldering galaxy (orange) with a Balmer break in the JADES bands (circles), the model spectra being generated by {\tt Bagpipes}. This comes about because the dusty continuum and emission lines mimic the Lyman break and rest-frame UV of a high-redshift galaxy, with the Balmer break feature being closely replicated by a gradually flattening dusty continuum that plateaus in the rest-frame NIR. Although this requires the dusty source to be at a very specific redshift with finely-tuned emission lines and dust content $A_\mathrm{V}$, it is at the very least possible that some of our smouldering candidates are low-redshift dusty interlopers. Well-chosen additional medium band photometry (diamonds) that targets the \OIII, \Ha\ and [\ion{S}{III}] emission lines for the low-redshift solution can firmly break this photometric degeneracy between low-redshift dusty (dark red) and high-redshift smouldering (yellow) galaxies.}
\label{fig:dusty_vs_smouldering}
\end{figure}

It is possible for the (primarily wide band) photometry of a low-redshift dusty galaxy to closely resemble that of a high-redshift smouldering galaxy with a Balmer break. We show a clear example in Fig.~\ref{fig:dusty_vs_smouldering}, where the $z = 1.55$ dusty galaxy (red) and $z = 6.05$ smouldering galaxy (orange) SEDs have been generated using {\tt Bagpipes}. Focusing just on the JADES photometry (red/orange circles), both systems appear to drop out in F090W, have roughly flat photometry redward of the break (the F115W, F150W, F200W filters), before exhibiting a second prominent break (F277W) and flat photometry thereafter (F335M, F356W, F410M, F444W). Hence such systems would be difficult to conclusively distinguish between, their photometry being close to identical within observational errors. 

For the smouldering galaxy, these photometric features come about because of the Lyman break, a flat rest-frame UV, and Balmer break, respectively. Similar to the famous example of a $z \sim 16$ candidate \citep{Donnan2023} now known to be a dusty galaxy at $z = 4.9$ \citep{ArrabalHaro2023}, the dusty continuum and line emission of the $z = 1.55$ dusty galaxy result in photometry that mimics that of a high-redshift Lyman-break galaxy with a flat rest-frame UV. Furthermore, owing to the declining dust attenuation at longer wavelengths, the once red dusty continuum begins to plateau (in the rest-frame NIR), resulting in photometry mimicking a Balmer break. Hence it is possible for low-redshift dusty galaxies to be mistakenly identified as high-redshift smouldering galaxies in this way. That being said, we do acknowledge that this degeneracy is quite contrived, requiring a dusty galaxy with very specific emission line strengths and dust attenuation $A_\mathrm{V}$, and also at a very specific redshift such that the various emission lines are in the appropriate filters. Additional medium band photometry (diamonds) at appropriate wavelengths (in this case F162M, F250M) can lift this degeneracy, either being greatly elevated (dark red) by dusty line-emission (\Ha\ and [\ion{S}{III}]) or tracing the lower-lying UV continuum level (yellow) in the smouldering galaxy.

\subsection{Alternate causes of weak line-emission} \label{subsec:weak_lines}

Smouldering galaxies are defined and selected to have weak emission lines. This can be taken to indicate that these systems generally have low levels of active star formation (i.e.\@ low sSFR). However, as seen in Section~\ref{subsec:flares}, this cannot be guaranteed to be the case. Actively star-forming galaxies (i.e.\@ moderate-to-high sSFR) can also exhibit weak emission lines, provided that the ionising photon escape fraction is large and/or the star formation takes place behind a substantial veil of dust \citep[also pointed out by][]{Faisst2024}. Hence such systems will also be selected via our colour selection procedure and in principle could also comprise part of our sample of smouldering galaxies.

\begin{figure}
\centering
\includegraphics[width=\linewidth] {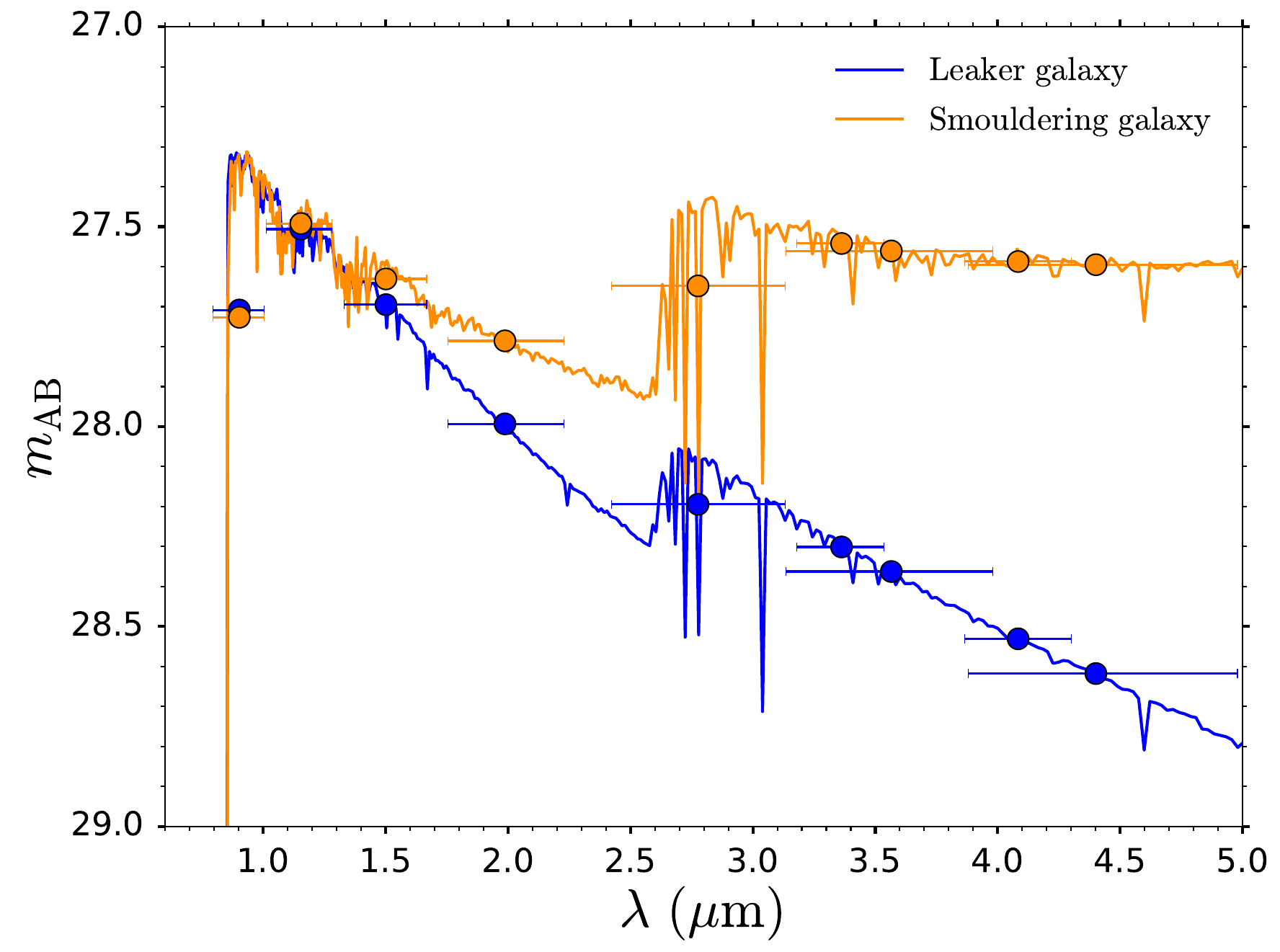}\\[3ex]
\includegraphics[width=\linewidth]{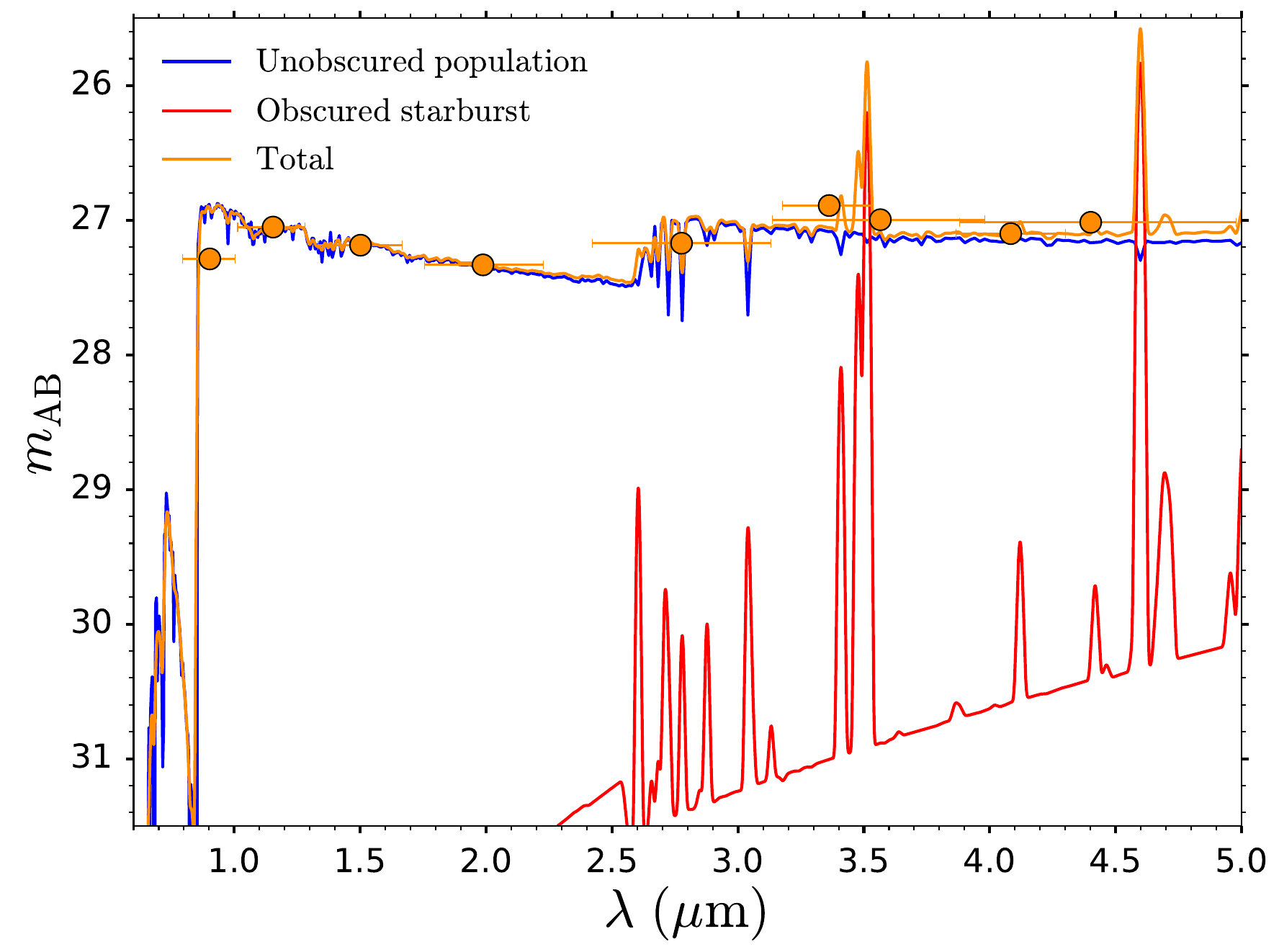}
\caption{SEDs and photometry for a range of $z = 6$ systems that exhibit weak emission lines for different reasons, generated by {\tt Bagpipes}. Top panel: The spectrum of a 25~Myr old instantaneous starburst (orange) and a 5~Myr old instantaneous starburst with ionising photon escape fraction $f_\mathrm{esc} = 1$ (an extreme case). In principle it should be possible to distinguish between these scenarios based on the UV slope and strength of the Balmer break, though in practice this will be more challenging if the escape fraction is lower or the leaker galaxy harbours an older stellar population. Bottom panel: The spectrum of a 25~Myr old unobscured instantaneous starburst (blue), an $A_\mathrm{V} = 2.5$ heavily dust-obscured 2.5~Myr old instantaneous starburst (red) and the resulting combined spectrum (orange) and photometry. Hence it is possible for the weak line-emitting smouldering galaxies we select to harbour a substantial amount of star formation (10 per cent of the unobscured mass in this example), if this takes place in heavily dust-obscured environments. NIRSpec constraints on the Balmer decrement, or longer wavelength observations with MIRI/ALMA would help to distinguish this scenario from the low sSFR case.} 
\label{fig:weak_lines}
\end{figure}

In the case of high escape fraction, the nebular line and continuum emission is weak, and thus the stellar continuum emission should dominate. Since the massive stars that dominate the spectra of young stellar populations are very hot, the SED of these leaky, high escape fraction systems should be very blue, as shown in the top panel of Fig.~\ref{fig:weak_lines} for a 5~Myr old instantaneous starburst with $f_\mathrm{esc} = 1$ (admittedly an extreme case) at $z=6$. Due to the steep blue slopes, especially in the rest-frame UV, but also the rest-frame optical, these systems may fail to satisfy our colour cuts, with the offset between the wide bands likely exceeding our requirement ($\leq 0.15$~mag) and potentially the medium--wide-band comparison also failing ($\leq 0.1$~mag). These selection issues notwithstanding, our sample of smouldering galaxies generally have relatively flat UV colours (see Tables~\ref{tab:z6_candidates} and \ref{tab:z7_candidates}), especially compared to the colours expected (F115W$-$F200W $\lesssim-0.5$) for a young (age $<$ 2.5~Myr) high escape fraction system. Moreover, we would expect these young systems to have a very weak Balmer break (provided there is no older underlying stellar population dominating the SED in the rest-frame optical), in contrast to the prominent Balmer breaks generally seen in our smouldering sample. 

In the case of heavily-obscured star formation, it is difficult to distinguish this from unobscured low star formation activity using NIRCam photometry alone. As shown in the bottom panel of Fig.~\ref{fig:weak_lines} \citep[and similar to the scenario described in][]{Faisst2024}, if the resulting dust-obscured nebular line-emission is weak, then the dusty continuum (red) will likely be far too faint to identify, being outshone by unobscured continuum emission from an older underlying stellar population (blue). Here we show a 25~Myr old unobscured population against the emission of a $A_\mathrm{V} = 2.5$ dust-obscured 2.5~Myr old instantaneous starburst with 10 per cent of the mass of the older population. Hence substantial star formation can be hidden in this way in the NIRCam photometry. Though from the FLARES simulations we expect such heavily dust-obscured star formation to be relatively rare in the dwarf galaxy regime ($M_* < 10^{9}~\mathrm{M}_\odot$). Nevertheless, MIRI photometry/spectroscopy at longer wavelengths, where the obscuration is weaker and thus where the dust-enshrouded continuum and line emission becomes more prominent, is key to distinguish between these scenarios. Additionally, as pointed out by \citet{Faisst2024}, ALMA can provide constraints on the dust and gas content in these galaxies, helping to establish whether the weak line-emission is due to a lack of ongoing star formation or substantial dust obscuration. Finally, in the case of weak-but-detectable line-emission, NIRSpec can provide constraints on the dust content through measurements of the Balmer decrement \Ha/\Hb\ \citep[see e.g.\@][]{Sandles2024} for our $z \sim  6$ sample.

\section{Conclusions} \label{sec:conclusions}

We develop a photometric search method to identify smouldering galaxies at $5 < z <  8$. Defined by their weak emission lines, these galaxies generally have low sSFR and may even be in a (temporary) quiescent state. Deep medium band photometry is essential to reliably find these systems, enabling a clear identification of their lack of line emission, thus revealing the lack of ongoing star formation (i.e.\@ relatively low sSFR) in these faint, relatively quiescent dwarf galaxies in the EoR. Thus we use the deep (${\sim}29\text{--}30$~AB~mag, $5\sigma$) public NIRCam imaging from the JADES second data release to conduct our photometric search, concentrating on smouldering galaxy candidates at $z \sim 6$ ($5.3 < z < 6.6$) and $z \sim 7$ ($6.8 < z < 7.8$). We require the photometry to be level in pairs (F335M, F356W), (F410M, F444W) of medium--wide-band filters, thus demanding the rest-frame optical line emission (\OIII\ + \Hb, \Ha) to be weak in our smouldering galaxy candidates.

We compare and contrast the star formation properties (inferred by fitting the JADES photometry with the SED-fitting code {\tt Bagpipes}) of our smouldering galaxy sample against those of the control sample, which are galaxies at the same redshift range that do not satisfy our medium--wide-band colour-selection criteria, i.e.\@ stronger line-emitters. As expected from our colour selection procedure, our smouldering galaxy candidates are generally inferred to have much weaker emission lines (\Ha\ and \OIII\ + \Hb) than the control sample, with a median rest-frame equivalent width of ${\sim}100$~\AA\ vs.\@ ${\sim}600$~\AA. This separation in equivalent width between the smouldering and control samples is far less pronounced at $z \sim 7$, due to the lack of direct constraints on the \Ha\ emission from NIRCam imaging. Following on from their weaker emission lines, smouldering galaxies are generally inferred to have much lower (${\sim}1$~dex) $\mathrm{sSFR}_{10}$ than the control sample, with e.g.\@ $10^{-8.80}$~yr$^{-1}$ vs.\@ $10^{-7.50}$~yr$^{-1}$ at $z \sim 6$. 

We find that smouldering galaxies comprise a considerable fraction of the EoR dwarf galaxy population, with smouldering fractions of 0.05--0.35 at $M_* \sim 10^{8\text{--}9}~\mathrm{M}_\odot$. This is in line with the predictions of simulations, which find that the smouldering or quiescent fraction increases with decreasing stellar mass. This reflects the fragile nature of these first dwarf galaxies, the star formation in their shallow potential wells easily snuffed out by feedback-driven winds triggered by secular or merger-driven starbursts. However, current simulations greatly disagree on whether these smouldering galaxies are rare (constituting ${\sim}1$\% of dwarf galaxies in IllustrisTNG) or common (${\sim}50$\%, FLARES and SERRA) in the EoR. Thus observational constraints on the smouldering fraction and smouldering comoving number density, which we find to be ${\sim}10^{-4}\text{--}10^{-5}$~dex$^{-1}$~Mpc$^{-3}$ (uncorrected for incompleteness), should greatly aid in our understanding of the primordial baryon cycle, informing the next-generation of cosmological simulations of galaxy formation and evolution in the EoR. However, this observational--theoretical comparison is not without its challenges, due to the inherent faintness of these smouldering galaxies resulting in considerable incompleteness in the observations below ${\sim}10^{8.5}~\mathrm{M}_\odot$, precisely the mass regime where the current simulations disagree the most.

We further verify our colour selection method for identifying smouldering galaxies by applying it to the FLARES simulations, representing the ideal case of perfect photometry where the inherent properties of the galaxies are fully known. We indeed preferentially select systems with weak emission lines, which generally also have low sSFRs. However, this is not a guarantee, as old starbursts (close to 10~Myr), systems with high ionising escape fractions and galaxies with heavily dust-obscured star formation can also exhibit weak emission lines, despite non-negligible $\mathrm{sSFR}_{10}$. We find that smouldering galaxies exhibit (subtly) distinct colours in the rest-frame UVJ colour plane, crucially being located in the star-forming region of this diagram, rather than the quiescent region or linear extension thereof. Thus, owing to the young age of the Universe, together with the likely bursty star formation in these first dwarf galaxies, traditional colour-selection methods like the UVJ diagram likely miss a large fraction of the quiescent and smouldering population in the EoR. 

Follow-up NIRSpec spectroscopy will place the definitive constraints on the line emission (or lack thereof) in our smouldering galaxy candidates. Furthermore, deep PRISM continuum spectroscopy would further constrain the stellar population parameters, which together with the emission line measurements would provide the best possible insights on the (potentially bursty) star-formation histories of these smouldering systems. These observational constraints would serve as a final valuable tool to confront against models, which are required to reproduce the inferred histories within a physically motivated framework describing the gas flows in these systems. Thus smouldering galaxies can shed a unique light on the primordial baryon cycle in the first dwarf galaxies, their star formation fragile, like a candle in the wind. 

\section*{Acknowledgements}

We thank the referee for their useful comments which helped to improve this article. JAAT thanks Tobias Looser and Viola Gelli for valuable discussions that helped shape the methodology and interpretation of results in this article. JAAT, CJC, NA and QL acknowledge support from the ERC Advanced Investigator Grant EPOCHS (788113). JAAT also  acknowledges support from the Simons Foundation and JWST program 3215. Support for program 3215 was provided by NASA through a grant from the Space Telescope Science Institute, which is operated by the Association of Universities for Research in Astronomy, Inc., under NASA contract NAS 5-03127. DA and TH acknowledge support from STFC in the form of PhD studentships. CCL acknowledges support from a Dennis Sciama fellowship
funded by the University of Portsmouth for the Institute of
Cosmology and Gravitation. LTCS is supported by an STFC studentship. APV acknowledges support from the Carlsberg Foundation (grant no CF20-0534). The Cosmic Dawn Center (DAWN) is funded by the Danish National Research Foundation under grant No.140. SMW thanks STFC for support through ST/X001040/1.

This work is based on observations made with the NASA/ESA \emph{Hubble Space Telescope} (\emph{HST}) and NASA/ESA/CSA \emph{James Webb Space Telescope} (\emph{JWST}) obtained from the Mikulski Archive for Space Telescopes (MAST) at the Space Telescope Science Institute (STScI), which is operated by the Association of Universities for Research in Astronomy, Inc., under NASA contract NAS 5-03127 for \emph{JWST}, and NAS 5–26555 for \emph{HST}. The \emph{HST} and {JWST} data used in this work are from the JADES public data release DR1 (http://dx.doi.org/10.17909/8tdj-8n28) and DR2 (http://dx.doi.org/10.17909/z2gw-mk31), with the photometric catalog also utilising data from JEMS (https://dx.doi.org/10.17909/fsc4-dt61). These data releases and catalogs are in turn based on \emph{JWST} observations associated with programs 1180, 1210, 1895, 1963 and 3215.

This research made use of Astropy,\footnote{http://www.astropy.org} a community-developed core Python package for Astronomy \citep{astropy2013, astropy2018}.

\section*{Data Availability}

The \emph{HST} and \emph{JWST} data used in this work are from the JADES public data release, available at https://archive.stsci.edu/hlsp/jades. The FLARES simulation results are publicly available at: https://flaresimulations.github.io/. Any remaining data underlying the analysis in this article will be shared on reasonable request to the first author.



\bibliographystyle{mnras}
\bibliography{main.bib} 




\appendix

\section{Stellar mass recovery for smouldering galaxies} \label{app:masses}

In this section we briefly investigate how well our {\tt Bagpipes} SED-fitting procedure can recover the stellar masses of smouldering galaxies. For this, we apply our main smouldering colour selections, Equations~\ref{eq:MW1}--\ref{eq:supplementary}, to the $z=6$ FLARES galaxies, as was done in the main body of the text. As the stellar masses are known in the simulation (here adopting a 30~kpc aperture), these can be compared to the masses derived following the {\tt Bagpipes} SED-fitting procedure, to test for potential biases in our stellar mass recovery. Mimicking our observational setup in the JADES fields, we utilise the (F090W, F115W, F150W, F200W, F277W, F356W, F444W) wide band filters, together with the (F335M, F410M) medium bands. We fit using the same {\tt Bagpipes} configuration outlined in Section~\ref{subsec:bagpipes} (assuming a double power law SFH parameterisation), fixing the redshift to $z=6$, and assuming a SNR = 10 in each band. Note that we do not perturb the FLARES photometry by this noise,  it simply constitutes our error budget in the fitting procedure. We carry out this process for 10 (or as many as available, if less than this) randomly-selected FLARES smouldering galaxies in 0.25~dex-wide mass bins, from $M_* = 10^{8\text{--}11}~\mathrm{M}_\odot$.

\begin{figure}
\centering
\includegraphics[width=\linewidth]{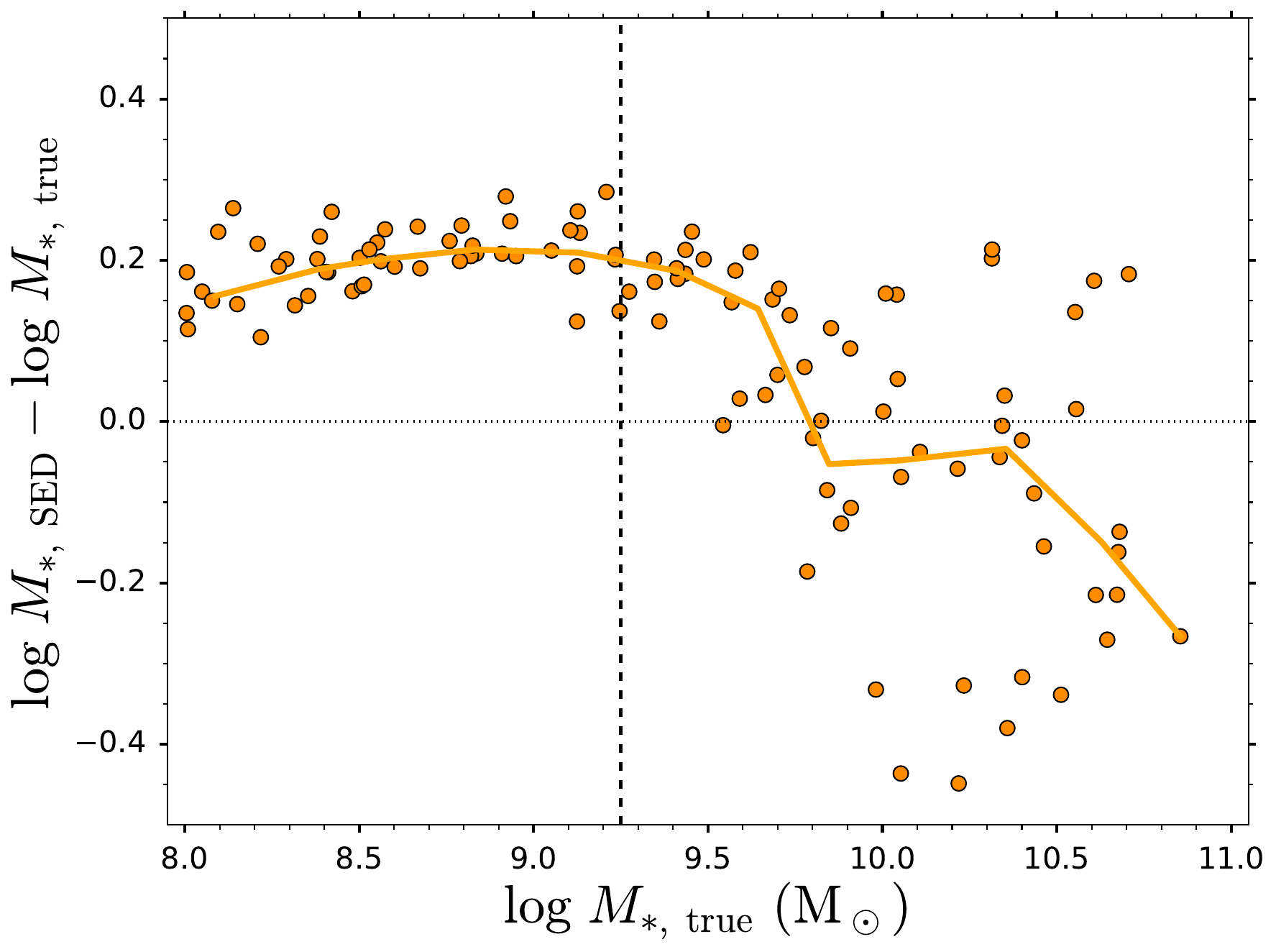}
\caption{The recovery of stellar masses for $z=6$ smouldering galaxies in the FLARES simulations, for individual galaxies (orange circles) and the median in each 0.25~dex-wide mass bin (orange line). Stellar masses derived from our {\tt Bagpipes} SED-fitting procedure ($\log\, M_{*,\  \mathrm{SED}}$) tend to be ${\sim}0.2$~dex overestimates compared to the known stellar mass in the simulation ($\log\, M_{*,\  \mathrm{true}}$) in the mass regime spanned by our observations (left of the vertical dashed line).}
\label{fig:mass_recovery}
\end{figure}

We show the outcome of this procedure in Fig~\ref{fig:mass_recovery}. Stellar mass offsets between the SED-derived masses ($\log\, M_{*,\  \mathrm{SED}}$) and the known masses directly from the simulation ($\log\, M_{*,\ \mathrm{true}}$) for individual FLARES smouldering galaxies are shown by orange circles. The median offset in each 0.25~dex-wide mass bin is traced by the orange lines. We find that in the stellar mass regime spanned by our observations ($M_* \sim 10^{8\text{--}9}~\mathrm{M}_\odot$, i.e.\@ left of the vertical dashed line), the masses of smouldering galaxies in the FLARES simulations tend to be slightly overestimated via our SED-fitting procedure by ${\sim}0.2$~dex. Above this mass threshold, the typical mass offset declines, with the SED-based masses tending to be underestimates at $M_* \gtrsim 10^{10}~\mathrm{M}_\odot$. This mass-dependent trend is in qualitative agreement with \citet{Cochrane2025}, who studied the recovery of stellar masses for general high-redshift galaxies in much greater detail than is done here. They find that the stellar masses are most strongly overestimated for strong emission-line galaxies, where more flux density in the photometry fit is attributed to a Balmer break relative to line emission than in the intrinsic SED, thus increasing the stellar mass inferred. Stellar masses are less overestimated (or perhaps even slightly underestimated) for galaxies with lower EW$_\mathrm{H\alpha}$, like our smouldering galaxies. The relatively small bias in stellar mass recovery for smouldering galaxies would result in a slight shift (to the left, if our simulated results are applicable to the real data) of the stellar mass bins for the comoving number densities presented in Fig.~\ref{fig:smouldering_number_density}. On the other hand, the differing biases in mass recovery for smouldering and control galaxies indicated by \citet{Cochrane2025}, could add further complication to the interpretation of the smouldering fractions presented in Fig.~\ref{fig:smouldering_fraction}.

\bsp	
\label{lastpage}
\end{document}